\documentclass[fleqn,usenatbib]{mnras}

\usepackage{newtxtext,newtxmath}

\usepackage[T1]{fontenc}

\DeclareRobustCommand{\VAN}[3]{#2}
\let\VANthebibliography\thebibliography
\def\thebibliography{\DeclareRobustCommand{\VAN}[3]{##3}\VANthebibliography}

\usepackage{graphicx}	
\usepackage{subcaption}
\usepackage{amsmath}	
\usepackage{enumitem}
\usepackage{pdflscape}

\def\hii{H\,\textsc{ii}}

\def\oii{O\,\textsc{ii}}
\def\oiii{O\,\textsc{iii}}
\def\sii{S\,\textsc{ii}}

\def\ciii{C\,\textsc{iii}}
\def\civ{C\,\textsc{iv}}

\def\heii{He\,\textsc{ii}}
\def\nii{N\,\textsc{ii}}

\def\neiii{Ne\,\textsc{iii}}

\def\ariii{Ar\,\textsc{iii}}

\def\met{$12+\log(\mathrm{O/H})$}
\def\metsl{$12+\log(\mathrm{O/H})_{\mathrm{SL}}$}
\def\mette{$12+\log(\mathrm{O/H})_{T_{\mathrm{e}}}$}

\def\nempg{50}
\def\nempgp{38}
\def\nempgm{24}
\def\nempgovl{11}
\def\nspec{4047}
\def\nspecm{2118}
\def\nspecp{1929}
\def\nsrc{$\sim$2500} 
\def\nmstar{2547}

\def\metminempg{6.7}
\def\metmaxempg{7.3}
\def\zminempg{1.2}
\def\zmaxempg{9.1}

\def\amstk{$z\sim0$ high-sSFR stack}

\usepackage{lineno}

\title[Metallicity Calibration and MZR]{
JADES: the mass-metallicity relation at $z=1-10$. New calibrations, extremely metal-poor galaxies, and chemical diversity
}

\author[Y. Isobe, M. Curti, R. Maiolino et al.]{Yuki Isobe,$^{1,2,3}$\thanks{E-mail: yi264@cam.ac.uk (YI)}
Mirko Curti,$^{4}$
Roberto Maiolino,$^{1,2,5}$
Qiao Duan,$^{1,2}$
William McClymont,$^{1,2}$
D\'avid Pusk\'as,$^{1,2}$
\newauthor
Francesco D'Eugenio,$^{1,2}$
Pierluigi Rinaldi,$^{6}$
James A. A. Trussler,$^{7}$
Jan Scholtz,$^{1,2}$
Tobias J. Looser,$^{7}$
\newauthor
Erica Nelson,$^{8}$
Xihan Ji,$^{1,2}$
Danial Langeroodi,$^{1,2,9}$
Sandro Tacchella,$^{1,2}$
Gareth C. Jones,$^{1,2}$
Ignas Juod\v{z}balis,$^{1,2}$
\newauthor
Robert G. Pascalau,$^{1,2}$
Tiger Yu-Yang Hsiao,$^{10,11}$
Hannah \"Ubler,$^{12}$
William M. Baker,$^{9}$
Andrew J. Bunker,$^{13}$
\newauthor
Stefano Carniani,$^{14}$
St\'ephane Charlot,$^{15}$
Emma Curtis-Lake,$^{16}$
Sophia Geris,$^{1,2}$
Maria Koller,$^{1,2}$
Jianwei Lyu,$^{17}$
\newauthor
Brant Robertson,$^{18}$
Christina C. Williams,$^{19}$
and
Zihao Wu,$^{7}$
\\
$^{1}$Kavli Institute for Cosmology, University of Cambridge, Madingley Road, Cambridge, CB3 0HA, UK\\
$^{2}$Cavendish Laboratory, University of Cambridge, 19 JJ Thomson Avenue, Cambridge, CB3 0HE, UK\\
$^{3}$Waseda Research Institute for Science and Engineering, Faculty of Science and Engineering, Waseda University, 3-4-1, Okubo, Shinjuku, Tokyo 169-8555, Japan\\
$^{4}$INAF, Osservatorio di Astrofisica e Scienza dello Spazio, Via P. Gobetti 93/3, I-40129 Bologna, Italy\\
$^{5}$Department of Physics and Astronomy, University College London, Gower Street, London WC1E 6BT, UK\\
$^{6}$Space Telescope Science Institute, 3700 San Martin Drive, Baltimore, Maryland 21218, USA\\
$^{7}$Center for Astrophysics $|$ Harvard \& Smithsonian, 60 Garden St., Cambridge MA 02138 USA\\
$^{8}$Department for Astrophysical and Planetary Science, University of Colorado, Boulder, CO 80309, USA\\
$^{9}$DARK, Niels Bohr Institute, University of Copenhagen, Jagtvej 155A, 2200 Copenhagen, Denmark\\
$^{10}$Department of Astronomy, The University of Texas at Austin, 2515 Speedway, Austin, Texas 78712, USA\\
$^{11}$Cosmic Frontier Center, The University of Texas at Austin, Austin, TX 78712, USA\\
$^{12}$Max-Planck-Institut f\"ur extraterrestrische Physik (MPE), Gie{\ss}enbachstra{\ss}e 1, 85748 Garching, Germany\\
$^{13}$Department of Physics, University of Oxford, Denys Wilkinson Building, Keble Road, Oxford OX1 3RH, UK\\
$^{14}$Scuola Normale Superiore, Piazza dei Cavalieri 7, I-56126 Pisa, Italy\\
$^{15}$Sorbonne Universit\'e, CNRS, UMR 7095, Institut d'Astrophysique de Paris, 98 bis bd Arago, 75014 Paris, France\\
$^{16}$Centre for Astrophysics Research, Department of Physics, Astronomy and Mathematics, University of Hertfordshire, Hatfield AL10 9AB, UK\\
$^{17}$Steward Observatory, University of Arizona, 933 N. Cherry Avenue, Tucson, AZ 85721, USA\\
$^{18}$Department of Astronomy and Astrophysics, University of California, Santa Cruz, 1156 High Street, Santa Cruz CA 96054, USA\\
$^{19}$NSF National Optical-Infrared Astronomy Research Laboratory, 950 North Cherry Avenue, Tucson, AZ 85719, USA\\
}

\date{Accepted XXX. Received YYY; in original form ZZZ}

\pubyear{\the\year{}}

\begin{document}
\label{firstpage}
\pagerange{\pageref{firstpage}--\pageref{lastpage}}
\maketitle

\begin{abstract}
We present gas-phase metallicities of star-forming galaxies at $z=1$--10 with deep \textit{JWST}/NIRSpec 
spectra from the JADES full data release, Dark Horse, and OASIS programmes.
We stack $\sim$1500 medium-resolution spectra, yielding detections of 
the [\oiii]$\lambda$4363 auroral line
down to $12+\log(\mathrm{O/H})=7.0$ to
derive stack-based strong-line calibrations 
over the metallicity range  $12+\log(\mathrm{O/H})=7.0$--8.7.
At a fixed metallicity, our stacks exhibit [\oiii]$\lambda$5007/H$\beta$ and [\oiii]$\lambda$5007/[\oii]$\lambda\lambda$3726,3729 values generally lower than 
calibrations based on high-$z$ individual auroral-line emitters, suggesting an observational bias towards higher excitation introduced when requiring auroral line detections in individual spectra.
Based on our new calibrations, we obtain canonical mass-metallicity relations (MZRs) at z$=$1--10, identifying a decrease in metallicities from $z\sim0$
to z$\sim$4--10, without significant change in slope.
Moreover, we identify \nempg\ promising candidates of extremely metal-poor galaxies (EMPGs) with $12+\log(\mathrm{O/H})=\metminempg$--\metmaxempg\ (1--4\% solar metallicity) at $z=\zminempg$--\zmaxempg.
The MZRs of EMPGs are characterised by a large scatter,
with those having lower metallicities generally exhibiting lower sSFRs, opposite of what expected from the local Fundamental Metallicity Relation.
These results support a stochastic star-formation history involving gas consumption/ejection and metal-poor inflow, strongly affecting metallicities of low-mass galaxies.
Furthermore, we identify two Little Red Dots in our EMPG candidates, both exhibiting broad H$\alpha$ and prominent Ly$\alpha$, offering insights into the early black-hole growth in extremely metal-poor environments.

\end{abstract}




\section{Introduction} \label{sec:intro}

Understanding the early stages of star formation hinges on the study of chemically pristine galaxies.
The first generation of stars (Population III; PopIII stars) forms from primordial gas composed only of hydrogen (H) and helium (He).
Subsequent stellar populations contain elements heavier than He (so-called metals), enriched by earlier generations of stars.
Metal-poor gas is less efficient at cooling, which can lead to stellar initial mass functions (IMFs) that are more top-heavy than those of local galaxies \citep[e.g.,][]{Hirano2015,Stacy2016,Chon2024}.
Observations of extremely metal-poor galaxies\footnote{Defined to have $12+\log(\mathrm{O/H})\le7.3$ (i.e., 4\% $Z_{\odot}$) in this paper. Note that the metallicity threshold widely adopted at $z\sim0$ is 10\% $Z_{\odot}$ \citep[e.g.,][]{Kunth2000,Izotov2012,Sanchez2016,Kojima2020}. However, we need to optimise the threshold at high $z$ \citep[see][]{Hsiao2025,Trussler2026}, where metallicities below 10\% $Z_{\odot}$ are no longer rare \citep[e.g.,][]{Curti2024,Morishita2024,Isobe2026}.} \citep[EMPGs; e.g.,][]{Kojima2020,Isobe2022} may hint at this initial mode of star formation.

The unprecedented sensitivity of Near-Infrared Spectrograph \citep[NIRSpec;][]{Jakobsen2022} on the \textit{James Webb Space Telescope} (\textit{JWST}; \citealt{jwst_new}) allows us to investigate gas-phase metallicities, quantified by \met, at high redshifts.
The most metal-poor galaxies observed at $z\sim0$ exhibit $12+\log(\mathrm{O/H})=6.8$--7.0 \citep[i.e., $\sim0.01$--0.02$\,Z_{\odot}$;][]{Izotov2018,Kojima2020,Scholte2026}, while \textit{JWST} has recently identified EMPGs at $z>3$ whose inferred metallicities are even lower
\citep[e.g.,][]{Asada2026,Cai2025,Chemerynska2024,Cullen2025,Hsiao2025,Mowla2024,Nakajima2025,Vanzella2023,Vanzella2024,Vanzella2025,Willott2025,Maiolino2025},
possibly even $12+\log(\mathrm{O/H})<6$ \citep{Morishita2025}.
The NIRSpec integral-field unit reveals satellite components whose inferred metallicities are even lower than their main galaxies at $z=8.5$--10.2 \citep{Pascalau2026,Koller2026}.
Remarkably, \citet{Ubler2026} and \citet{Maiolino2026} identify a $z=10.6$ system with H$\gamma$ and \heii$\,\lambda$1640 emission without any metal lines,
suggesting a pristine system whose stellar mass is dominated by PopIII stars \citep{Rusta2026,Jeon2026}.
Intriguingly, even some Little Red Dots (LRDs) potentially hosting overmassive black holes have narrow line ratios that suggest very low metallicity \citep{Maiolino2025,Ivey2026}, which provides new insights into black hole growth in near-pristine environments.

Accurate metallicities for metal-poor galaxies provide a better understanding of early galaxy evolution and the baryon cycle.
Galaxies are thought to evolve as stars produce metals, some of which are expelled from galaxies via outflows, while the accretion of metal-poor gas triggers new star formation, which shapes the scaling relation between stellar mass ($M_{*}$) and metallicity \citep[mass-metallicity relation; MZR; e.g.,][]{Lequeux1979,Tremonti2004,Maiolino2019,BakerMaiolino2023}.
Local observations have established a primary correlation in the MZR, along with a secondary anti-correlation between metallicity and star formation rate \citep[SFR; e.g.,][]{Ellison2008,Mannucci2010}.
Taking this SFR dependence into account yields a tighter scaling relation between metallicity, $M_{*}$, and SFR \citep[so-called Fundamental Metallicity Relation, FMR;][]{Mannucci2010,Andrews2013,Curti2020,Baker2023}, which suggests a balance between star formation, metal production, inflow, and outflow.
However, recent studies report that low-mass galaxies ($\log(M_{*}/M_{\odot})\lesssim9$), both at $z\sim0$ \citep{Laseter2025} and at high redshift \citep[e.g.,][]{Nakajima2023,Langeroodi2023,Curti2024,Nishigaki2025b}, exhibit lower metallicities than those predicted by the FMR, suggesting a breakdown of this picture.
Characterising EMPGs on the MZR and FMR planes will shed light on the physical mechanisms driving the metallicity deficits.

Measuring electron temperature ($T_{\mathrm{e}}$) and electron density ($n_{\mathrm{e}}$) is essential for deriving ion abundance ratios with the physically motivated ``direct-$T_{\mathrm{e}}$'' method \citep[e.g.,][]{Peimbert1967,Osterbrock1989}.
However, auroral lines that are sensitive to $T_{\mathrm{e}}$ are typically very faint.
Indeed, only two high-$z$ EMPG candidates have detections of the [\oiii]$\lambda$4363 auroral line \citep{Mowla2024,Cullen2025} and thus metallicities based on the direct-$T_{\mathrm{e}}$ method (\mette, hereafter).
Metallicities of the other EMPG candidates are derived from empirical
calibrations
between the ratios of strong emission lines (such as $\mathrm{R3}\equiv[$\oiii]$\lambda$5007/H$\beta$) and 
metallicities, generally referred to as the ``strong-line'' method.
Table \ref{tab:idx} defines metallicity-sensitive indices used for the strong-line method.
The strong-line method has been improved with deep local observations \citep[e.g.,][]{Pagel1979,Pettini2004,Nagao2006,Maiolino2008,Marino2013,Curti2020,Nakajima2022} and with photoionisation models \citep[e.g.,][]{McGaugh1991,Zaritsky1994,Kewley2002} over decades, as described in many reviews \citep{Maiolino2019,Kewley2019_review}.
With the advent of \textit{JWST}, it became possible to calibrate the strong-line method exploiting auroral line emitters directly at $z\gtrsim2$--10 \citep{Sanders2024,Sanders2025,Chakraborty2025,Cataldi2025}.

One key challenge is that these samples are mostly limited to relatively metal-rich ($Z\gtrsim5$\% $Z_{\odot}$) galaxies.
To avoid sample scarcity and metallicity uncertainty, \citet{Sanders2025} quantify a valid metallicity range within $12+\log(\mathrm{O/H})=7.3$--8.6.
Beyond the valid range, the metallicity measurement inevitably introduces significant systematics arising from both the uncertainty in the calibration itself and in the extrapolation.
One major way of extrapolation is to simply adopt the same function calibrated within the valid range \citep[e.g.,][]{Vanzella2025,Hsiao2025,Asada2026}, while no physical model underlies this extrapolation.
Another possibility is to use photoionisation models that match the low-metallicity end of the observed calibration \citep[e.g.,][]{Morishita2025,Nakajima2025}, while the adopted ionisation parameter is quite high ($\log(U)=-0.5$).

An additional issue is that galaxies with detectable auroral lines may be biased towards systems with bright emission lines.
At $z\sim0$, \citet{Sanders2021} point out the possibility that requiring auroral-line detections biases the sample to high excitation and large equivalent widths (EW) of emission lines.
Indeed, \citet{Nakajima2022} separate local low-metallicity galaxies into subsamples with large restframe EWs of H$\beta$ (EW$_{0}$(H$\beta)>200$ \AA) and small EW$_{0}$(H$\beta)$ ($<$100 \AA), and reported that the R3 and O32 values of the large-EW subsample are significantly higher than those of the small-EW subsample.
\citet{Langeroodi2026} find that this trend persists even including high-$z$ sources.
They propose a practical approach using EW$_{0}$(H$\beta)$ values to mitigate the systematics in metallicity measurements, while it may become difficult to measure EWs for fainter, more metal-poor sources, in that the stellar continuum is often undetected.

This paper presents an alternative approach: spectral stacking to enable auroral-line detections without requiring them for individual sources.
Several studies have put effort on spectral stacking of $z\sim0$ galaxies \citep[e.g.,][]{Brown2016,Curti2017,Bian2018} to detect auroral lines, which are used for recalibrating the strong-line method.
Thanks to the progress of the NIRSpec observations, it is timely to carry out this analysis for high-$z$ galaxies.
Using the large NIRSpec dataset of the \textit{JWST} Advanced Deep Extragalactic Survey \citep[JADES;][]{Bunker2024,DEugenio2025}, combined with our established spectral stacking technique \citep[][]{Isobe2025},
we have demonstrated that we robustly detect [\oiii]$\lambda$4363 from the stacks of hundreds of spectra at $z=4$--7, whose \mette\ are down to 7.45 \citep{Isobe2026}.
Leveraging the full NIRSpec dataset of JADES Data Release 4 \citep[DR4;][]{Curtis-Lake2025_DR4,Scholtz2025_DR4} and deep JADES Dark Horse (DH) data targeting faint sources \citep{DEugenio2025_DH}, we expect
[\oiii]$\lambda$4363 detections from stacks of even more metal-poor systems.
The current high-$z$ strong-line calibrations are not well constrained at $12+\log(\mathrm{O/H})\lesssim7.3$ as discussed by \citet{Sanders2025}, highlighting the requirement of performing spectral stacking to extend the valid metallicity range.

The paper is organised as follows.
We describe our data and sample in Section \ref{sec:datsamp}, analysis in Section \ref{sec:ana}, our metallicity calibrations in Section \ref{sec:calib}, high-$z$ EMPG candidates in Section \ref{sec:empg}, SFR-$M_{*}$ relations in Section \ref{sec:msfr}, results and discussions of the MZR and FMR in Section \ref{sec:mzrfmr}, and conclusions in Section \ref{sec:con}.
We assume a standard $\Lambda$CDM cosmology with parameters of $\Omega_{0}=0.315$ and $H_{0}=67.4\ \mathrm{km\ s^{-1}\ Mpc^{-1}}$ \citep{Planck2020}.
Throughout the paper, we use the solar abundance ratios of \citet{Asplund2021}, where the gas-phase solar metallicity $Z_{\odot}$ is $12+\log(\mathrm{O/H})=8.69$.

\section{Data and Sample} \label{sec:datsamp}
\begin{table}
	\centering
	\caption{Definition of strong-line indices. The pair of two wavelengths connected by a comma indicate the sum of the emission-line fluxes at each wavelength.
    $^{\mathrm{a}}$: \citet{Laseter2024},
    $^{\mathrm{b}}$: \citet{Chakraborty2025},
    $^{\mathrm{c}}$: \citet{Cataldi2025},
    $^{\mathrm{d}}$: \citet{Scholte2025}
    }
	\label{tab:idx}
	\begin{tabular}{lc}
		\hline
        Index & Definition\\
        \hline
        log(R3) & log([\oiii]$\lambda$5007/H$\beta$)\\
        log(R2) & log([\oii]$\lambda\lambda$3726,3729/H$\beta$)\\
        log(R23) & log(([\oii]$\lambda\lambda$3726,3729\,+\,[\oiii]$\lambda$4959,5007)/H$\beta$)\\
        log(O32) & log([\oiii]$\lambda$5007/[\oii]$\lambda\lambda$3726,3729)\\
        $\hat{\mathrm{R}}_{\mathrm{Laseter}}$\,$^{\mathrm{a}}$ & $0.47\times\log(\mathrm{R2})+0.88\times\log(\mathrm{R3})$\\
        $\hat{\mathrm{R}}_{\mathrm{Chakraborty}}$\,$^{\mathrm{b}}$ & $0.18\times\log(\mathrm{R2})+0.98\times\log(\mathrm{R3})$\\
        $\tilde{\mathrm{R}}_{\mathrm{Cataldi}}$\,$^{\mathrm{c}}$ & $0.46\times\log(\mathrm{R2})+0.88\times\log(\mathrm{R3})$\\
        log(Ne3) & log([\neiii]$\lambda$3869/H$\gamma$)\\
        log(RO2Ne3) & log(([\oii]$\lambda\lambda$3726,3729\,+\,[\neiii]$\lambda$3869)/H$\delta$)\\
        $\widehat{\mathrm{RNe}}$\,$^{\mathrm{d}}$ & $0.47\times\log(\mathrm{[\text{\oii}]\lambda\lambda3726,3729}/\mathrm{H}\gamma)+0.88\times\log(\mathrm{Ne3})$\\
        log(Ne3O2) & log([\neiii]$\lambda$3869/[\oii]$\lambda\lambda$3726,3729)\\
        log(N2) & log([\nii]$\lambda$6583/H$\alpha$)\\
        log(O3N2) & $\log(\mathrm{R3})-\log(\mathrm{N2})$\\
        log(S2) & log([\sii]$\lambda\lambda$6716,6731/H$\alpha$)\\
        log(O3S2) & $\log(\mathrm{R3})-\log(\mathrm{S2})$\\
        log(Ar3) & log([\ariii]$\lambda$7135/H$\alpha$)\\
        \hline
	\end{tabular}
\end{table}

We analyse NIRSpec spectroscopic data obtained with the NIRSpec micro-shutter array (MSA; \citealt{Jakobsen2022}; \citealt{Ferruit2022}) and related value-added catalogues of the following \textit{JWST} programmes in the GOODS-S and GOODS-N fields: the JADES \citep[PIDs 1180, 1181, 1210, 1286, 1287, and 3215;][]{Bunker2024,DEugenio2025,Eisenstein2023b,Curtis-Lake2025_DR4,Scholtz2025_DR4,Robertson2026_DR5}, the JADES DH\footnote{The DH programme involved re-designed observations from the allocated 3215 UltraDeep programme, which were re-planned due to the original observations being affected by detector shorts.} \citep[][]{DEugenio2025_DH}, and the Observing All phases of StochastIc Star formation survey (OASIS; PID 5997, PIs: T. Looser \& F. D'Eugenio; Looser et al. in prep.).

\subsection{JADES NIRSpec data} \label{subsec:dr4}
We use the complete NIRSpec dataset of the JADES DR4 \citep[][]{Curtis-Lake2025_DR4,Scholtz2025_DR4}.
Since these DR4 papers describe the JADES observations and data reduction, we summarise only the main points here.
The JADES observed 5190 targets in the GOODS-S and GOODS-N fields with the low-resolution prism ($R\sim100$; R100, hereafter) and most of these were also observed with the three medium-resolution gratings ($R\sim1000$; R1000, hereafter): F070LP-G140M, F170LP-G235M, and F290LP-G395M.
Long exposure times of $\gtrsim20$ hours were allocated for $\sim700$ sources, and the remaining $>4400$ sources were observed with exposure times of $\sim1$--3 hours per setting \citep{Curtis-Lake2025_DR4}, and the integration times for R100 were generally longer than those for R1000 by a factor of 1--4 \citep{DEugenio2025}.
The observed data were reduced with the pipeline constructed by the European Space Agency (ESA) NIRSpec Science Operations Team \citep{Ferruit2022} and the NIRSpec Guaranteed Time Observations (GTO) Team \citep{Oliveira2018}.
The standard pipeline conservatively accounts for correlated noise by incorporating error propagation and employing variance-conserving resampling \citep{Dorner2016}.
The DR4 one-dimensional spectra were extracted with both the default full microshutter aperture of 5 pixels (0.5'') and the small box-car aperture of 3 pixels.
Although the 3-pixel extraction provides higher $S/N$ for compact sources \citep{Scholtz2025_DR4}, which may be more optimal for high-$z$ EMPGs, we use the 5-pixel extraction aperture for more conservative selection of EMPG candidates.
The DR4 provides a flux catalogue\footnote{\url{https://jades.herts.ac.uk/search/}} based on the spectral fitting softwares \textsc{ppxf} \citep{Cappellari2023} and \texttt{QubeSpec}\footnote{\url{https://github.com/honzascholtz/Qubespec}} for the R100 and R1000 data, respectively.
In addition, the flux catalogue lists five quality flags of redshifts, where Flags `A', `B', and `C' indicate redshifts from emission lines and/or the continuum while Flags `D' and `E' mean tentative and no redshifts, respectively.
We use 3297 sources with robust redshift flags of `A', `B', and `C', discarding sources with Flags `D' and `E'.

\subsection{DH NIRSpec data} \label{subsec:dh}
We summarise the DH observations and dataset here (\citealt{DEugenio2025_DH} for more details).
The DH observations were designed to demonstrate NIRSpec/MSA `dense shutter spectroscopy', i.e. dramatically increasing the multiplex of MSA grating spectroscopy by deliberately allowing a large number of spectral overlaps
on the detector. The sample consists of all $z>3$ candidates in the MSA footprint with faint continuum and falling inside an operable micro-shutter slit.
The DH observed $\sim850$ sources in the GOODS-S field with long exposure times of 9--10 hours for the two R1000 gratings of F170LP-G235M and F290LP-G395M.
The data reduction procedure follows the standard methods of the JADES data reduction with the 5-pixel (0.5'') aperture.
The emission-line fluxes are measured using the emission-line fitting procedure for the DR4 R1000 data \citep{Scholtz2025_DR4}.
The final catalogue comprises 519 sources with robust redshifts from multiple emission lines.

\subsection{OASIS NIRSpec data} \label{subsec:oasis}
Here we briefly describe the OASIS observations and dataset; full details will be presented in Looser et al. (in prep.). OASIS observed 332 targets in the GOODS-S field using the R100 prism with a total exposure time of 28 hours.
The emission-line fluxes were measured using the emission-line fitting procedure for the DR4 R100 data \citep{Scholtz2025_DR4}.
The final catalogue comprises 261 sources with robust redshifts.

\subsection{Photometry catalogue and inferred galaxy properties} \label{subsec:phot}

Deep imaging of the GOODS-S and GOODS-N fields was obtained with \textit{JWST}/Near-Infrared Camera (NIRCam) as part of the JADES survey.
The JADES Data Release 5 (DR5) NIRCam dataset builds on the primary JADES programmes (PIDs 1180, 1181, 1210, 1286, 1287 and 4540), supplemented with additional public NIRCam imaging from various community surveys in the GOODS fields \citep{Johnson2026}\footnote{
The JADES Origins Field (PID 3215),
MIRI Deep Imaging Survey \citep[MIDIS; PID 1283;][]{Ostlin2025},
First Reionization Epoch Spectroscopically Complete Observations \citep[FRESCO; PID 1895;][]{Oesch2023},
the \textit{JWST} Extragalactic Medium-band Survey \citep[JEMS; PID 1963;][]{Williams2023},
the Next Generation Deep Extragalactic Exploratory Public (NGDEEP) Survey \citep[PID 2079][]{Bagley2024},
PANORAMIC \citep[PID 2514][]{Williams2025},
Complete NIRCam Grism Redshift Survey (CONGRESS; PID 3577),
the Bias-free Extragalactic Analysis for Cosmic Origins with NIRCam \citep[BEACON; PID 3990;][]{Morishita2025_beacon},
the Public Observation Pure Parallel Infrared Emission-Line Survey (POPPIES; PID 5398),
Observing All phases of StochastIc Star formation survey (OASIS; PID 5997; Looser et al. in preparation),
Slitless Areal Pure-Parallel HIgh-Redshift Emission Survey \citep[SAPPHIRES; PID 6434;][]{Sun2025b},
and the Director’s Discretionary transient follow-up program (PID 6541).
}.
The DR5 catalogues further incorporate imaging of the \textit{Hubble Space Telescope} (\textit{HST}) from the Hubble Legacy Fields mosaics, including the Advanced Camera for Surveys (ACS)/F435W, F606W, F775W, F814W, F850LP, and the Wide Field Camera 3 (WFC3)/F105W, F125W, F140W, and F160W \citep{Illingworth2016,Whitaker2019}.
\citet{Robertson2026_DR5} conduct source detections and photometric measurements to create the JADES DR5 NIRCam photometry catalogue.

This rich JADES DR5 photometric dataset enables us to perform spectral energy distribution (SED) fitting to infer galaxy properties.
Full details of the model setup and prior descriptions are presented in the JADES DR5 stellar population catalogue paper by \citet{Duan2026}, so only the main points are described below.

We use the \texttt{Prospector} \citep[][]{Johnson2019,Johnson2021} spectral energy distribution (SED) modeling code to infer galaxy properties, and following a similar methodology as in \citet{Tacchella2022} and \citet{Simmonds2024b}. Our model consists of 18 free parameters and covers a wide range of galaxy properties, including stellar mass, star-formation history (SFH), nebular emission, dust attenuation and emission, and contributions of active galactic nuclei (AGN).
Note that, as we only use the NIRCam, ACS, and WFC3 photometry and exclude the MIRI data, the contributions from dust emission and AGN are negligible.
Our \texttt{Prospector} models adopt the MIST stellar isochrones \citep{choi2016, dotter2016}, the MILES stellar spectral library \citep{sanchez2006, miles_2011}, and a \citet{chabrier2003imf} IMF.

We fix the redshift to spectroscopic values. We adopt the stellar mass function prior \citep{wang2023}, where lower prior probabilities are assigned to higher-mass galaxies to avoid spurious high-mass solutions. The stellar metallicity is linked to the stellar mass following the \citet{gallazzi2005} mass-metallicity relation, adopting a Gaussian prior with a mass-dependent mean and a scatter that increases toward lower stellar masses. Note that stellar metallicities do not significantly impact stellar mass measurements. We use a non-parametric SFH, specifically the continuity SFH model \citep{Leja2019}, with 7 age bins.
The first age bin spans the lookback time from 0 to 20--30 Myr, which depends on redshift (\citealt{Duan2026} for more details).
For this model, we implement the physically motivated star-forming main sequence (SFMS) SFH prior introduced in \citet{Duan2026}. This prior assumes that galaxies form and evolve along the main sequence, with substantial scatter around SFMS to capture sudden starburst and quenching events. The gas-phase metallicity follows a uniform prior between $-2 < \log(Z_{\rm gas}/Z_\odot) < 0.5$.

Dust attenuation follows the two-component model of \citet{charlot2000}, where the diffuse component optical depth $\tau_{\mathrm{dust,2}}$ follows gaussian distributions with $\mathcal{N} (\mu, \sigma) = \mathcal{N}(0.3, 1)$ truncated to $[0, 4]$, and the birth-cloud to diffuse ratio $\tau_{\mathrm{dust,1}}/\tau_{\mathrm{dust,2}} \sim \mathcal{N}(1, 0.3)$ truncated to $[0, 2]$. The attenuation curve slope of the diffuse component is parameterised by a \citet{Calzetti2000} power-law index $n$, drawn from a uniform prior over $[-1.2, 0.4]$.

We crossmatch our spectroscopic sources with the JADES DR5 photometric sources with a maximum angular separation of 0.5''.
We also require that the spectroscopic redshifts have reliable quality flags of 'A', 'B', or 'C'. A detailed description
of the construction and homogenisation of these spectroscopic samples can be
found in \citet{Puskas2025a} and \citet{Puskas2025b}.
This crossmatch connects the spectroscopic measurements and the global galaxy properties such as $M_{*}$ and SED-based SFRs (SFR$_{\mathrm{SED}}$).
In this paper, we use SFRs averaged over 20 Myr (SFR$_{20}$) and 100 Myr (SFR$_{100}$).

\subsection{Parent sample} \label{subsec:parent}
From the sources with robust redshifts, we select $z=1$--10 sources that have H$\beta$ detections with $S/N>3$.
We exclude Type-1 AGN with the significant H$\alpha$ broadening identified in the R1000 data: 30 reported from JADES \citep{Juodzbalis2025} and 9 from DH \citep{DEugenio2025_DH}.
Note that these studies do not survey $\sim$30\% of our spectroscopic sources, indicating that our spectroscopic sample contains some unidentified Type-1 AGN.
However, given that the number of the reported Type-1 AGN is only $\lesssim1$\% of the surveyed sources, the impact of unidentified Type-1 AGN on the stacked results and on the overall sample distribution must be negligible.

We obtain \nspec spectra as our parent sample, which consists of \nspecp\ spectra of the R100 data and \nspecm\ spectra of the R1000 data.
Crossmatching our parent sample to the JADES DR5 photometric sources (Section \ref{subsec:phot}), we find that the number of unique JADES DR5 photometric sources in our parent sample is \nmstar.

\section{Analysis} \label{sec:ana}
\begin{figure*}
	\centering
    \includegraphics[width=\textwidth]{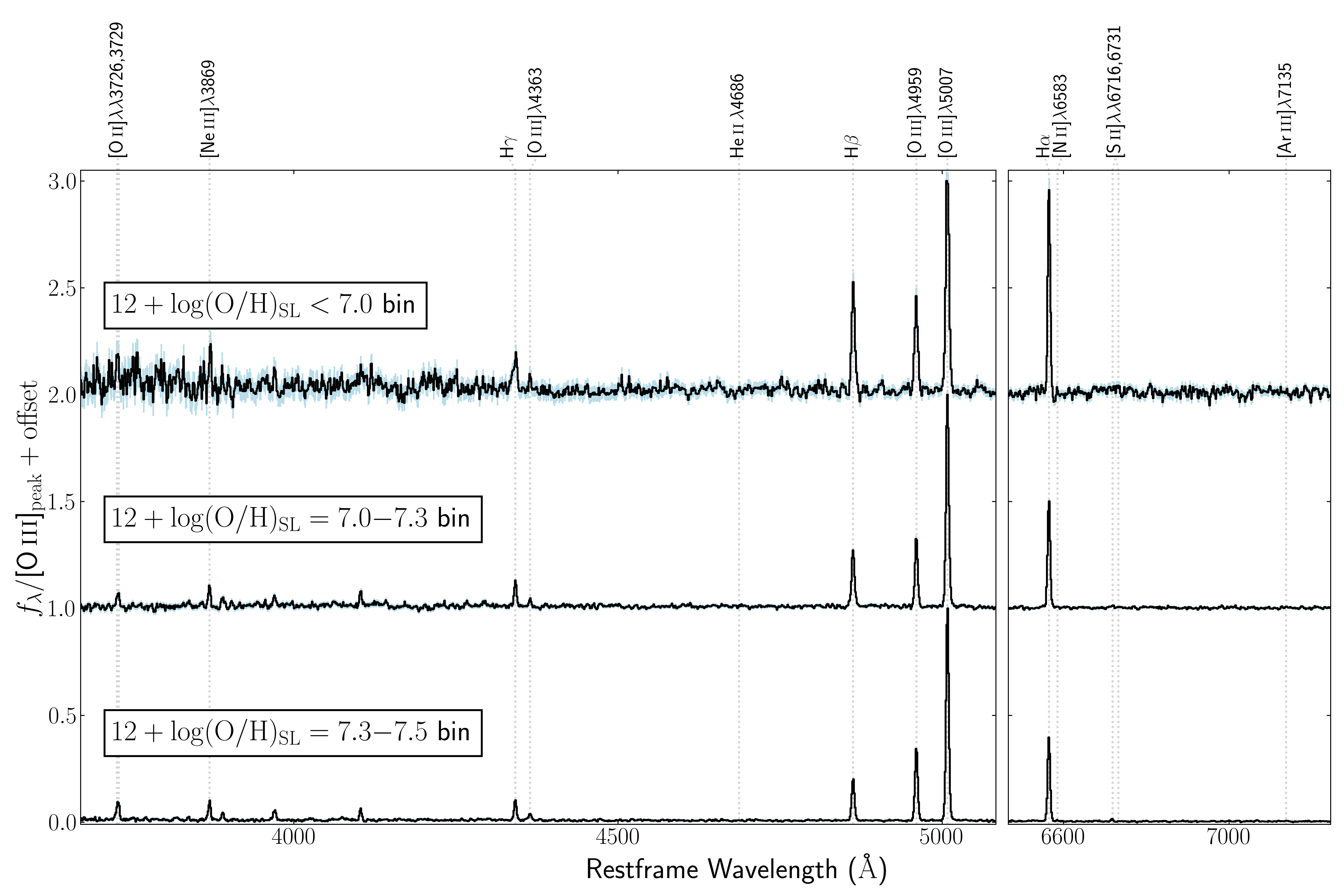}
    \caption{
    Stacked spectra in bins of strong-line based metallicities of $12+\log(\mathrm{O/H})_{\mathrm{SL}}<7.0$ (31 sources; top), $12+\log(\mathrm{O/H})_{\mathrm{SL}}=7.0$--7.3 (105 sources; middle), and $12+\log(\mathrm{O/H})_{\mathrm{SL}}=7.3$--7.5 (135 sources; bottom).
    The \metsl\ values for these stacks are based on \citet{Cataldi2025}'s calibration, while we recalibrate the strong-line method by measuring $T_{\mathrm{e}}$-based metallicities of the stacks.
    All these stacks have the detection of [\oiii]$\lambda$4363 with $S/N>3$.
    }
    \label{fig:stkspec}
\end{figure*}

\subsection{Dust reddening of individual sources} \label{subsec:ebv}

We measure nebular colour excesses $E(B-V)$ from Balmer decrements.
We use H$\alpha$/H$\beta$ ratios for sources whose H$\alpha$ is covered by NIRSpec, otherwise we use H$\gamma$/H$\beta$ ratios.
We refer to the theoretical ratios of Balmer lines under the assumption of Case B recombination at $T_{\mathrm{e}}=15000$ K and $n_{\mathrm{e}}=100$ cm$^{-3}$ ($\mathrm{H}\alpha/\mathrm{H}\beta=2.86$, $\mathrm{H}\gamma/\mathrm{H}\beta=0.47$; \citealt{Osterbrock1989}).
We assume \citet{Cardelli1989}'s extinction law to correct emission-line fluxes for dust reddening, following other high-$z$ strong-line calibrations from the literature \citep{Sanders2024,Sanders2025,Cataldi2025}.
Note that the shape of \citet{Cardelli1989}'s extinction curve agrees with the mean nebular attenuation curve of $z\sim2$ galaxies \citep{Reddy2020}, while a dedicated study will be required to investigate nebular attenuation curves at higher $z$ \citep[][also \citealt{Markov2025,Markov2025b,Maheson2025,Shivaei2025}]{Sanders2025b}.
No correction is applied to sources lacking Balmer decrement measurements, while it is worth noting that the metallicity indices used for our \metsl\ measurements, such as R3 and Ne3O2 (Section \ref{subsec:met}), are based on emission lines with small wavelength separations and thus insensitive to variations in $E(B-V)$.
We further verify that the fraction of such sources remains below 10\% in all stellar-mass bins (0.5 dex width) over the range of $\log(M_{*}/M_{\odot})=6.5$--10.0, indicating a negligible impact on the overall sample distribution.

\subsection{Strong-line based metallicity measurement} \label{subsec:met}
We measure gas-phase metallicities via the strong-line method ($12 + \log(\mathrm{O/H})_\mathrm{SL}$, hereafter)
following the procedure developed by \citet{Curti2024}.
Emission-line fluxes are first corrected for dust attenuation
using the $E(B-V)$ values derived from the Balmer decrement (Section~\ref{subsec:ebv})
and the extinction law of \citet{Cardelli1989}. The dereddened fluxes are then passed to the
metallicity fitting framework.

We combine up to four metallicity-sensitive indices simultaneously when their constituent emission
lines are detected at $S/N > 3$: R3, $\tilde{R}_{\mathrm{Cataldi}}$, O32, and Ne3O2 (Table \ref{tab:idx}).
These indices are all based on oxygen or neon lines and are therefore insensitive to deviation and potential
anomalies in other chemical abundance patterns, such as strong N/O enhancement \citep[e.g.][]{Cameron2023, Isobe2023b, Topping2024a}.
For objects where R3 is the sole available diagnostic, the branch degeneracy is broken via a
hierarchical set of constraints. First, we exploit $3\sigma$ upper limits on
$[\mathrm{O\,\textsc{ii}}]$ and $[\mathrm{N\,\textsc{ii}}]$ --- used as bounds on R2 and N2,
respectively --- to restrict the allowed metallicity range to either the upper or lower branch of
the calibration. In cases where this is insufficient to resolve the degeneracy, we additionally
impose a last-resort prior based on the gas-phase metallicity inferred from SED fitting
(Section~\ref{subsec:phot}): if the posterior still straddles the turnover of the calibration, the
SED-based metallicity estimate is used to select the lower or upper branch accordingly.

We initially compute $12 + \log(\mathrm{O/H})_\mathrm{SL}$ for all individual sources in our
parent sample using the \citet{Cataldi2025} calibrations, applying a linear extrapolation below
$12 + \log(\mathrm{O/H}) = 7$ with the slope fixed at the value of the calibration at that
boundary. These preliminary metallicities are used solely to assign individual galaxies to
metallicity bins for spectral stacking (Section~\ref{subsec:subsamp}). After deriving our
stack-based calibrations (Section~\ref{sec:calib}), we recompute
$12 + \log(\mathrm{O/H})_\mathrm{SL}$ for all sources adopting the new calibrations, and it is
these recalculated values that are used throughout the rest of the paper.

For each galaxy, the metallicity is inferred via a Markov Chain Monte Carlo (MCMC) approach using
the \textsc{emcee} ensemble sampler \citep{ForemanMackey2013}, with 48 walkers, a burn-in phase
of 600 steps, and 3000 production steps thinned by a factor of 20. The log-likelihood function
assumes a Gaussian likelihood for each diagnostic, with a total uncertainty equal to the
quadrature sum of the observational uncertainty and the intrinsic scatter of the calibration.
A flat (uniform) prior is adopted over the allowed metallicity range $[Z_\mathrm{lo}, Z_\mathrm{hi}]$,
defined by the intersection of the validity domains of all diagnostics entering the fit. The
reported metallicity value is the median of the resulting posterior probability distribution
function (PDF), with the $1\sigma$ confidence interval defined by the 16th and 84th percentiles.

We apply several quality-assessment criteria to the posterior distributions. Measurements with a
$1\sigma$ confidence interval wider than 1~dex are considered poorly constrained and excluded from
further analysis. In addition, we flag and exclude metallicity measurements whose posterior PDF
exhibits a bimodal shape --- identified via a peak-finding algorithm applied to a Gaussian kernel density estimation of
the posterior samples --- and whose median value falls near a boundary of the prior. We further
flag posteriors that are leaning against the lower or upper boundary of the prior range, defined
as having more than 15~per~cent of the posterior samples within 0.02~dex of either bound.

\subsection{Subsample for spectral stacking} \label{subsec:subsamp}
\begin{figure*}
	\centering
    \includegraphics[width=\textwidth]{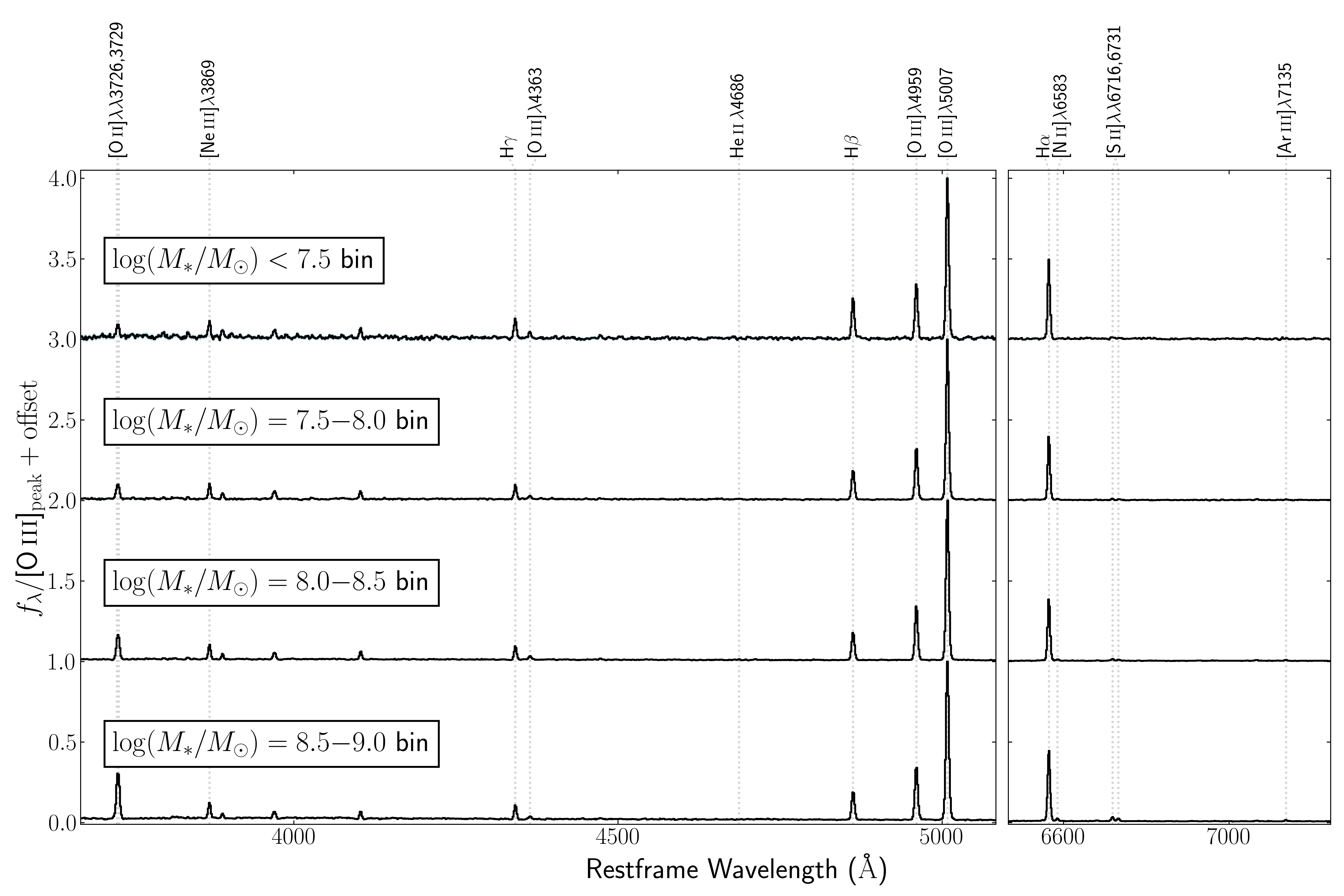}
    \caption{Same as Figure \ref{fig:stkspec} but for the stacked spectra of $\log(M_{*}/M_{\odot})<7.5$ (133 sources; top), $\log(M_{*}/M_{\odot})=7.5$--8.0 (322 sources; second top), $\log(M_{*}/M_{\odot})=8.0$--8.5 (517 sources; second bottom), and $\log(M_{*}/M_{\odot})=8.5$--9.0 (482 sources; bottom).
    All these stacks have the detection of [\oiii]$\lambda$4363 with $S/N>3$.
    }
    \label{fig:stkspec_mbin}
\end{figure*}

We construct subsamples of our R1000 sources for spectral stacking and the stack-based metallicity calibrations that focus on low metallicities.
There are two possible ways to construct low-metallicity samples without requiring auroral lines: selecting sources with low \metsl\ ($Z$-bin sample, hereafter) or low stellar masses ($M$-bin sample, hereafter).
Given that many simulations predict a large scatter of the MZR especially below $M_{*}\sim10^{8}\,M_{\odot}$ \citep[e.g.,][]{McClymont2025c,Pallottini2025,Liu2025}, the $Z$-bin sample and the $M$-bin sample may be significantly biased to the sources below and above the MZR, respectively.
We thus use both the $Z$-bin sample and the $M$-bin sample for our analysis.

We construct the $Z$-bin sample with the \metsl\ based on the calibrations of \citet{Cataldi2025} in bins of \metsl\,$<$\,7.0, $=$7.0--7.3, and $=$7.3--7.5 (Section \ref{subsec:met}).
To ensure the robustness of the metallicity measurements, we remove the following sources:
\begin{itemize}[leftmargin=*]
    \item H$\alpha$/H$\beta<2.5$, which is significantly below the Case B value.
    While the assumption of density-bounded nebulae or optically-thick, excited \hii\ clouds can decrease H$\alpha$/H$\beta$ \citep[][see also \citealt{Ivey2026a}]{Yanagisawa2024b,McClymont2025}, the galaxies discussed in these studies still exhibit H$\alpha$/H$\beta>2.5$ \citep[see also][]{Scarlata2024}.
    Explaining values below this threshold would require more extreme physical conditions.
    \item AGN candidates probed with emission lines.
    As we already remove the Type-1 AGN reported by \citet{Juodzbalis2025} and \citet{DEugenio2025_DH} from our parent sample (Section \ref{subsec:parent}), here we exclude Type-2 AGN candidates.
    Following the procedure of \citet{Scholtz2025}, we select Type-2 AGN candidates based on the N2 vs. R3 diagram \citep[a.k.a. the BPT diagram;][]{Baldwin1981} and the S2 vs. R3 diagram \citep{VO87} with the demarcation lines of \citet{Scholtz2025}, the N2 vs. \heii$\lambda$4686/H$\beta$ diagram with a demarcation line of \citet{Shirazi2012}, the \ciii]$\lambda\lambda$1907,1909/\heii$\lambda$1640 vs. \civ$\lambda\lambda$1549,1551/\ciii]$\lambda\lambda$1907,1909 diagram with a demarcation line of \citet{Hirschmann2023}.
    In addition, we select Type-2 AGN candidates with strong [\oiii]$\lambda$4363 by using demarcation lines on the [\oiii]$\lambda$4363/H$\gamma$ vs. O32 diagram, the [\oiii]$\lambda$4363/H$\gamma$ vs. Ne3O2 diagram, and the [\oiii]$\lambda$4363/H$\gamma$ vs. [\oiii]$\lambda$4363/[\oiii]$\lambda$5007 proposed by \citet{Mazzolari2024} \citep[also][]{Jones2025}.
    As we focus on restframe-optical emission-line properties, we do not exclude obscured AGN candidates exhibiting significant excess in NIR photometry while not classified as Type-1 or Type-2 AGN \citep{Rinaldi2025}.
    \item Evolved galaxies (e.g., with a Balmer break), which are not fully excluded with the R1000 data only but identified with the R100 data because the R100 prism has higher continuum sensitivity than the R1000 gratings.
    We select the evolved galaxies by visual inspection because the quantification with the Balmer break index \citep[e.g.,][]{Kriek2006,Curtis-Lake2023} does not sufficiently work for our sample galaxies with faint Balmer breaks (also Looser et al. in prep.).
    This selection misses a population of galaxies with both a Balmer break and emission lines, suggesting a rejuvenating SFH such as A2744-YD4 \citep[][]{Witten2025}, while we note that the observed R23 and O32 values of A2744-YD4 cannot be simultaneously reproduced by photoionisation models \citep{Witten2025}, suggesting the need for a focused study of metallicity measurements in such rejuvenating galaxies.
\end{itemize}

We construct the $M$-bin sample by selecting sources with the $M_{*}$ measurements (Section \ref{subsec:phot}) in four bins of $\log(M_{*}/M_{\odot})<7.5$, $=$7.5--8.0, $=$8.0--8.5, and $=$8.5--9.0.
It is worth noting that \citet{Geris2026} identify the H$\alpha$ broadening in the JADES NIRSpec stack with $\log(M_{*}/M_{\odot})=9.1$--9.8, suggesting a non-negligible population of hidden broad-line AGN in the sample of galaxies with $M_{*}\gtrsim10^{9}\ M_{\odot}$.
Table \ref{tab:fund} lists the numbers of the sources and median properties in the bins of the $Z$-bin and $M$-bin samples.
We use 1488 unique R1000 sources for these stacks.

\subsection{Spectral stacking} \label{subsec:stk}

We produce stacked spectra in the same manner as \citet[][also see \citealt{Isobe2026,Geris2026,Zucchi2026}]{Isobe2025}.
We stack the R1000 data to resolve H$\gamma$ and [\oiii]$\lambda$4363, which are essential for the direct-$T_{\mathrm{e}}$ method.
We perform Gaussian fitting to [\oiii]+H$\beta$ for each source to obtain redshifts and [\oiii]$\lambda$5007 fluxes.
We use \texttt{spectres} \citep{Carnall2017} to resample the spectra into a common restframe wavelength grid within 1200--7500 \AA, whose spectral pixel size is half of the full-width half maximum of the NIRSpec line-spread function (LSF) for point sources \citep{deGraaff2024}.
The spectral pixel size is derived for the median redshift of the stacked sample.

We normalise the individual spectra with respect to their [\oiii]$\lambda$5007 fluxes and calculate the median flux value of the individual spectra at each common spectral pixel.
This spectral stacking method reduces the influence of a small number of bright outliers, thereby enabling the resulting spectrum to represent general properties of the stacked sample.
Indeed, \citet{Isobe2026} show that this stacking method produces stacked spectra whose [\oiii]$\lambda$5007/H$\beta$ value is consistent with the median [\oiii]$\lambda$5007/H$\beta$ ratio of the stacked sources, suggesting little bias towards a bright galaxy population.
The error spectrum for the stack is estimated by calculating the standard deviation of 1000 stacked spectra at each wavelength, each constructed by perturbing the individual spectra according to their measurement errors.
We limit the analysis of each stacked spectra to the wavelength range that is contributed by more than 70\% of the input spectra, so that even the most widely separated emission lines in each stacked spectrum originate from averaging over similar parent galaxy samples.
We refer to this wavelength range as a valid restframe wavelength ($\lambda_{\mathrm{rest}}$) range, which is shown in Table \ref{tab:fund}.
We confirm that the valid $\lambda_{\mathrm{rest}}$ range of each stack adequately covers all the necessary emission lines from [\oii]$\lambda\lambda$3726,3729 to [\ariii]$\lambda$7135.
Figure \ref{fig:stkspec} shows a representative stacked spectrum for the \metsl\,$<7.0$ bin, exhibiting a clear signal at the location of the [\oiii]$\lambda$4363 emission line.

\subsection{Emission line analysis of the stacks} \label{subsec:emis}

\begin{table}
	\centering
	\caption{References on transition probabilities and collision strengths.}
	\label{tab:atom}
	\begin{tabular}{lcc}
		\hline
        Ion & Transition probability & Collision Strength\\
        \hline
		H$^{0}$ & \citet{Storey1995} & $\cdots$ \\
		O$^{+}$ & \citet{FroeseFischer2004} & \citet{Kisielius2009} \\
		O$^{2+}$ & \citet{FroeseFischer2004} & \citet{Lennon1994} \\
		S$^{+}$ & \citet{Rynkun2019} & \citet{Tayal2010} \\
        \hline
	\end{tabular}
\end{table}

We measure emission line fluxes and derive nebular properties of the stacked spectra in the same manner as \citet{Isobe2026}.
To summarise, we fit a Gaussian for each emission line with the same redshift and intrinsic velocity width convolved by the velocity width of \citet{deGraaff2024}'s LSF.
We model continua with $\lambda<3640$ and $>3820$ \AA\ with independent power-law functions, which are linearly connected at $3640\leq\lambda\leq3820$ \AA.
We find the best-fit model and the flux errors based on the $\chi^{2}$ minimisation.
We set a line detection threshold at $S/N>3$, and put upper limits at $3\sigma$ for undetected lines.
We treat [\oii] of the \metsl\,$<7.0$ stack as a non detection since the spectrum becomes noisy around [\oii].
Table \ref{tab:emis} summarises the measured emission line fluxes, highlighting that all stacks have a detection of [\oiii]$\lambda$4363.

We use PyNeb \citep{Luridiana2015} to obtain the nebular colour excess $E(B-V)$ from the Balmer decrements, electron temperature of [\oiii] ($T_{\mathrm{e}}$[\oiii]) from the [\oiii]$\lambda$4363/[\oiii]$\lambda$5007 ratio, and electron density of [\sii] ($n_{\mathrm{e}}$[\sii]) from the [\sii]$\lambda$6716/[\sii]$\lambda$6731 ratio.
Table \ref{tab:atom} summarises the atomic data that we use in this paper.
We assume Case B recombination and the dust extinction law of \citet{Cardelli1989} for consistency with the analysis of our individual sources (Section \ref{subsec:ebv}).
We find the best $E(B-V)$ value that has the least $\chi^{2}$ value based on H$\beta$/H$\alpha$, H$\gamma$/H$\alpha$, and H$\gamma$/H$\beta$ ratios.
As the values of $E(B-V)$, $T_{\mathrm{e}}$[\oiii], and $n_{\mathrm{e}}$[\sii] slightly depend on each other, we adopt an iterative procedure until the $T_{\mathrm{e}}$ value converges within $<$10 K, which takes at most 4 steps.
For the stacks without the detection of [\sii], we assume $n_{\mathrm{e}}$[\sii$]=300$\,cm$^{-3}$, which is typical for the singly-ionised region of high-$z$ galaxies \citep[e.g.,][]{Isobe2023b,Topping2025b,Harikane2025b}.
We perform dust correction using the obtained $E(B-V)$ value and \citet{Cardelli1989}'s extinction curve.

We derive ion abundance ratios of O$^{+}$/H$^{+}$ from [\oii]/H$\beta$ and O$^{2+}$/H$^{+}$ from [\oiii]$\lambda$5007/H$\beta$.
We adopt $T_{\mathrm{e}}$[\oiii] for O$^{2+}$ and $T_{\mathrm{e}}$[\oii] for O$^{+}$.
We assume an empirical $T_{\mathrm{e}}$[\oii]-$T_{\mathrm{e}}$[\oiii] relation recently calibrated at high redshifts \citep{Cataldi2025}.
We regard $12+\log(\mathrm{O/H})$ as $12+\log(\mathrm{(O^{+}+O^{2+})/H^{+}})$ because O$^{3+}$ and higher order oxygen ion abundances are negligible in most star-forming regions \citep[e.g.,][]{Izotov2006,Flury2020,Berg2021} or even AGN \citep{Dors2020}.
We assume $n_{\mathrm{e}}$[\sii] for all ion abundances (i.e., uniform gas density structure), following the conventional direct-$T_{\mathrm{e}}$ method \citep[e.g.,][]{Osterbrock1989,Izotov2006,Berg2019} for consistency with other metallicity calibrations \citep[e.g.,][]{Nakajima2022,Cataldi2025,Sanders2025}.
We note that, using multiple emission lines and assuming multi-density zones, several studies point out that the contribution of the dense gas component to [\oiii]$\lambda$4363 fluxes can increase the total metallicity \citep[e.g.,][]{Marconi2024,Harikane2025b,Martinez2025,Moreschini2026}.
A dedicated study is warranted to evaluate the impact of the expected metallicity differences in faint galaxies, where metallicity estimates rely on a more limited set of emission lines.

To estimate the measurement errors on $E(B-V)$, $T_{\mathrm{e}}$[\oiii], $n_{\mathrm{e}}$[\sii], and \met, we calculate the 16th and 84th percentiles of $E(B-V)$, $T_{\mathrm{e}}$[\oiii], $n_{\mathrm{e}}$[\sii], and \met\ values repeatedly calculated for 1000 sets of line fluxes by perturbing them according to their errors under the assumption of a normal distribution.
Even in this case, we explore the $n_{\mathrm{e}}$[\sii] values within the range of 4--50000 cm$^{-3}$.
For the stacks without [\oii] detection, we derive measurement values of $\mathrm{O^{2+}/H^{+}}$ and 3$\sigma$ upper limits on $\mathrm{O^{+}/H^{+}}$.
We find that the sum of the $\mathrm{O^{2+}/H^{+}}$ value and the $\mathrm{O^{+}/H^{+}}$ upper limit is only $<0.1$ dex larger than the $\mathrm{O^{2+}/H^{+}}$ value, indicating that the O$^{+}$ abundance is negligible compared to the O$^{2+}$ abundance.
We thus consider $\mathrm{O^{2+}/H^{+}}$ as O/H and include the increase when the $\mathrm{O^{+}/H^{+}}$ upper limit is added in the upper error on O/H.

Table \ref{tab:emis} lists the derived values of $E(B-V)$, $T_{\mathrm{e}}$[\oiii], $n_{\mathrm{e}}$[\sii], and \met\ for each stack.
We highlight that the lowest \met\ value of our stacks is down to \met\,$=7.0$, which corresponds to only 2\% solar metallicity.
Within the literature on metallicity high-$z$ calibrations based on individual auroral-line emitters \citep{Sanders2024,Sanders2025,Cataldi2025,Chakraborty2025}, only one data point below this metallicity looks to be included in \citet{Sanders2024}.
Even when the metallicity range is extended to 7.2, the number of such data points appears to be no more than four as in \citet{Sanders2025}.

\section{Stack-based Metallicity Calibrations} \label{sec:calib}

\subsection{Stack-based metallicity indices} \label{subsec:stkidx}

We derive all the strong-line indices listed in Table \ref{tab:idx} from the dust-corrected line ratios.
For indices that include undetected emission lines, we adopt the following treatment for consistency with the O/H calculation (Section \ref{subsec:emis}).
For the indices that include the sum calculation, the measured value is computed by setting the flux of undetected lines to zero.
The increase when the upper limits of the undetected lines are added is included in the upper error.
For the indices that include the product calculation (including sums in log scale), the value calculated from the upper limits of the undetected lines and the measured values of the other lines is used as the appropriate upper or lower limit.
The derived indices are shown in Table \ref{tab:idxstk}.

To supplement metal-rich measurements, we use the emission line ratios of the stacks of galaxies at $z=0.027$--0.25 \citep{Andrews2013} with sSFR values higher than those of the star-formation main sequence (SFMS) at $z>1$ \citep{McClymont2025c} based on the \textsc{thesan-zoom} simulations \citep{Kannan2025}.
The median sSFR value of these stacks is 1.8 Gyr$^{-1}$, which is comparable to that of the observed SFMS at $z\sim3$--4 \citep[e.g.,][]{Clarke2024,Simmonds2025}.
Hereafter, we refer to these stacks as \amstk.
We derive the strong-line indices of the \amstk s in the same manner as our stack measurements.
We calculate median values and 16th-84th percentile ranges of the \amstk\ measurements in O/H bins with 0.2-dex intervals.

We combine our stack measurements and the binned measurements of the \amstk s to obtain stack-based metallicity calibrations in a metallicity range of \met\,$=7.0$--8.7, where our stacks and the \amstk s have \met\,$<8$ and $>8$, respectively.

\subsection{Extension towards lower metallicities} \label{subsec:photmod}
\begin{figure*}
	\centering
    \includegraphics[width=\textwidth]{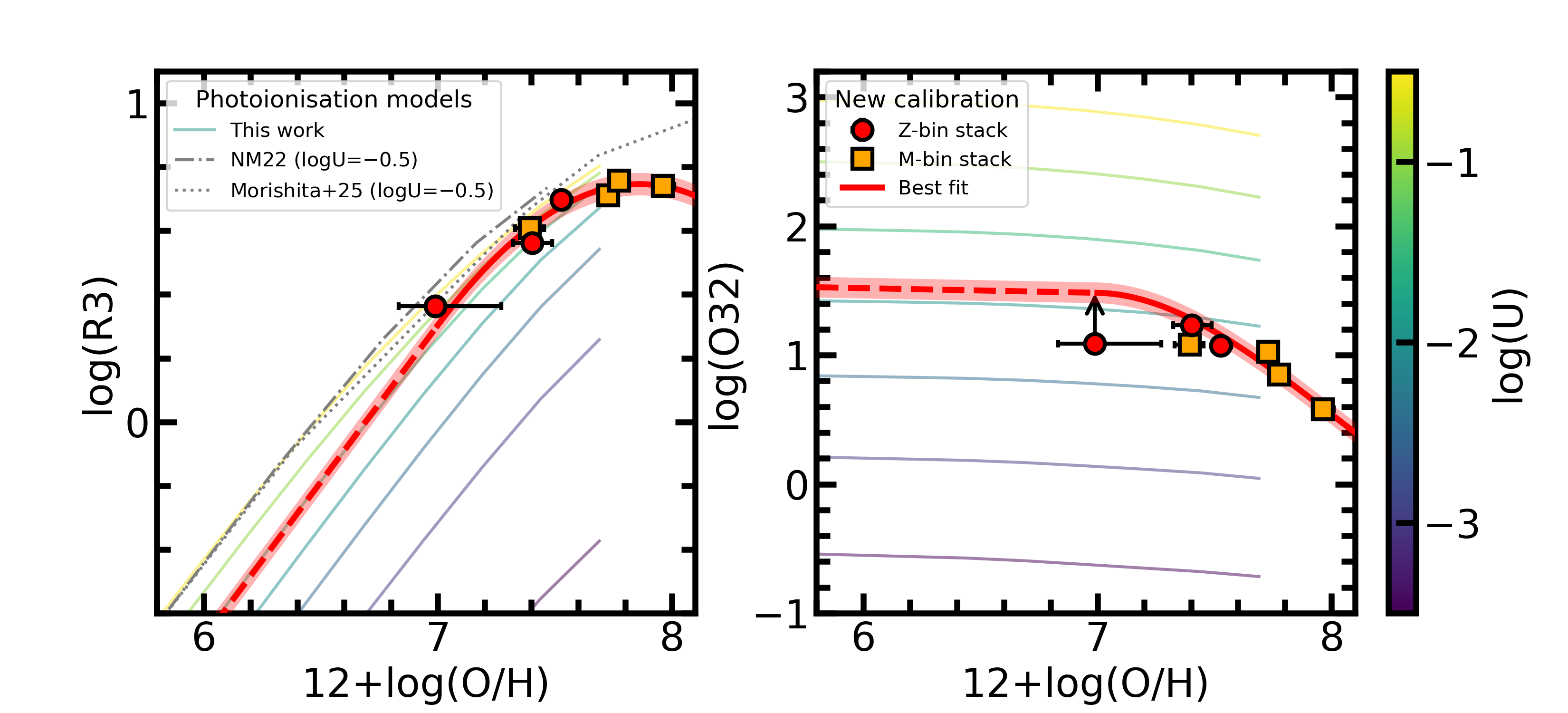}
    \caption{Relations between metallicities and R3 (left) and O32 (right).
    We grid our photoionisation models with 0.5 dex intervals in $\log(U)=(-3.5)$--($-0.5$), which are shown by the solid lines colour-coded by $\log(U)$.
    Our R3 model with $\log(U)=-0.5$ agrees well with the models from the literature with the same $\log(U)$ value \citep{NakajimaMaiolino2022,Morishita2025}.
    These models are almost parallel especially at $12+\log(\mathrm{O/H})\lesssim7$.
    We fix the slope of our calibrations below \met\,$=7$ (red dashed line) to that of our photoionisation models.
    }
    \label{fig:calib_ph}
\end{figure*}

To extend our calibrations below \met\,$=7$, we construct photoionisation models using \textsc{cloudy} \citep{Ferland2013}.
We model young, metal-poor starbursts with the Binary Population and Spectral Synthesis \citep[BPASS;][]{Stanway2018}, assuming an instantaneous burst of star formation with a stellar age of 1 Myr,
a \citet{Salpeter1955} IMF with an upper mass cut of 100 $M_{\odot}$,
a hydrogen density of 300 cm$^{-3}$, and stellar metallicities fixed to the gas-phase metallicity.
We assume metal-to-oxygen abundance ratios to be the solar abundance ratios \citep{Asplund2021}.
We compute the photoionisation models on a grid with 0.25 dex intervals in the range of $5.69\le12+\log(\mathrm{O/H})\le7.69$ and with 0.5 dex intervals in $-3.5\le\log(U)\le-0.5$.

Figure \ref{fig:calib_ph} shows R3 and O32 indices of our photoionisation models as a function of metallicity.
The R3 values of our models agree well with the photoionisation models from the literature (\citealt{Morishita2025}; \citealt{NakajimaMaiolino2022} used in \citealt{Nakajima2025}).
In particular, the O32 values are almost independent of \met\ at a fixed $\log(U)$, in line with the observations of low-metallicity galaxies \citep{Nakajima2022,Langeroodi2026,Rosales-Ortega2026}.
Additionally, \citet{Sanders2025} report that the median O32 indices of high-$z$ galaxies in bins of metallicity are gradually flattened towards lower metallicities, suggesting that the median $\log(U)$ value does not drastically evolve in the low-metallicity regime.
We thus decide to extrapolate our calibrations with a fixed $\log(U)$ value.
Given that the R3 and O32 indices scale with \met\ almost linearly below \met\,$\sim7$ with a similar slope for different $\log(U)$ models, we extrapolate our calibrations by fixing the slopes of all the metallicity indices to those of the photoionisation models.
As $\log(U)\sim(-2.0)$--($-1.5$) reproduces most of the indices based on oxygen or neon lines at the low-metallicity end, we obtain the slopes from the models of \met\,$=5.69$ and 6.69 with $\log(U)=-2$.

\subsection{Polynomial function fitting} \label{subsec:formula}
\begin{table*}
	\centering
	\caption{Coefficients of our calibrations (Equation \ref{equ:calib}) and the root-mean square (RMS) of the residual of each index.}
	\label{tab:calib}
	\begin{tabular}{lcccccc}
		\hline
        Index & $c_{0}$ & $c_{1}$ & $c_{2}$ & $c_{3}$ & $c_{4}$ & RMS\\
        \hline
        log(R3) & 0.3034 & 0.9842 & $-$0.4987 & $-$0.0550 & $\cdots$ & 0.04\\
        log(R2) & $-$1.1562 & 1.0204 & 0.7884 & $-$0.4693 & $\cdots$ & 0.08\\
        log(R23) & 0.4405 & 0.9852 & $-$0.4853 & $\cdots$ & $\cdots$ & 0.04\\
        log(O32) & 1.4831 & $-$0.0363 & $-$1.3509 & 0.4446 & $\cdots$ & 0.08\\
        $\hat{\mathrm{R}}_{\mathrm{Laseter}}$ & $-$0.2600 & 1.3456 & $-$0.1156 & $-$0.2471 & 0.0021 & 0.05\\
        $\hat{\mathrm{R}}_{\mathrm{Chakraborty}}$ & 0.0978 & 1.1481 & $-$0.3815 & $-$0.1155 & $\cdots$ & 0.04\\
        $\tilde{\mathrm{R}}_{\mathrm{Cataldi}}$ & $-$0.2670 & 1.3354 & $-$0.0146 & $-$0.3730 & 0.0435 & 0.05\\
        log(Ne3) & $-$0.4213 & 0.9792 & $-$0.4516 & $-$0.0912 & $\cdots$ & 0.11\\
        log(RO2Ne3) & $-$0.0095 & 0.9829 & 0.0128 & $-$0.1018 & $\cdots$ & 0.09\\
        $\widehat{\mathrm{RNe}}$ & $-$0.7158 & 1.3412 & $-$0.0991 & $-$0.2961 & $\cdots$ & 0.06\\
        log(Ne3O2) & 0.2777 & $-$0.0410 & $-$0.9684 & 0.2569 & $\cdots$ & 0.07\\
        log(Ar3) & $-$2.4807 & 1.0438 & $-$0.1070 & $-$0.1361 & $\cdots$ & 0.03\\
        log(N2) & $-$2.4104 & 1.0109 & $-$0.0243 & $\cdots$ & $\cdots$ & 0.04\\
        log(O3N2) & 2.7564 & $-$0.0267 & $-$0.5867 & $\cdots$ & $\cdots$ & 0.04\\
        log(S2) & $-$1.9475 & 0.9996 & $-$0.1049 & $\cdots$ & $\cdots$ & 0.12\\
        log(O3S2) & 2.3512 & $-$0.0154 & $-$0.5747 & $\cdots$ & $\cdots$ & 0.12\\
        \hline
	\end{tabular}
\end{table*}

We fit the following polynomial function to the relation between the derived strong-line index
and \met:
\begin{equation}
    \mathrm{Index}=
    \begin{cases}
    \sum_{n=0}c_{n}x^{n} & \text{when}\ x\ge0 \\
    c_{0}+c_{1}x & \text{when}\ x<0
    \end{cases}
    \label{equ:calib}
\end{equation}
where $x=12+\log(\mathrm{O/H})-Z_{\mathrm{switch}}$ and  $Z_{\mathrm{switch}}=7$.
We fix the $c_{1}$ value to the slope of the photoionisation models.

We define the best-fit parameters as those that maximise the following log-likelihood function ($\ln{\mathcal{L}}$):
\begin{equation}
    \ln{\mathcal{L}}=\sum_{i}\bigl[(\ln{\mathcal{L})_{i,\mathrm{det}}}+(\ln{\mathcal{L})_{i,\mathrm{uplim}}}+(\ln{\mathcal{L})_{i,\mathrm{lowlim}}}\bigr],
    \label{equ:llh}
\end{equation}
where $(\ln{\mathcal{L})_{i,\mathrm{det}}}$ is
log-likelihood function of the $i$th stack for the measurement value of the index,
$(\ln{\mathcal{L})_{i,\mathrm{uplim}}}$ for that with the upper limit, and $(\ln{\mathcal{L})_{i,\mathrm{lowlim}}}$ for that with the lower limit.
Assuming an asymmetric Gaussian distribution based on the upper and lower errors of the indices, we define $(\ln{\mathcal{L})_{\mathrm{det}}}$ as follows:
\begin{equation}
    (\ln{\mathcal{L})_{i,\mathrm{det}}}=-\frac{1}{2}\Bigl(\frac{\mathrm{Index}_{i,\mathrm{obs}}-\mathrm{Index}_{i,\mathrm{cal}}}{\sigma_{i,\mathrm{obs}}}\Bigr)^{2}-\ln(\sqrt{2\pi}\sigma_{i,\mathrm{obs}}),
    \label{equ:llhdet}
\end{equation}
where $\sigma_{i,\mathrm{obs}}$ is the upper error or the lower error when the index predicted by the calibration ($R_{i,\mathrm{cal}}$) is larger or smaller than the observed index ($R_{i,\mathrm{obs}}$), respectively.

To impose a large penalty when the observed upper limit on the index ($\mathrm{Index}_{i,\mathrm{uplim}}$) is smaller than $\mathrm{Index}_{i,\mathrm{cal}}$ or when the observed lower limit on the index ($\mathrm{Index}_{i,\mathrm{lowlim}}$) is larger than $\mathrm{Index}_{i,\mathrm{cal}}$, we define $(\ln{\mathcal{L})_{i,\mathrm{uplim}}}$ and $(\ln{\mathcal{L})_{i,\mathrm{lowlim}}}$ as follows:
\begin{equation}
    (\ln{\mathcal{L})_{i,\mathrm{uplim}}}=
    \begin{cases}
    -10^{5} & \text{when}\ \mathrm{Index}_{i,\mathrm{uplim}}<\mathrm{Index}_{i,\mathrm{cal}} \\
    0 & \text{when}\ \mathrm{Index}_{i,\mathrm{uplim}}\ge \mathrm{Index}_{i,\mathrm{cal}}
    \end{cases}
    \label{equ:llhuplim}
\end{equation}
\begin{equation}
    (\ln{\mathcal{L})_{i,\mathrm{lowlim}}}=
    \begin{cases}
    -10^{5} & \text{when}\ \mathrm{Index}_{i,\mathrm{lowlim}}>\mathrm{Index}_{i,\mathrm{cal}} \\
    0 & \text{when}\ \mathrm{Index}_{i,\mathrm{lowlim}}\le \mathrm{Index}_{i,\mathrm{cal}}
    \end{cases}
    \label{equ:llhlowlim}
\end{equation}

To determine which order of polynomials we adopt, we fit the second- to fifth-order polynomials and calculate the Akaike Information Criterion (AIC) for each polynomial, which quantifies the trade-off between goodness of fit and model complexity.
The AIC is defined as $\mathrm{AIC}=2k-2\ln{\hat{\mathcal{L}}}$, where $k$ is the number of free parameters and $\ln{\hat{\mathcal{L}}}$ is the maximum log-likelihood function.
Obtaining the minimum AIC values among the orders of the polynomials we fit (AIC$_{\mathrm{min}}$), we exclude the fitting results with $\Delta\mathrm{AIC}\equiv \mathrm{AIC}-\mathrm{AIC_{min}}>10$, which are often regarded as a lack of statistical support of a good fit \citep[e.g.,][]{Burnham2004}.
From the remaining results, we choose the order with the best-fit function without unphysical inflections at the edge of the fitted region.
Table \ref{tab:calib} lists the best-fit parameters of our calibrations together with the root-mean square (RMS) of the residual of each index \citep{Curti2020}.

Figure \ref{fig:calib_ph} highlights the low-metallicity end of our R3 and O32 calibrations.
The extrapolated regimes below $12+\log(\mathrm{O/H})<7$ (red dashed line) are parallel to the photoionisation models by design (Section \ref{subsec:photmod}).
Our R3 calibration provides a simple scaling at this regime, whereby R3 is approximately equal to the metallicity expressed as a percentage of the solar value.
For instance, $\mathrm{R3}=1$ corresponds to $\approx1$\% $Z_{\odot}$, whereas $\mathrm{R3}=0.1$ corresponds to $\approx0.1$\% $Z_{\odot}$.

\subsection{Comparison with the calibrations from the literature} \label{subsec:calibcomp}
\begin{figure*}
	\centering
    \includegraphics[width=\textwidth]{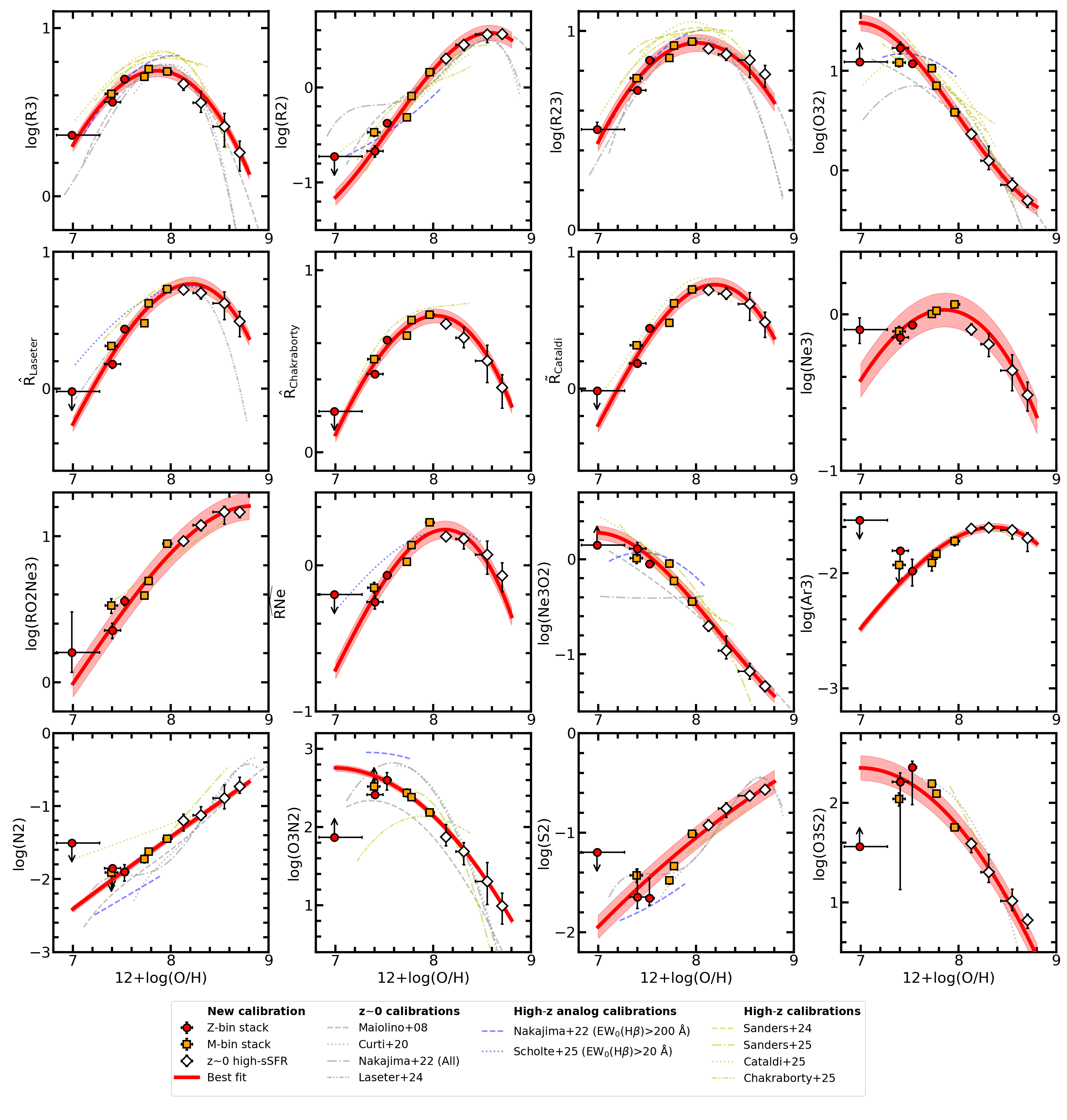}
    \caption{Relations between metallicities and strong-line diagnostics of our $Z$-bin stacks (red circles) and $M$-bin stacks (orange squares).
    We extend our calibration to higher metallicities by adding median values of the \amstk s \citep[white diamonds;][]{Andrews2013} whose sSFRs are higher than those of the star-formation main sequence simulated at $z>1$ \citep{McClymont2025a}.
    The red solid curves with the $1\sigma$ uncertainty (red shades) are the best-fit polynomial functions to our $Z$-bin stacks, $M$-bin stacks, and the median values of the $z\sim0$ high-sSFR stacks.
    For a comparison, we plot the $z\sim0$ calibrations \citep{Maiolino2008,Curti2020,Nakajima2022,Laseter2024}, the high-$z$ analogue calibrations \citep{Nakajima2022,Scholte2025}, and the high-$z$ calibrations \citep{Sanders2024,Sanders2025,Cataldi2025,Chakraborty2025}.
    }
    \label{fig:calib}
\end{figure*}

\begin{figure*}
	\centering
    \includegraphics[width=\textwidth]{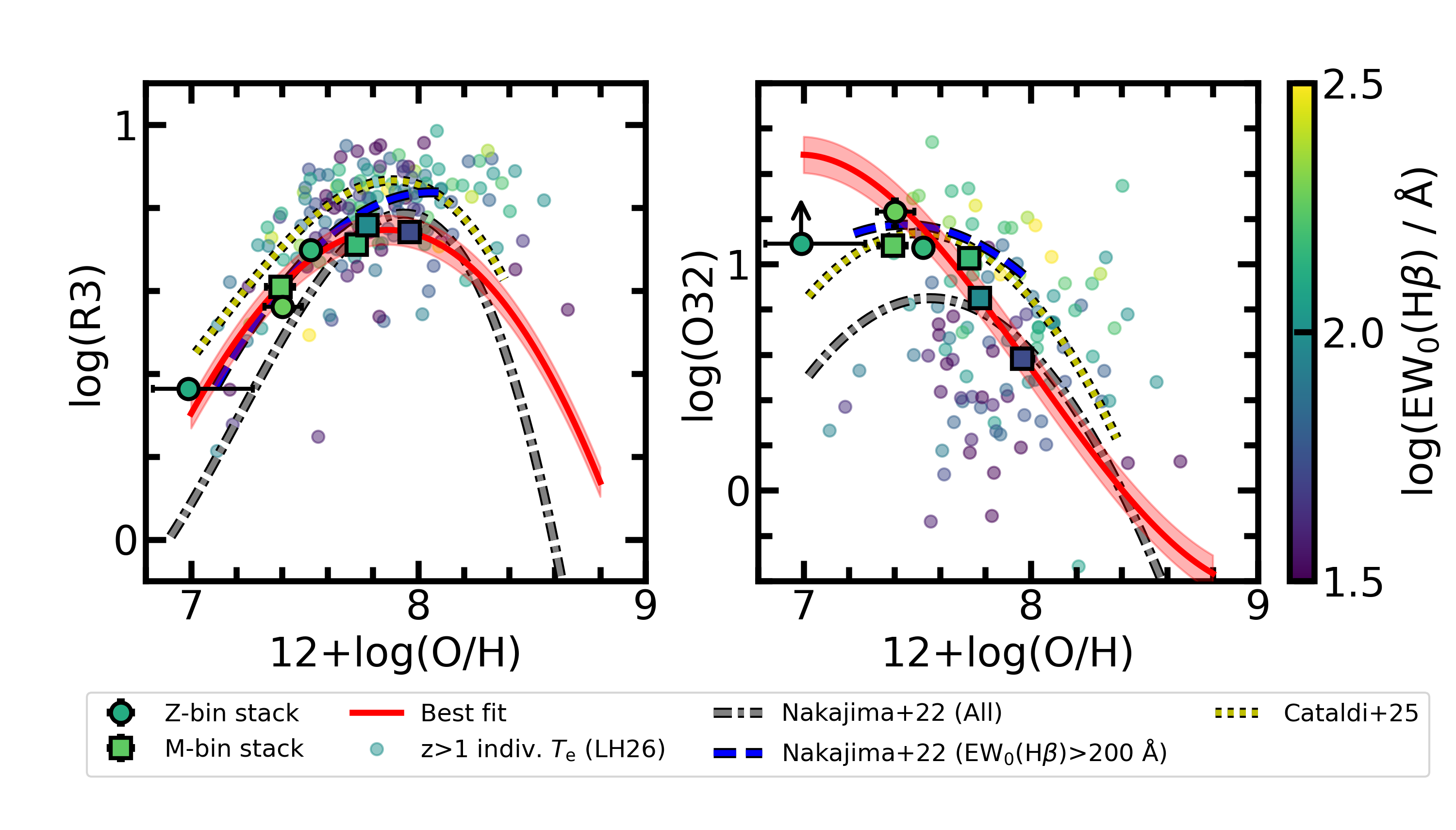}
    \caption{Same as Figure \ref{fig:calib} but highlight R3 (left) and O32 (right). Our $Z$-bin stacks (circles), $M$-bin stacks (squares), and individual auroral-line emitters at $z>1$ (dots) are colour-coded by their H$\beta$ equivalent widths.}
    \label{fig:calib_ew}
\end{figure*}

Figure \ref{fig:calib} shows the best-fit functions of the calibrations and the measurements of the stacks used for the calibrations.
We find that our calibrations exhibit lower values of R3, R23, O32, $\hat{\mathrm{R}}_{\mathrm{Laseter}}$, $\hat{\mathrm{R}}_{\mathrm{Chakraborty}}$, $\tilde{\mathrm{R}}_{\mathrm{Cataldi}}$, and Ne3O2 compared to other high-$z$ calibrations based on individual auroral-line emitters \citep{Sanders2024,Sanders2025,Cataldi2025,Chakraborty2025} at a given metallicity.
In particular, the lower values of O32 and Ne3O2 suggest that the sources used in our stacks tend to have lower ionisation parameters than the individual auroral-line emitters at high $z$.

Figure \ref{fig:calib_ew} focuses on the R3 and O32 indices together with individual auroral-line emitters with $T_{\mathrm{e}}$ measurements at $z>1$ drawn from \citet{Langeroodi2026}, which include observational results from both ground-based telescopes \citep{Sanders2020,Revalski2024} and the \textit{JWST} \citep{Nakajima2023,Sanders2024,Morishita2024}.
In this figure, the data points of our stacks and the individual auroral-line emitters are colour-coded by their H$\beta$ EWs (EW$_{0}$(H$\beta$)).
We find that the high-$z$ calibration based on individual sources \citep[represented by][]{Cataldi2025} overlaps with the individual auroral-line emitters whose EW$_{0}$(H$\beta$) values are generally larger than those of our stacks, especially within a metallicity range of \met\,$\sim7.5$--8.0.
This is in line with our finding that our stacks exhibit lower O32 values than the current calibrations, given that O32 values and ionisation parameters correlate with sSFRs \citep[e.g.,][]{Nakajima2014,Sanders2016,Kashino2019} and thus EW$_{0}$(H$\beta$).

These differences may arise from the faint auroral-line detection required for the $T_{\mathrm{e}}$ measurement in individual high-$z$ sources,
which can bias the sample towards higher ionisation parameter and larger EWs (see Section \ref{sec:intro}).
On the other hand, we construct the stacked samples based on H$\beta$ emitters (see Sections \ref{subsec:parent} and \ref{subsec:subsamp}), for which we do not require faint line detections.
In this sense, we can say that our stack-based calibrations are better applicable to the bulk of the galaxy population of high-$z$ emission-line galaxies.

Compared to the $z\sim0$ calibrations \citep{Maiolino2008,Curti2020,Nakajima2022}, our calibrations generally have higher R3 and R23 values, especially at the low-metallicity branch of \met\,$\lesssim7.6$.
This suggests that the general population of high-$z$ metal-poor star-forming galaxies still exhibit slightly higher excitation, although not as high as the individual auroral line emitters at high $z$.
Our $\mathrm{\hat{R}_{Laseter}}$ calibration matches \citet{Laseter2024}'s calibration at the low-metallicity branch, in contrast to \citet{Scholte2025}'s calibration, which agrees with the high-$z$ calibration of \citet{Sanders2025}.
Notably, the R3 and R23 indices of our calibrations align with those of the $z\sim0$ calibrations with large EWs \citep{Nakajima2022} at $12+\log(\mathrm{O/H})\lesssim7.6$.
In addition, the $z\sim0$ calibrations with large EWs provide R2, O32, and Ne3O2 comparable to those of our stacks within the same metallicity range, indicating a similar ionisation structure.
However, our calibrations exhibit higher N2 and lower O3N2 than the $z\sim0$ calibrations with large EWs, which can be explained by mild enhancement of N/O in singly-ionised regions at high $z$ \citep{Cataldi2025b,Cameron2026}.

\subsection{Metallicity recalculation and H$\alpha$-based SFR} \label{subsec:sfr}

We recalculate \metsl\ values of the spectroscopic sources in our parent sample with our stack-based calibrations in the same manner as described in Section \ref{subsec:met}.
Compared to \citet{Cataldi2025}'s calibrations, our calibrations provide higher metallicities at \met\,$\lesssim7.8$ and lower metallicities at \met\,$\gtrsim7.8$.
This is because R3 and $\tilde{\mathrm{R}}_{\mathrm{Cataldi}}$ are more sensitive to \met\ than O32 and Ne3O2 at \met\,$\lesssim7.8$, where our calibrations provide higher \met\ at a fixed R3 and $\tilde{\mathrm{R}}_{\mathrm{Cataldi}}$.
Conversely, O32 and Ne3O2 are more sensitive to \met\ than R3 and $\tilde{\mathrm{R}}_{\mathrm{Cataldi}}$ at \met\,$\gtrsim7.8$, where our calibrations provide lower \met\ at a fixed O32 and Ne3O2.

The 16th-84th percentile of \metsl\ for our R1000 sources spans from 7.5 to 8.3, indicating that most of $z>1$ sources have metallicities much lower than the solar value.
Such a low metallicity can impact the determination of H$\alpha$-based SFRs (SFR$_{\mathrm{H}\alpha}$, hereafter).
We derive SFR$_{\mathrm{H}\alpha}$ from the dust-corrected H$\alpha$ luminosities ($L_{\mathrm{H\alpha}}$) using the following equation:
\begin{equation}
    \log(\mathrm{SFR}_{\mathrm{H}\alpha}/M_{\odot}\,\mathrm{yr}^{-1})=\log(L_{\mathrm{H\alpha}}/\mathrm{erg\,s^{-1}})+C,
    \label{equ:sfr}
\end{equation}
where $C$ is a conversion factor that depends on the number of H-ionising photons from newborn stars, which is predicted to increase with lower stellar metallicities and more top-heavy IMFs at a given SFR \citep[e.g.,][]{Hao2011,Madau2014,Reddy2018,Shapley2023,KorhonenCuestas2025}.

Throughout the paper, we adopt $C=-41.64$ for the BPASS model with $Z=0.002$ (14\% $Z_{\odot}$) reported by \citet{Eldridge2017}.
It should be noted that the $C$ values can change only by $~0.1$ dex within a wide metallicity range of $\sim2$\%--60\% $Z_{\odot}$, which corresponds to $12+\log(\mathrm{O/H})\sim7.0$--8.5.
For the sources without H$\alpha$ covered by NIRSpec, we use the brightest available Balmer line and assume Case B recombination.

\section{High-Redshift EMPG Candidates} \label{sec:empg}
\begin{figure*}
	\centering
    \includegraphics[width=\textwidth]{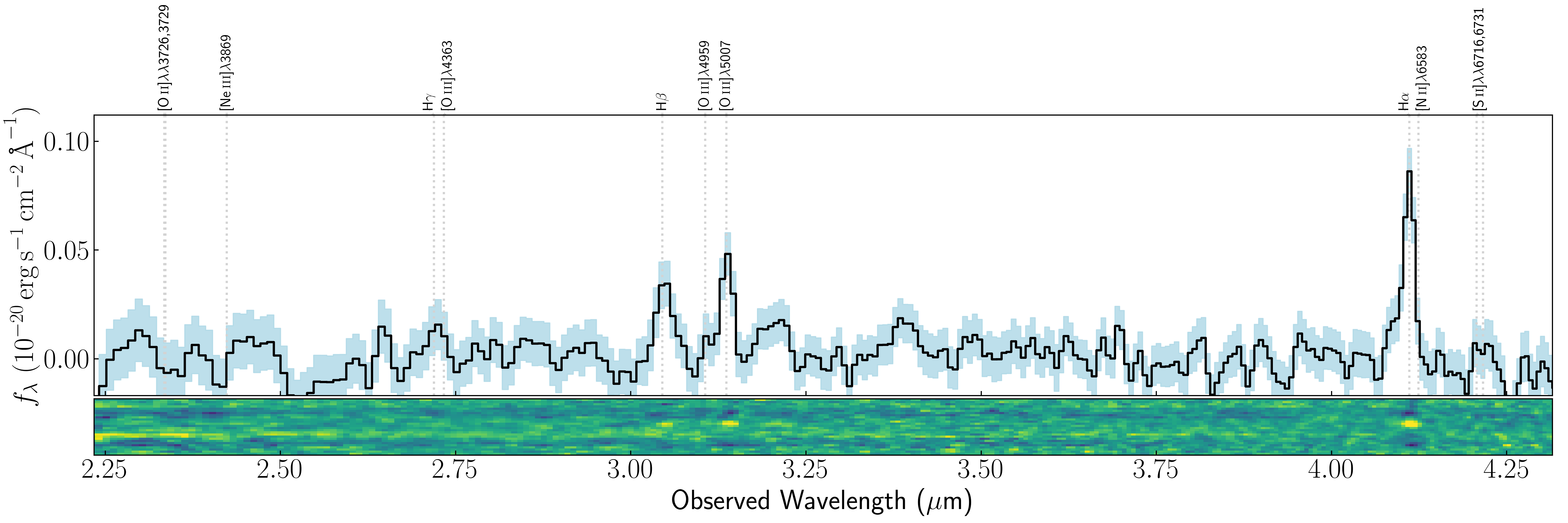}
    \caption{One-dimensional R100 spectrum (top) and two-dimensional spectrum (bottom) of our most metal-poor galaxy candidate (JADES DR4 Unique ID: goods-s-ultradeep\_127079 at $z=5.262$). The strong H$\alpha$ and H$\beta$ lines, the weak [\oiii]$\lambda$5007, and the absence of [\oii]$\lambda\lambda$3726,3729 suggest \metsl\,$=6.72^{+0.13}_{-0.13}$ based on our calibrations.}
    \label{fig:empgspec}
\end{figure*}

\subsection{EMPG candidate selection} \label{subsec:empgslct}
From the \nsrc\ sources with the R100 and/or R1000 data, we select high-$z$ EMPG candidates with \metsl\,$\le7.3$ based on our stack-based calibrations.
To preselect candidates to secure robustness of metallicity measurements, we basically adopt the same criteria as we use to construct the $Z$-bin stacks in Section \ref{subsec:subsamp}.
For sources with only R100 data, we remove Type-2 AGN candidates only based on the S2 vs. R3 diagram because the R100 data do not resolve the emission lines necessary for the other Type-2 AGN diagnostics.
In addition, we remove the R100 sources at $z<4$ to ensure that the [\oiii]$\lambda\lambda$4959,5007 doublet and H$\beta$ are sufficiently resolved.
We perform a visual inspection for the preselected candidates, and finally select the \nempgp\ most promising candidates from the R100 data and \nempgm\ from the R1000 data.
We identify \nempgovl\ candidates selected from both the R100 and R1000 data and one candidate selected from two different tiers, which means that \nempg\ candidates are the unique sources.

Table \ref{tab:empg} lists basic properties of the \nempg\ EMPG candidates.
Figure \ref{fig:empgspec} shows the spectrum of the most metal-poor candidate with \metsl\,$=\metminempg$, exhibiting strong H$\beta$ comparable to [\oiii]$\lambda$5007.
It is worth mentioning that \citet{Trussler2026} select EMPG candidates using the JADES deep medium-band photometry, and only one of these photometric EMPG candidates, goods-s-mediumjwst\_182875, has the JADES NIRSpec spectra.
We confirm that goods-s-mediumjwst\_182875 has $12+\log(\mathrm{O/H})=7.19$ and satisfies our EMPG selection criteria.

In addition, we serendipitously identify two LRDs in our EMPG candidates.
Given their distinctive characteristics, we focus on our EMPG candidates excluding the two LRDs up to Section \ref{subsec:empgmzr}, and discuss the LRDs separately in Section \ref{subsec:lrd}.

\subsection{Literature EMPG candidates} \label{subsec:litempg}
\begin{figure*}
	\centering
    \includegraphics[width=\textwidth]{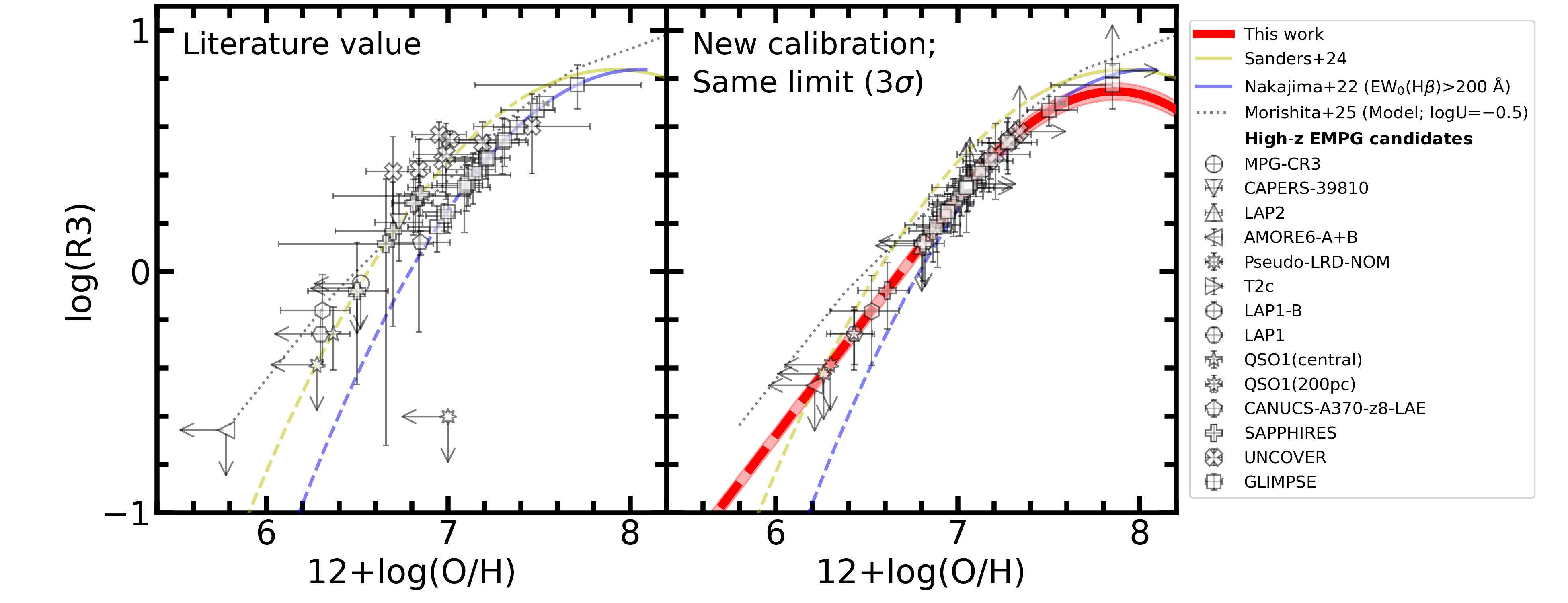}
    \caption{R3 vs. metallicity. (Left) The values simply drawn from the literature. (Right) The values that we uniformly recalculate with our new calibration and the same detection limit ($3\sigma$). We plot EMPG candidates from the literature without $T_{\mathrm{e}}$ measurements or \metsl\ values based on our calibrations (Section \ref{subsec:litempg}). }
    \label{fig:calib_empg}
\end{figure*}

We compile the following EMPG candidates from the literature:
MPG-CR3 at $z=3.193$ \citep{Cai2025},
The Cliff at $z=3.548$ \citep{Ivey2026},
CAPERS-39810 at $z=3.654$ \citep{Cai2026},
LAP2 at $z=4.19$ \citep{Vanzella2025},
AMORE6-A+B at $z=5.7253$ \citep{Morishita2025},
Pseudo-LRD-NOM at $z=5.96$ \citep{Caputi2026},
T2c at $z=6.146$ \citep{Vanzella2024},
LAP1-B at $z=6.625$ \citep{Nakajima2025},
LAP1 at $z=6.639$ \citep{Vanzella2023},
Abell2744-QSO1 at $z=7.04$ \citep[hereafter QSO1;][]{Maiolino2025},
CANUCS-A370-z8-LAE at $z=8.203$ \citep{Willott2025},
Hebe at $z=10.6$ \citep{Ubler2026},
sources from SAPPHIRES \citep{Hsiao2025},
UNCOVER\footnote{Ultradeep NIRSpec and NIRCam ObserVations before the Epoch of Reionization \citep{Bezanson2024}} \citep{Chemerynska2024}, and GLIMPSE \citep{Fujimoto2025b,Asada2026},
satellite galaxies of SMACS0723\_4590 at $z=8.45$ and RX2129\_11027 at $z=9.51$ \citep[referred to as `satellites' in this paper;][]{Koller2026},
EXCELS-63107 at $z=8.271$ \citep{Cullen2025}, and
Firefly Sparkle at $z=8.304$ \citep{Mowla2024}.

We draw $M_{*}$ values of these EMPG candidates from the literature listed above, except for QSO1 \citep[dynamical mass as an upper limit on $M_{*}$;][]{Juodzbalis2026} and Hebe \citep{Rusta2026,Maiolino2026}.
We also take SFR values based on H$\alpha$ or H$\beta$ from the literature by scaling the conversion factor to match the value we adopt ($C=-41.64$; Section \ref{subsec:sfr}), except for MPG-CR3, CAPERS-39810, LAP2, the SAPPHIRES sources, the UNCOVER sources, T2c, GLIMPSE-16043, LAP1-B, LAP1, Firefly Sparkle, and Hebe, which do not have SFR(H$\alpha$) measurements or the $C$ values used for SFR(H$\alpha$).
For these sources, we uniformly derive SFR(H$\alpha$) values from the reported Balmer line fluxes in the same manner as in Section \ref{subsec:sfr}.

Among these EMPG candidates, only EXCELS-63107 and Firefly Sparkle have $T_{\mathrm{e}}$ measurements.
In addition, The Cliff, Hebe, and the satellites already have \metsl\ values or upper limits based on our calibrations.
For the remaining EMPG candidates, we uniformly adopt our R3 calibration.
For the sources whose metallicity values are based on only [\oiii]$\lambda$5007/H$\alpha$, we estimate R3 values by assuming Case B recombination with H$\alpha$/H$\beta=2.86$ (Section \ref{subsec:ebv}).
While different works adopted different confidence levels for providing metallicity upper limits (from 3$\sigma$ to 2$\sigma$), here, for consistency, we set a common detection threshold of $S/N>3$ and give 3$\sigma$ upper limits for undetected lines in the same manner as for our sample (Section \ref{subsec:emis}).
As \citet{Chemerynska2024} do not report the actual R3 values used for the \metsl\ estimates, we derive the R3 values from the line flux table provided by \citet{Price2025}.
Table \ref{tab:empglit} lists the metallicity values of these EMPG candidates together with their $M_{*}$ and SFR values.

Figure \ref{fig:calib_empg} compares R3 and \met\ values from the literature to those uniformly based on our stack-based calibration and the same $3\sigma$ detection threshold.
We confirm that our recalculated values are generally higher than the literature values based on either \citet{Sanders2024}'s calibration or the photoionisation models \citep[represented by][]{Morishita2025} with a very high $\log(U)$ of $-0.5$ \citep[also][]{Nakajima2025}, but comparable to those based on \citet{Nakajima2022}'s large-EW calibration.
Note that this figure does not necessarily indicate which measurement is more reliable, as most of these candidates are too faint to constrain physical conditions such as ionisation parameters.
Rather, we argue that, irrespective of the method employed, \metsl\ values should be compared fairly using a consistent methodology, as well as uniform detection and confidence criteria.

\section{SFR-Stellar Mass relation} \label{sec:msfr}
\begin{figure*}
	\centering
    \includegraphics[width=\textwidth]{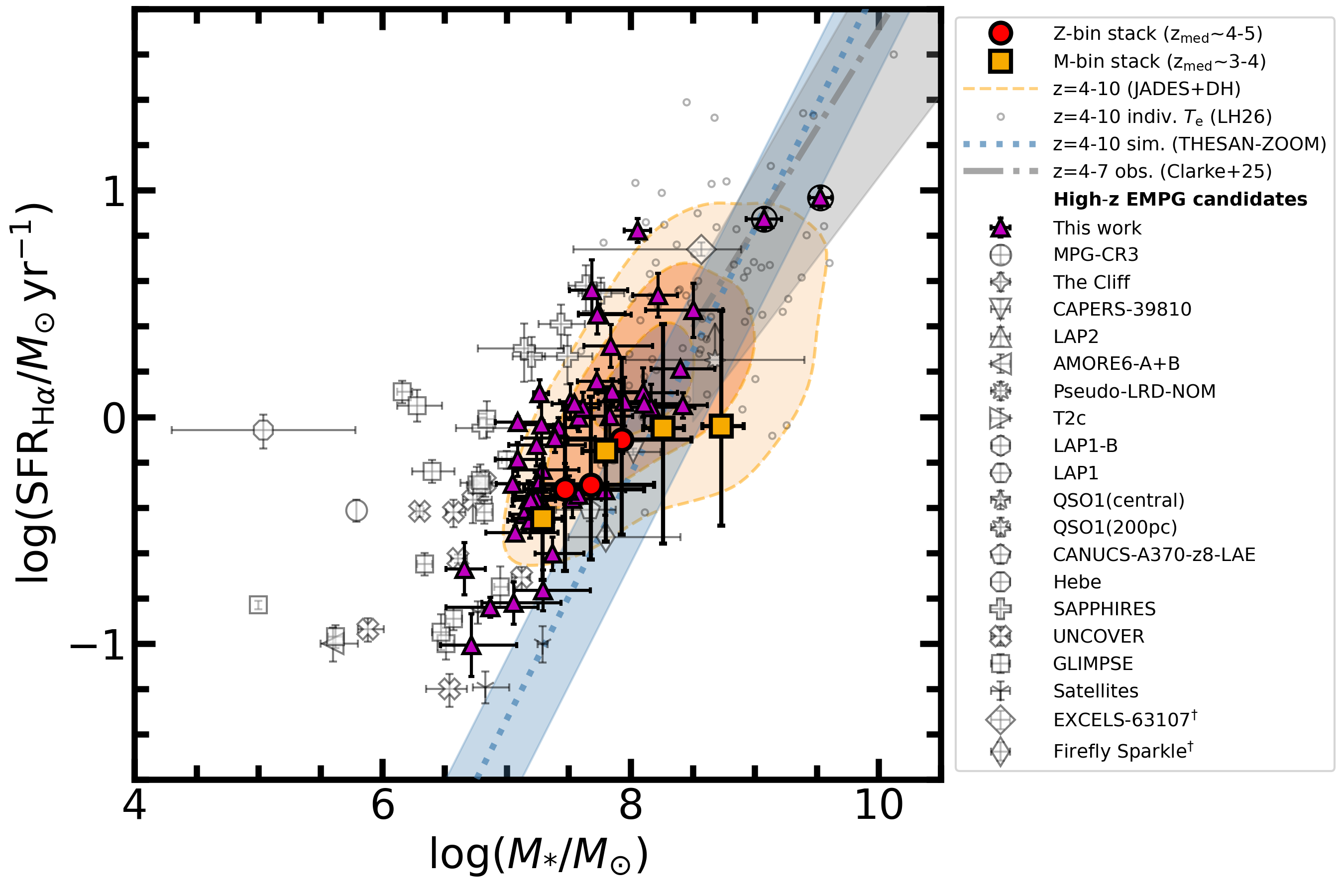}
    \caption{H$\alpha$-based SFR (SFR$_{\mathrm{H}\alpha}$) vs. $M_{*}$.
    The red circles with the error bars represent the median values with the 16th-84th percentiles of the sources used for the $Z$-bin stacks, and the oranges squares show the same measurements for the $M$-bin stacks.
    The distribution of the R1000 sources from JADES and DH (JADES+DH sources) at $z=4$--10 is shown by the orange dashed contour, whose levels are 16, 50, and 84 percentiles of the kernel density derived using the \texttt{gaussian\_kde} package. The small grey circles are the individual auroral-line emitters at $z=4$--10 \citep{Langeroodi2026}. The blue dotted line with the blue shade corresponds to the star-formation main sequence (SFMS) predicted by the \textsc{thesan-zoom} simulations at $z=7$ with the range of the SFMSs simulated by \textsc{thesan-zoom} across $z=4$--10 \citep{McClymont2025a}. The grey dashdot line with the grey shade is the observed SFMS at $5<z<6$ with the range between the SFMSs at $4<z<5$ and $6<z<7$ \citep{Clarke2025}. The magenta triangles are our EMPG candidates, while the two encircled magenta triangles are Little Red Dots (LRDs) serendipitously identified in our EMPG candidates (Section \ref{subsec:lrd}). The white points are the EMPG candidates from the literature (Section \ref{subsec:litempg}).  $^{\dagger}$: EMPGs with $T_{\mathrm{e}}$ measurements.}
    \label{fig:sfms}
\end{figure*}

Figure \ref{fig:sfms} shows the SFR-$M_{*}$ relations of the sample discussed in this paper.
We show all the R1000 sources from JADES and DH at $z=4$--10, which we refer to as the JADES+DH sources hereafter.
The JADES+DH sources are distributed around the star-formation main sequences (SFMSs) at similar redshifts based on both the observations \citep{Clarke2025} and the simulations \citep{McClymont2025a} at $\log(M_{*}/M_{\odot})\gtrsim8.5$.
Below this $M_{*}$, the JADES+DH sources exhibit slightly higher SFRs than the simulated SFMS, suggesting an observational bias of high-$z$ emission-line galaxies towards higher SFRs even with NIRSpec \citep[e.g.,][]{Clarke2025}.

A key point is that our stacks overlap with the JADES+DH distribution in the SFR-$M_{*}$ plot, especially at $\log(M_{*}/M_{\odot})\lesssim8$, indicating that our calibrations empirically include the potential observational bias of the NIRSpec sources at the low-mass end.
Compared to the current high-$z$ metallicity calibrations based on individual auroral-line emitters, most of which have $\log(M_{*}/M_{\odot})\gtrsim8$ \citep{Sanders2025}, our calibrations trace more representative populations of low-mass star-forming galaxies.

We find that our EMPG candidates generally lie on the low-mass end of the JADES+DH distribution ($\log(M_{*}/M_{\odot})\sim7$--8), suggesting that our EMPG candidates do not constitute a distinct population within our parent sample.
Compared to our EMPG candidates, the majority of the EMPG candidates from the literature exhibit lower stellar masses ($\log(M_{*}/M_{\odot})\lesssim7$) at similar SFRs, most of which are lensed sources such as AMORE6-A+B \citep{Morishita2025}.

\section{Mass-metallicity relation \& fundamental metallicity relation} \label{sec:mzrfmr}

\subsection{General redshift evolution} \label{subsec:gen}

\begin{figure*}
	\centering
    \begin{minipage}{\textwidth}
    \includegraphics[width=0.56\textwidth]{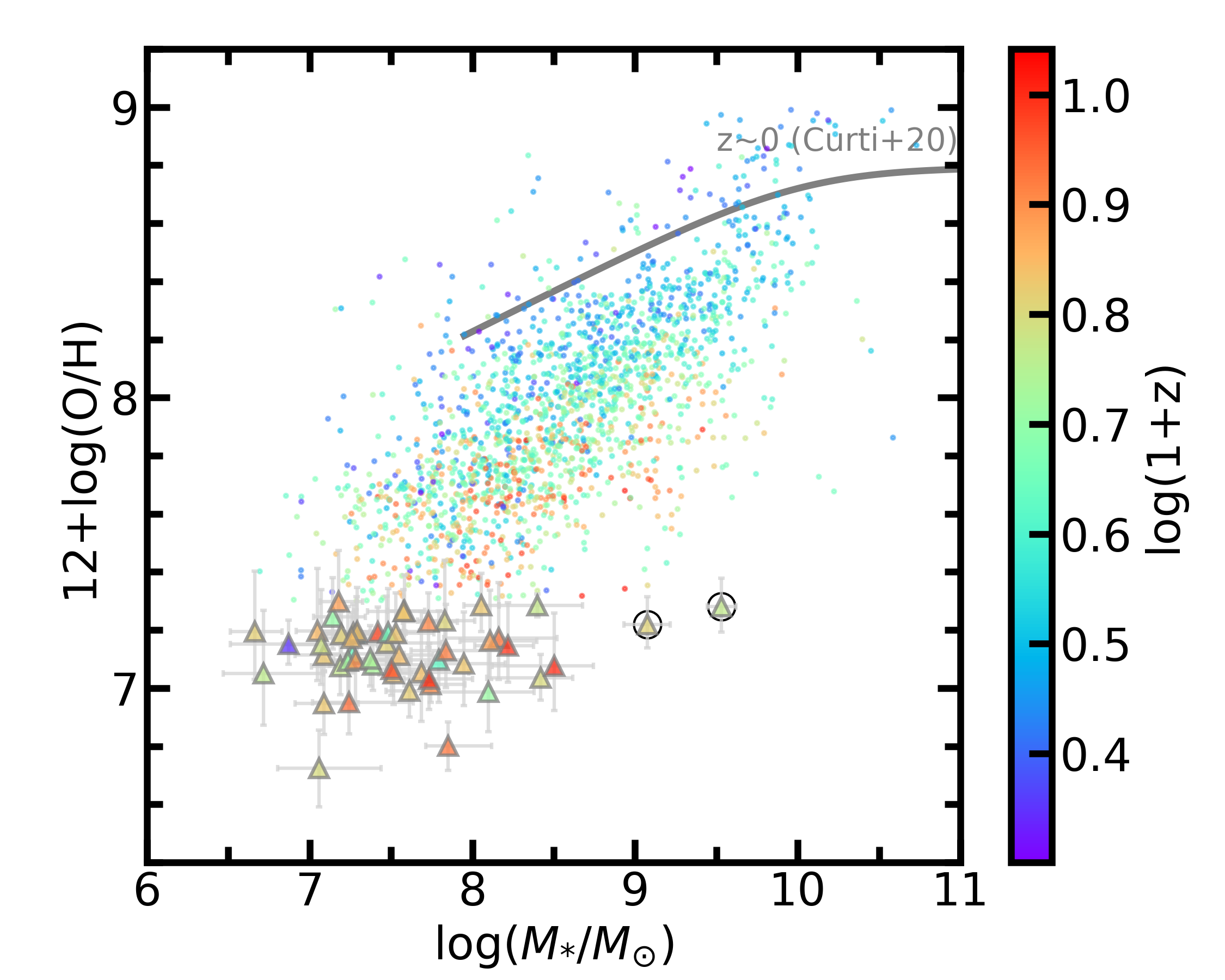}
    \includegraphics[width=0.44\textwidth]{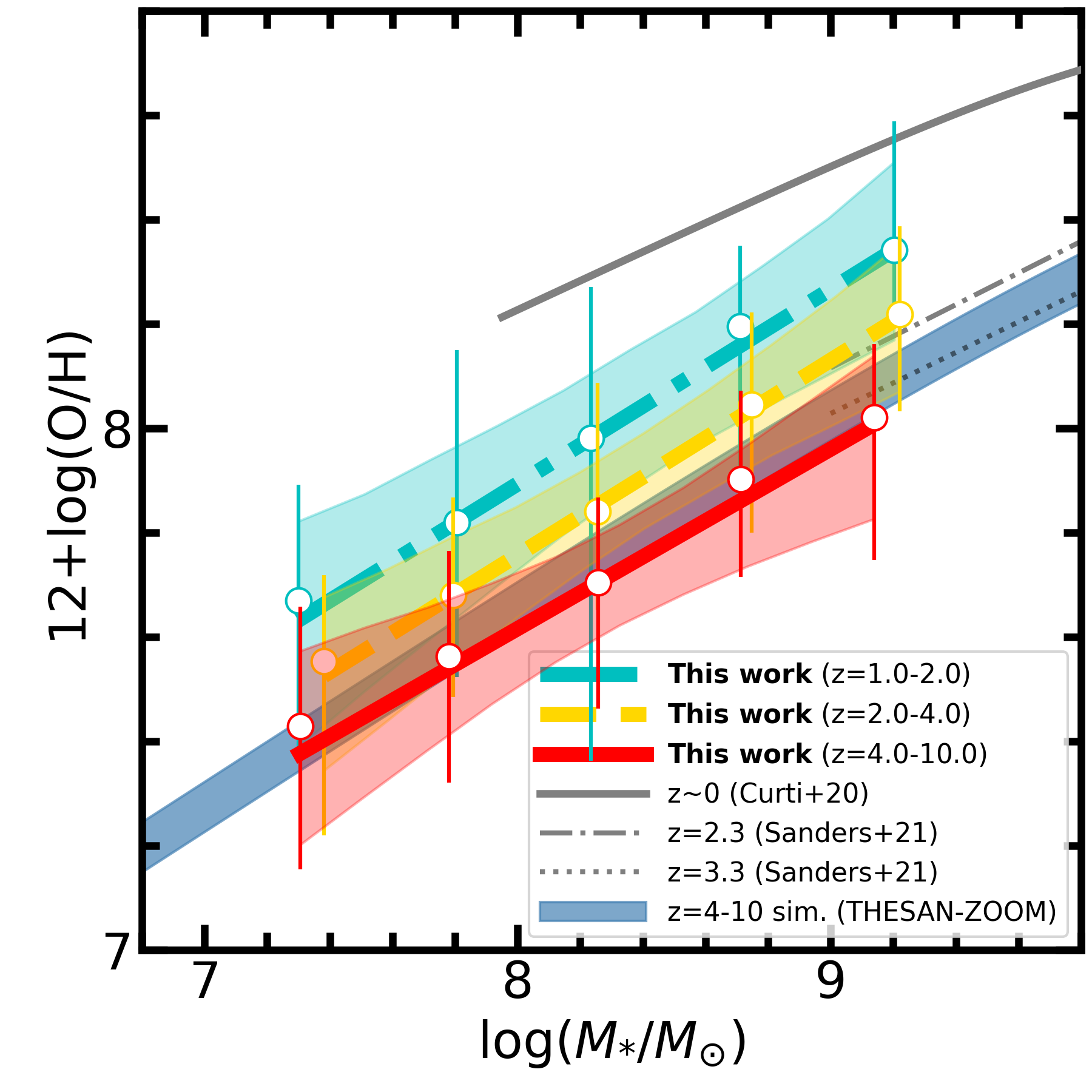}
    \end{minipage}
    \caption{Mass-metallicity relation of our JADES+DH sources at $z=1$--10. (Left) Actual distribution (dots) together with our EMPG candidates (triangles). These data points are colour-coded by $z$. The grey solid curve is the $z\sim0$ MZR reported by \citet{Curti2020}. (Right) Median values with 16th and 84th percentiles in bins of $M_{*}$ shown by the open circles in cyan ($z=1$--2), yellow ($z=2$--4), and red ($z=4$--10). The best-fit linear functions of these measurements with the uncertainties are shown by the cyan dashdot line with the cyan shade ($z=1$--4), the yellow dashed line with the yellow shade ($z=2$--4), and the red solid line with the red shade ($z=4$--10). The blue shaded region shows the MZR predicted by the \textsc{thesan-zoom} simulations at $z=4$--10 \citep{McClymont2025c}. We confirm the decrease in metallicity towards higher $z$. Our $z=4$--10 MZR matches the prediction by the \textsc{thesan-zoom} simulations at the same redshift.}
    \label{fig:mzrall}
\end{figure*}

\begin{table}
	\centering
	\caption{Best-fit parameters of the linear function (Equation \ref{equ:mzr}) fitted to the mass-metallicity relation of the JADES+DH sources within each redshift range.}
	\label{tab:mzrfit}
	\begin{tabular}{cccc}
		\hline
        Redshift & $z_{\mathrm{med}}$ & $\beta$ & $Z_{8}$\\
        \hline
		1--2 & 1.70 & $0.37^{+0.17}_{-0.13}$ & $7.89^{+0.13}_{-0.18}$ \\
		2--4 & 3.00 & $0.37^{+0.17}_{-0.13}$ & $7.76^{+0.09}_{-0.14}$ \\
		4--10 & 5.50 & $0.34^{+0.14}_{-0.18}$ & $7.62^{+0.11}_{-0.10}$ \\
        \hline
	\end{tabular}
\end{table}

\begin{figure*}
	\centering
    \includegraphics[width=\textwidth]{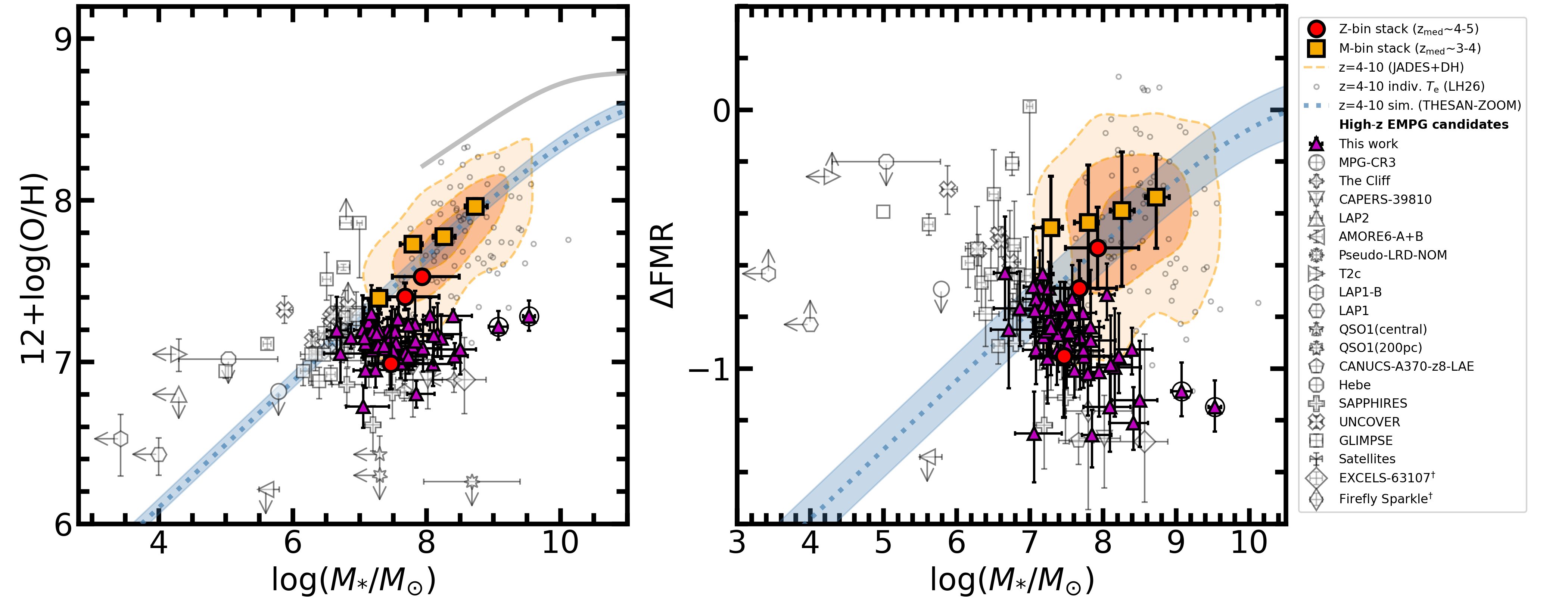}
    \caption{Mass-metallicity relation (left) and metallicity offset from \citet{Curti2020}'s $z\sim0$ Fundamental Metallicity Relation ($\Delta$FMR; right). The symbols are the same as those in Figure \ref{fig:sfms}, while the errors on the metallicities of the $Z$-bin and $M$-bin stacks are the errors measured from the stacked spectra.
    }
    \label{fig:mzrfmr}
\end{figure*}

The left panel of Figure \ref{fig:mzrall} shows MZRs of the JADES+DH sources at $z=1$--10 colour-coded by redshift.
As the metallicity decreases at a fixed $M_{*}$, the colour appears to shift from blue to red, suggesting lower metallicities towards higher $z$.
More quantitatively, we separate the sample with $\log(M_{*}/M_{\odot})=7$--9.5 into five $M_{*}$ bins of a 0.5-dex width, and calculate the median value and 16th and 84th percentiles in each bin.
These measurements are shown by the circles with the error bars in the right panel of Figure \ref{fig:mzrall}, and fitted with the following linear function used by \citet{Curti2024}:
\begin{equation}
    12+\log(\mathrm{O/H})=\beta[\log(M_{*}/M_{\odot})-8]+Z_{8}.
    \label{equ:mzr}
\end{equation}
We explore the best-fit values of $\beta$ and $Z_{8}$ by maximising the sum of the log-likelihood function defined by Equation \ref{equ:llhdet}.
The best-fit functions with their uncertainties based on Monte Carlo simulations with 1000 steps are shown by the lines with the shaded regions in the right panel of Figure \ref{fig:mzrall}.
Table \ref{tab:mzrfit} lists the median redshifts ($z_{\mathrm{med}}$) and the best-fit parameters with their errors based on the Monte Carlo simulations.

Comparing the results with the $z\sim0$ MZR \citep{Curti2020}, we confirm the decrease in metallicities from $z\sim0$, $z=1$--2, $z=2$--4 to $z=4$--10 as pointed out in numerous previous works \citep[e.g.,][]{Nakajima2023,Langeroodi2023b,Curti2024,Stanton2026}.
Indeed, the $Z_{8}$ values are almost linearly anticorrelated with $\log(1+z_{\mathrm{med}})$.
Fitting the linear function by maximising the log-likelihood function of Equation \ref{equ:llhdet}, we obtain $Z_{8}(z)=-0.71\times\log(1+z)+8.20$.
The $Z_{8}(z=0)$ value agrees well with the metallicity of \citet{Curti2020}'s $z\sim0$ MZR at $M_{*}=10^{8}\ M_{\odot}$.
Our MZR at $z=2$--4 is well connected to the $z=2.3$ MZR of \citet{Sanders2021} at $M_{*}\sim10^{9}\,M_{\odot}$.
In addition, our MZR at $z=4$--10 agrees well with that of the \textsc{thesan-zoom} simulations \citep{McClymont2025c}.
Detailed comparisons with more references on high-$z$ MZRs are presented in Appendix \ref{apsec:mzrref}.

We find that our MZR slopes do not significantly evolve from $z=1$--2, $z=2$--4 to $z=4$--10 with $\beta=0.34$--0.37, which matches the slope at the low-mass end of the $z\sim0$ MZR \citep[$\beta\rightarrow0.28$;][]{Curti2020}.
This suggests a small evolution of the MZR slopes across a wide $z$ range.
It has been argued that a physical mechanism of galactic winds determines the MZR slope at the metallicity equilibrium, where energy-driven winds predict a steeper slope \citep[$\beta\sim0.33$;][]{Dave2012,Lilly2013} than momentum-driven winds \citep[$\beta\sim0.17$;][]{Guo2016}.
Our MZR slopes agree well with the steeper slope, supporting the scenario of energy-driven winds.

\subsection{Low-mass end} \label{subsec:empgmzr}

\begin{figure*}
    \centering
    \begin{minipage}{\textwidth}
      \includegraphics[width=\textwidth]{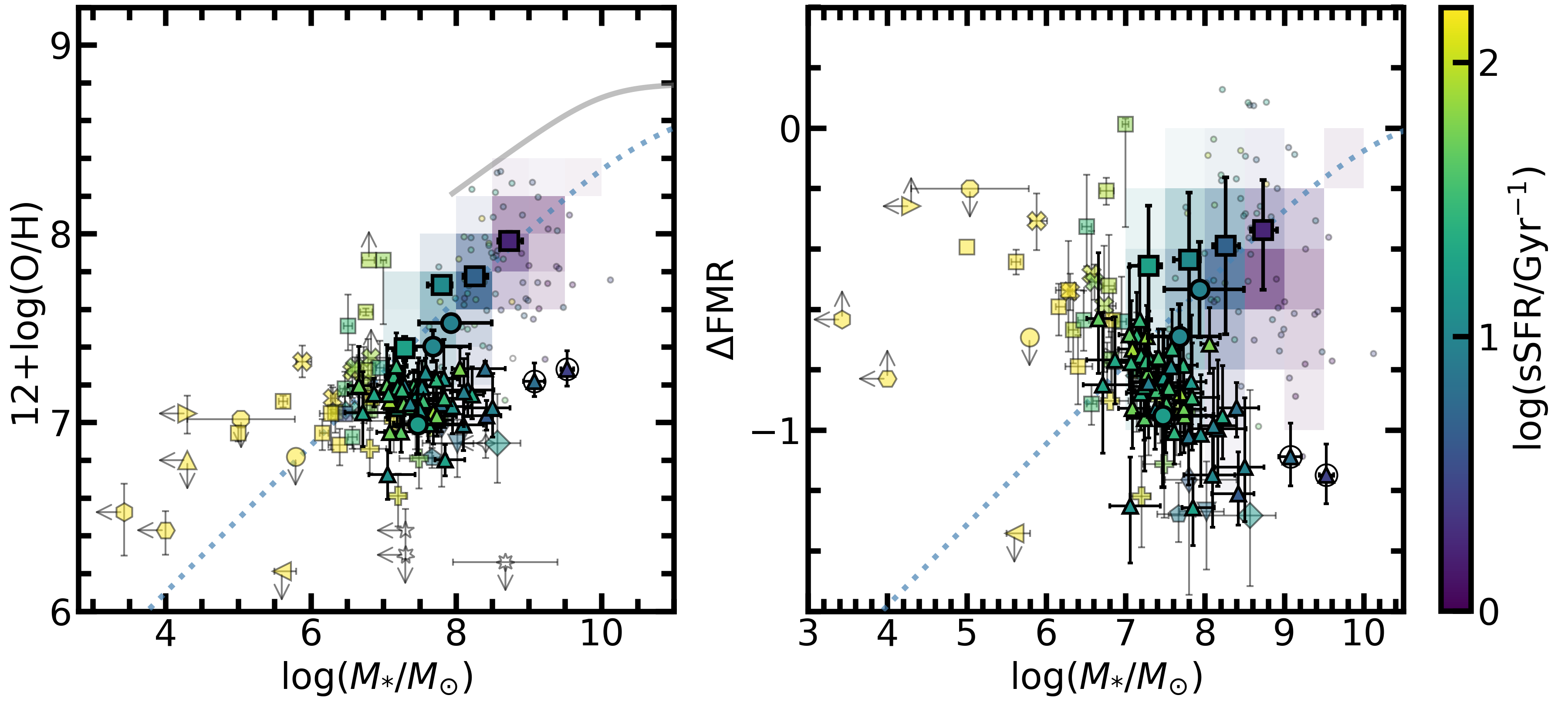}
    \end{minipage}
    \begin{minipage}{\textwidth}
      \includegraphics[width=\textwidth]{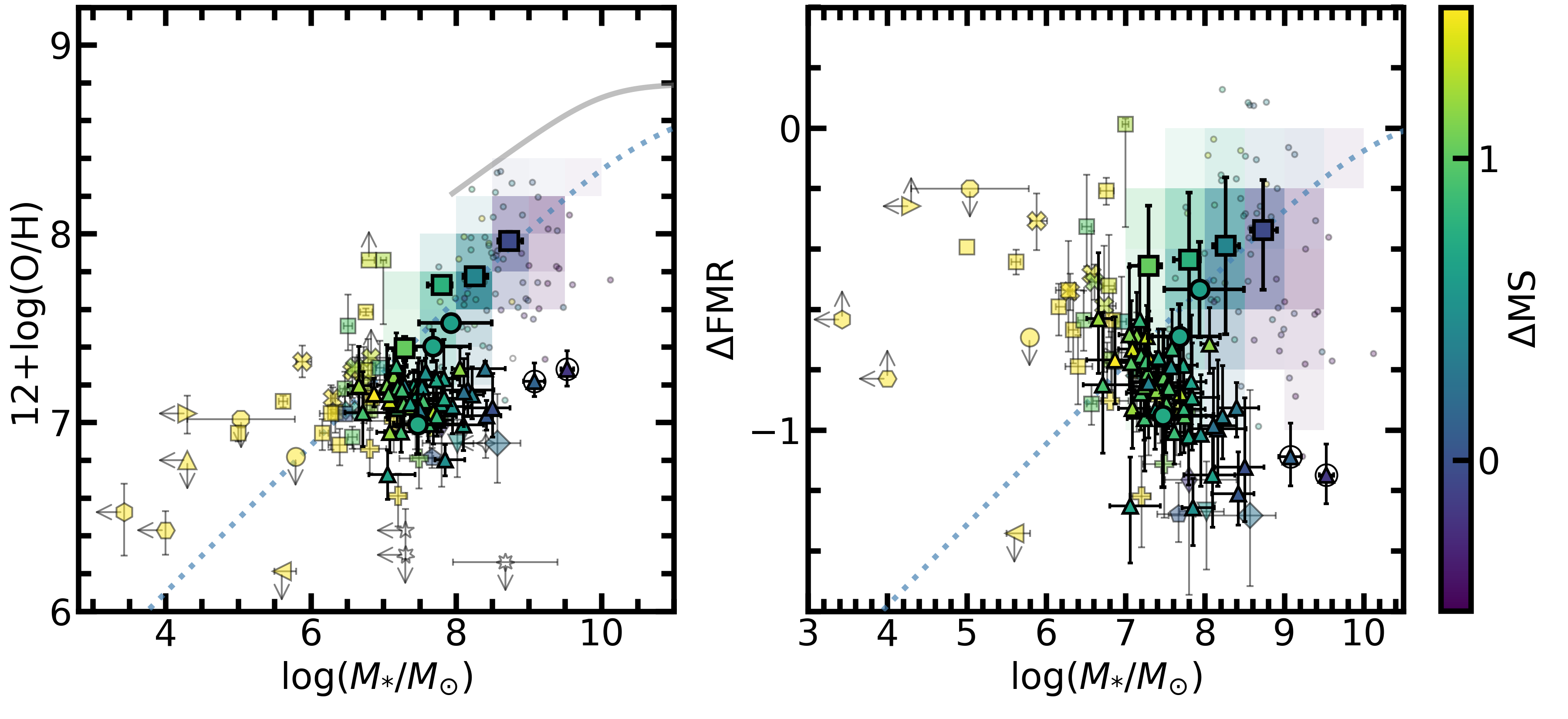}
    \end{minipage}
    \caption{Same as Figure \ref{fig:mzrfmr} but colour-coded by H$\alpha$-based sSFR (top) and the offset of H$\alpha$-based SFRs from the \textsc{thesan-zoom} simulations ($\Delta$MS; bottom). The blue dotted curves are the prediction by the \textsc{thesan-zoom} simulations \citep{McClymont2025a,McClymont2025c} at $z=7$. More metal-poor galaxies generally show lower sSFRs and $\Delta$MS values.}
    \label{fig:mzrfmr_cc}
\end{figure*}

\begin{figure*}
	\centering
    \includegraphics[width=\textwidth]{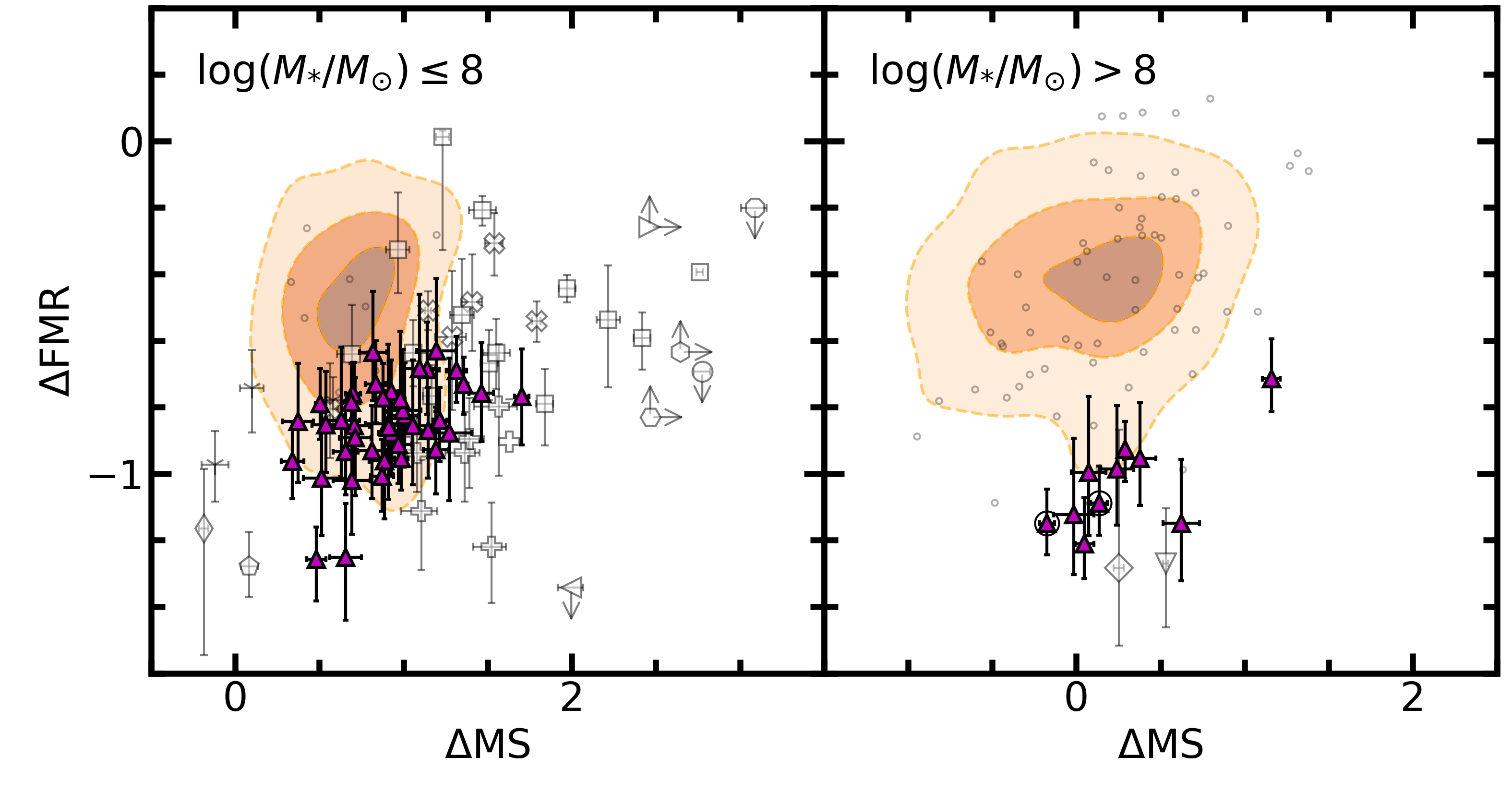}
    \caption{$\Delta$FMR vs. $\Delta$MS for sources with $\log(M_{*}/M_{\odot})\le8$ (left) and $\log(M_{*}/M_{\odot})>8$ (right). The symbols are the same as in Figure \ref{fig:mzrfmr}. Both EMPG candidates (magenta triangles and white points) and the JADES+DH sources at $z=4$--10 (orange contour) exhibit positive correlations between $\Delta$FMR and $\Delta$MS in the $\log(M_{*}/M_{\odot})\le8$ bin, while the correlations look weaker in the $\log(M_{*}/M_{\odot})>8$ bin.}
    \label{fig:dfmrdms}
\end{figure*}

\begin{figure*}
	\centering
    \includegraphics[width=\textwidth]{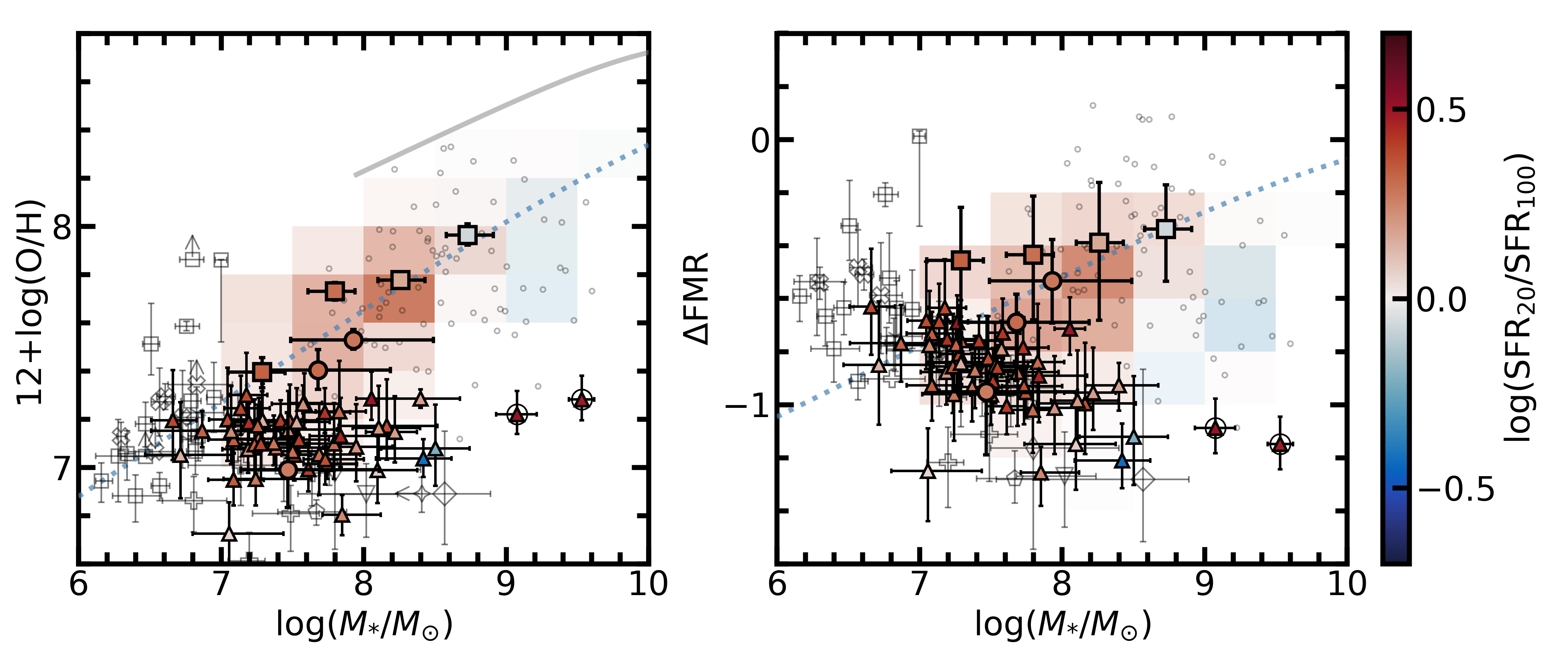}
    \caption{Same as Figure \ref{fig:mzrfmr} but we colour-code our EMPG candidates and the JADES+DH sources by SFR$_{20}$/SFR$_{100}$ (Section \ref{subsec:phot}). Most of our EMPG candidates have SFR$_{20}$/SFR$_{100}>0$.}
    \label{fig:mzrfmr_cc2}
\end{figure*}

\begin{figure}
	\centering
    \includegraphics[width=\columnwidth]{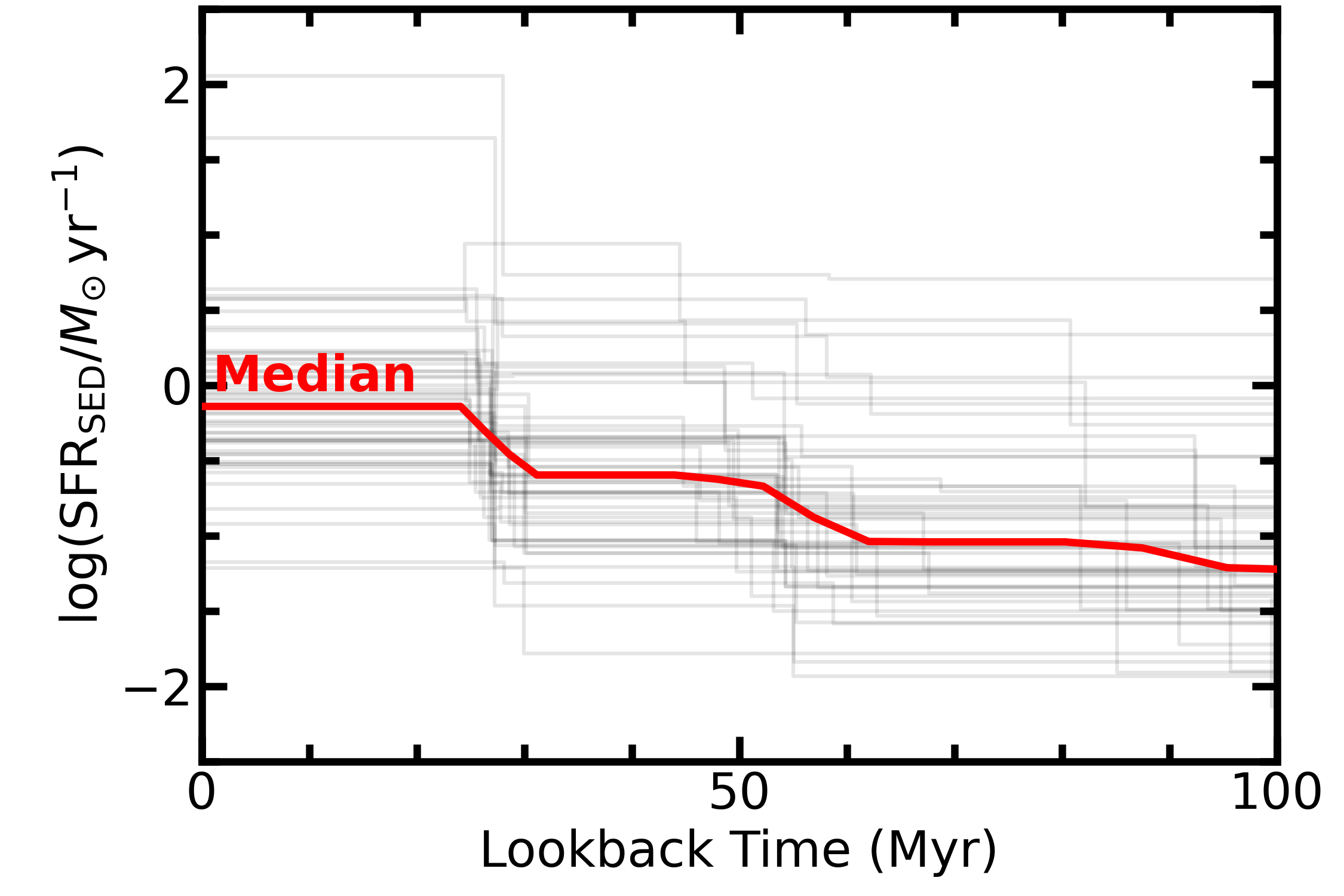}
    \caption{SFRs based on the SED fitting (SFR$_{\mathrm{SED}}$; Section \ref{subsec:phot}) of our EMPG candidates (grey thin lines) and their median value (red thick line) as a function of lookback time. Our EMPGs generally exhibit a rising star-formation history.}
    \label{fig:sfh}
\end{figure}

The magenta triangles in Figure \ref{fig:mzrfmr} show our EMPG candidates.
We find that our EMPG candidates generally have lower \met\ than the overall R1000 sources.
Although the exposure times of JADES, DH, and OASIS are long (Section \ref{sec:datsamp}), the observational limit on $M_{*}$ for the unlensed fields is at $\sim10^{7}\ M_{\odot}$.
Most our EMPG candidates exhibit higher $M_{*}$ values than the simulated MZR, likely because the simulated MZR predicts $M_{*}\lesssim10^{7}\ M_{\odot}$ for our metallicity threshold of \met\,$<7.3$ \citep[a.k.a. Malmquist bias;][]{Malmquist1922}.
Indeed, high-$z$ EMPG candidates from the literature show a large scatter in the MZR plot, some of which are above the simulated MZR.
Such EMPG candidates are mostly lensed sources such as LAP1 \citep{Vanzella2023}.

The right panel of Figure \ref{fig:mzrfmr} shows the offset of the observed \met\ values from those inferred from the $z\sim0$ FMR \citep{Curti2020} based on the SFR$_{\mathrm{H}\alpha}$ values ($\Delta$FMR, hereafter).
As reported by several studies on low-mass galaxies (see Section \ref{sec:intro}), we confirm that the R1000 sources at $z=4$--10 exhibit systematically negative $\Delta$FMR values.
Notably, our EMPG candidates have even lower $\Delta$FMR values, some of which reach $<-1$.
Combining the EMPG candidates from the literature, we identify a large scatter in the $\Delta$FMR-$M_{*}$ relations as well.

To discuss the origin of the large scatters, we colour-code the data points in Figure \ref{fig:mzrfmr} by sSFRs, which are shown in the top panels of Figure \ref{fig:mzrfmr_cc}.
We find that the EMPG candidates below the MZR tend to have lower sSFRs than those above the MZR.
We identify a similar trend in the $\Delta$FMR-$M_{*}$ plane.
In addition, we calculate the offset of SFR$_{\mathrm{H}\alpha}$ from the SFMS predicted by the \textsc{thesan-zoom} simulations ($\Delta$MS).
The bottom panels of Figure \ref{fig:mzrfmr_cc} illustrate that the EMPG candidates below the simulated MZR and the $\Delta$FMR-$M_{*}$ relation tend to have smaller $\Delta$MS values.
Figure \ref{fig:dfmrdms} shows the relation between $\Delta$FMR and $\Delta$MS of the EMPG candidates and the JADES+DH sources, separating them into those with $\log(M_{*}/M_{\odot})\le8$ and $\log(M_{*}/M_{\odot})>8$.
We find that the EMPG candidates exhibit a positive correlation between $\Delta$FMR and $\Delta$MS in the $\log(M_{*}/M_{\odot})\le8$ bin.
Moreover, we identify a positive correlation in the JADES+DH sources with $\log(M_{*}/M_{\odot})\le8$.
On the other hand, these correlations look weaker in the $\log(M_{*}/M_{\odot})>8$ bin.
These results suggest that more metal-poor galaxies have \textit{less} bursty star formation at the low-mass end.

This trend is opposite to the well-established $z\sim0$ observational work, which has reported that more metal-poor galaxies have higher sSFRs \citep[e.g.,][]{Andrews2013,Curti2020}.
However, these studies are based on more massive galaxies ($M_{*}\gtrsim10^{9}\,M_{\odot}$) than the high-$z$ sources that we discuss.
Indeed, a recent observational work on low-mass galaxies at $z\sim0$ \citep{Laseter2025} reports that galaxies with more negative $\Delta$FMR values tend to have lower sSFRs at a given $M_{*}$, suggesting that this trend may be specific to low-mass galaxies across a range of redshifts.

Two main scenarios have been proposed to explain more negative $\Delta$FMR values: inflows of metal-poor gas and outflows of metal-enriched gas \citep[e.g.,][]{Nishigaki2025b}.
It has been argued that the $\Delta$FMR decrease cannot be explained by inflows unless the pre-existing ISM is substantially lost \citep{Dalcanton2007,Tacchella2023a,Laseter2025}.
However, the \textsc{thesan-zoom} simulations \citep{McClymont2025a,McClymont2025c} predict such significant mass loss in low-mass galaxies with shallow potential wells. These simulations indicate a stochastic SFH characterised by cycles of starburst, gas consumption and ejection, temporary quenching, subsequent pristine inflow, and rejuvenated starburst.
In these simulations, the evolutionary tracks of galaxies deviate from the canonical MZR during phases of downward motion, indicating that these galaxies have recently undergone a gas-poor, mini-quenched phase, during which metal-poor gas inflow substantially dilutes their metallicity.
Since a shorter timescale of inflow is enough to produce a significant metallicity dilution for a smaller pre-existing gas reservoir, lower-mass galaxies are expected to recover their metallicities more rapidly than their $\Delta$MS values, leading to lower $\Delta$MS values than those of the galaxies on the canonical MZR.
In fact, both the evolutionary tracks and the bulk of simulated galaxies below the MZR show more negative $\Delta$MS values, which agree with the observational results.

In this scenario, galaxies below the MZR are thought to be observed at a stage when their inflow-triggered SFR is still increasing.
Consequently, the SFR measured over recent timescales is expected to exceed that averaged over longer timescales.
To test this, we colour-code our EMPG candidates by $\mathrm{SFR_{20}/SFR_{100}}$ in Figure \ref{fig:mzrfmr_cc2}, which are derived from the SED fitting (Section \ref{subsec:phot}).
We find that most candidates exhibit $\log(\mathrm{SFR_{20}/SFR_{100}})>0$.
Indeed, Figure \ref{fig:sfh} shows the SFHs of our EMPG candidates.
The median $\mathrm{SFR_{SED}}$ value increases with decreasing lookback time, indicating that more recent star formation is generally more active.
Such rising SFHs are in agreement with the inflow scenario.

Of course, this consistency does not rule out the enriched outflow scenario.
However, our results place constraints on such outflow models, requiring that $\log(\mathrm{SFR_{20}/SFR_{100}})$ remains positive even when some gas is lost through outflows.

\subsection{Metal-poor LRDs} \label{subsec:lrd}
\begin{figure*}
	\centering
    \includegraphics[width=\textwidth]{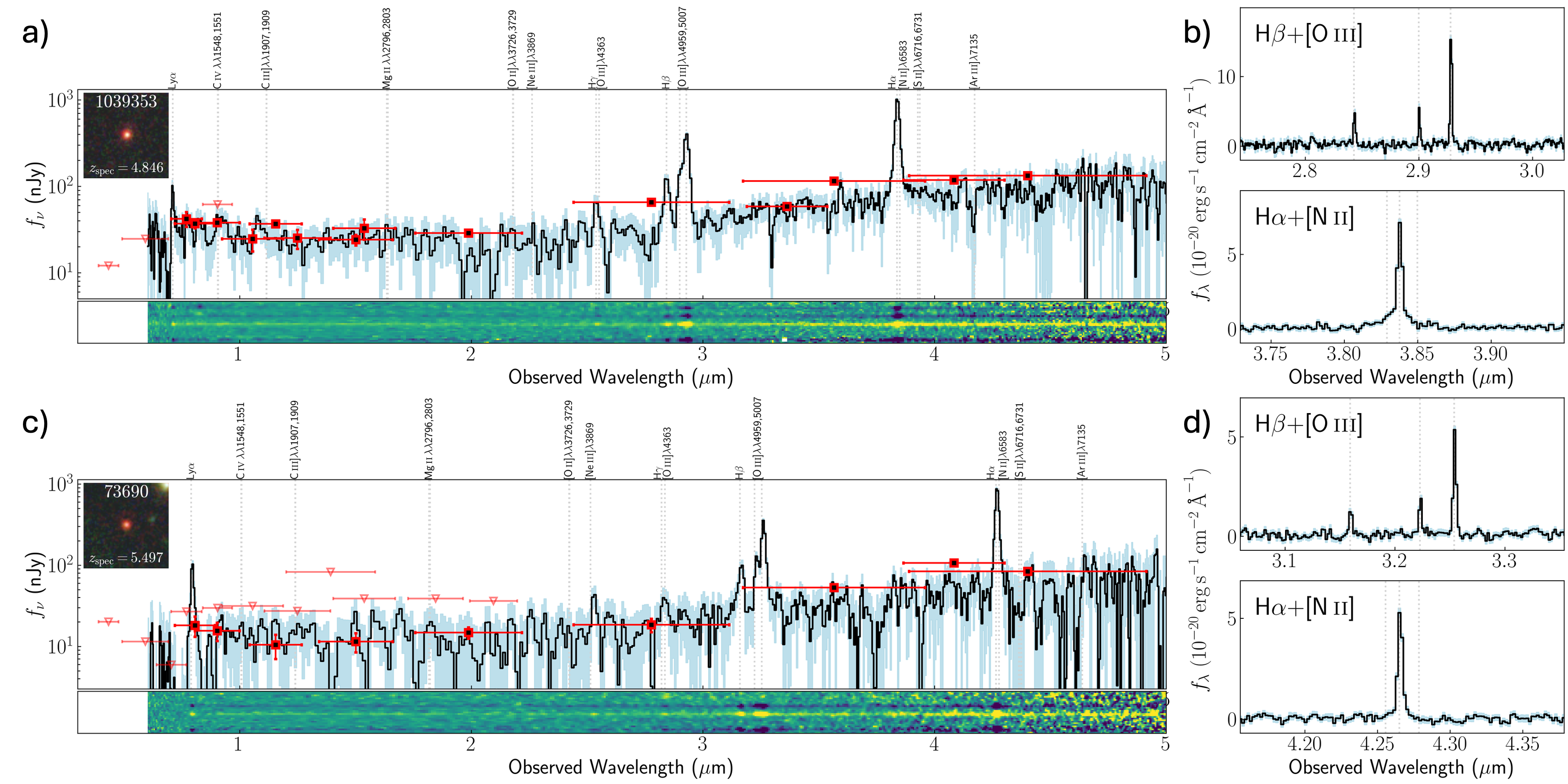}
    \caption{
    Spectra of LRDs identified in our EMPG candidates: goods-n-mediumjwst\_39353 and goods-s-mediumjwst\_73690.
    \textbf{a)}: R100 spectrum of goods-n-mediumjwst\_39353 with the Kron photometry of NIRCam, ACS, and WFC3 (red square: $S/N>3$; red triangle: 3$\sigma$ upper limit). The inset panel shows the 2''$\times$2'' cutout image of NIRCam (R: F444W; G: F200W; B: F115W).
    \textbf{b)}: R1000 goods-n-mediumjwst\_39353 around H$\beta$+[\oiii]$\lambda\lambda$4959,5007 (top) and H$\alpha$+[\nii]$\lambda\lambda$6548,6583 (bottom).
    \textbf{c)} and \textbf{d)}: Same as a) and b), respectively, but for goods-s-mediumjwst\_73690.
    Both sources exhibit blue UV colour and red optical colour that satisfy the selection criteria of LRDs proposed by \citet{Rinaldi2026}.
    These sources also have a broad component in H$\alpha$ but not in [\oiii], suggesting the presence of accreting black holes.
    }
    \label{fig:lrd}
\end{figure*}

From our EMPG candidates, we identify two sources whose restframe optical colour is significantly red: goods-n-mediumjwst\_39353 and goods-s-mediumjwst\_73690.
Figure \ref{fig:lrd} shows the spectra and photometric points of these sources.
Both sources are included in the updated census of LRDs in GOODS-N and GOODS-S by \citet{Rinaldi2026}, as they satisfy the more inclusive photometric criteria adopted in that work: $\mathrm{F150W}-\mathrm{F200W}<1$~mag and $\mathrm{F277W}-\mathrm{F444W}>0.5$~mag. Specifically, goods-s-mediumjwst\_73690 has $\mathrm{F150W}-\mathrm{F200W}=0.18$~mag and $\mathrm{F277W}-\mathrm{F444W}=1.16$~mag, while goods-n-mediumjwst\_39353 has $\mathrm{F150W}-\mathrm{F200W}=0.01$~mag and $\mathrm{F277W}-\mathrm{F444W}=0.88$~mag.
Crossmatching with the JADES DR5 morphology catalogue \citep{Carreira2026}, we find that both sources exhibit effective radii of F444W smaller than the size of the point-spread function in the same filter, indicating that they are extremely compact.
The cutout images generated using \texttt{Trilogy} \citep{Coe2012} also illustrate the compact morphology.
Hereafter, we refer to these two sources as EMP LRDs.

The right panels show the R1000 spectra of the EMP LRDs around H$\beta$+[\oiii]$\lambda\lambda$4959,5007 and H$\alpha$+[\nii]$\lambda\lambda$6548,6583.
We find that both EMP LRDs appear to have a broad component in H$\alpha$.
Following the methodology of \citet{Juodzbalis2025} and using the R1000 data, we fit the model that includes narrow components of [\nii]$\lambda\lambda$6548,6583 and narrow+broad components of H$\alpha$.
We fix the ratio of [\nii]$\lambda$6548/[\nii]$\lambda$6583 to 1/3, and assume that all the components share the same systemic velocity.
We constrain the narrow components to have the same velocity width.
We obtain $S/N$ of the broad component ($S/N_{\mathrm{Broad}}$) from the ratio of the broad line flux integrated within the full-width half maximum to the error added in quadrature over the same range.
We also fit the model without the broad H$\alpha$ component (i.e., narrow-component only model).
The fitting results are shown in Figure \ref{fig:lrd_br}.
We calculate the Bayesian information criterion (BIC) for the narrow-component only model (BIC$_{\mathrm{Narrow}}$) as well as the original model (BIC$_{\mathrm{Narrow+Broad}}$).
We obtain $\Delta\mathrm{BIC}\equiv\mathrm{BIC_{Narrow}}-\mathrm{BIC_{Narrow+Broad}}=6.9$--77 and $S/N_{\mathrm{Broad}}=6.4$--15 for the EMP LRDs, both exceeding 5 and thus satisfying the criteria used by \citet{Juodzbalis2025} for statistically supporting the presence of a broad-line component.
Note that these objects are not included in \citet{Juodzbalis2025}'s Type-1 AGN catalogue simply because they were observed more recently than the data analysed by that study.
On the other hand, the [\oiii]$\lambda$5007 lines do not exhibit such a significant broad component, suggesting that the observed broad components of H$\alpha$ are emitted from AGN broad-line regions.

Our EMP LRDs, which likely host growing black holes in a very low metallicity environment, are
similar to other known low-metallicity LRDs, such as QSO1 \citep{Maiolino2025} and The Cliff \citep{Ivey2026}. As discussed in \citet{Maiolino2025}, \citet{Juodzbalis2026}, and \citet{Ivey2026}, these low-metallicity LRDs likely trace a very early phase of black hole formation, also supporting a heavy seeding scenario for their origin, at least for these specific cases.
Indeed, as highlighted by the encircled magenta triangles in Figure \ref{fig:mzrfmr}, our EMP LRDs exhibit the two highest values of stellar masses beyond $\log(M_{*}/M_{\odot})>9$, which cause $\sim1$-dex lower metallicities than those of the JADES+DH sources at a given $M_{*}$.
\citet{Ivey2026} discuss that such metallicity tension can possibly be seen in QSO1 \citep{Maiolino2025,Juodzbalis2026} and The Cliff \citep{Ivey2026} as well as $z\sim0$ LRD analogues \citep{Lin2026}, given that reducing $M_{*}$ to match the canonical MZR would require even more extreme conditions such as higher black-hole-to-stellar mass ratios beyond 1 ($M_{\mathrm{BH}}/M_{*}>1$) or higher Eddington ratios beyond 100 ($\lambda_{\mathrm{Edd}}>100$).
A detailed analysis on $M_{\mathrm{BH}}$ and $M_{*}$ for our EMP LRDs will be presented in a separate paper.

In addition, both EMP LRDs exhibit strong Ly$\alpha$ even with relatively high $E(B-V)$ values of $\sim0.5$, one of which has been reported in the JADES Ly$\alpha$ emitter catalogue \citep{Jones2025a}.
Following the procedure of \citet{Jones2024} and using the R100 data, we fit Ly$\alpha$ together with the UV continuum modelled by a power-law continuum redder than Ly$\alpha$ and the zero flux bluer than Ly$\alpha$ convolved with a Gaussian of width $F_{\mathrm{R}}\sigma_{\mathrm{R}}$, where $\sigma_{\mathrm{R}}$ is the dispersion of the R100 LSF \citep{deGraaff2024} and $F_{\mathrm{R}}$ is a variable factor ($\ge1$) to match the spectra.
We obtain $F(\mathrm{Ly}\alpha)=5.7$--$7.9\times10^{-18}\,\mathrm{erg\,s^{-1}\,cm^{-2}}$ and high values of $F(\mathrm{Ly}\alpha)/F(\mathrm{H}\beta)=5.9$--12 without dust-reddening corrections, which already reach 25\%--50\% of the ratio inferred from Case B recombination ($T_{\mathrm{e}}=15000$ K and $n_{\mathrm{e}}=100$ cm$^{-3}$).
Such a high Ly$\alpha$/H$\beta$ ratio comparable to Case B is also reported in another metal-poor LRD, QSO1 \citep{Ji2026,Tang2026}, in contrast to the low Ly$\alpha$/H$\alpha=0.07$ reported in another LRD, UNCOVER-45924 \citep{Torralba2026}, with a relatively high metallicity of $12+\log(\mathrm{O/H})=7.15$--7.67 \citep{Ji2025b}.
Our EMP LRDs do not have significant Balmer breaks, which is in line with the recent report that the stack of LRDs with weak Balmer breaks exhibits a more prominent Ly$\alpha$ than those with strong Balmer breaks \citep{Ando2026}.
Details of the Ly$\alpha$ properties of the EMP LRDs will also be discussed in a separate paper.

\section{Conclusions} \label{sec:con}
We present gas-phase metallicities of \nsrc\ star-forming galaxies at $z=1$--10 with the NIRSpec R100 and R1000 data from the JADES, DH, and OASIS surveys.
The large number of deep R1000 spectra allows us to detect [\oiii]$\lambda$4363 from all our R1000 stacks in narrow bins of \metsl\ and $M_{*}$.
Our findings are summarised as follows:
\begin{itemize}
    \item 
    We derive $T_{\mathrm{e}}$-based metallicities for our stacks down to $12+\log(\mathrm{O/H})=7.0$, entering the regime where the high-$z$ strong-line method was not well calibrated ($12+\log(\mathrm{O/H})\gtrsim7.3$).
    \item 
    At a given metallicity, our stacks have EW$_{0}$(H$\beta$) lower than individual auroral-line emitters at z$>$1.
    These results suggest that requiring faint auroral-line detections biases the samples towards galaxy populations with higher excitation and more bursty star formation. Indeed, our stacks have R3, R23, O32, $\hat{\mathrm{R}}_{\mathrm{Laseter}}$, $\hat{\mathrm{R}}_{\mathrm{Chakraborty}}$, $\tilde{\mathrm{R}}_{\mathrm{Cataldi}}$, and Ne3O2 that are generally lower than those of high-$z$ calibrations from the literature based on individual auroral-line emitters.
    \item Using our $z=1$--10 stacks, supplemented with stacks of $z\sim0$ metal-rich galaxies (with sSFRs higher than the $z=1$ SFMS), we obtain stack-based strong-line calibrations 
    over the range of $12+\log(\mathrm{O/H})=7.0$--8.7.
    Our R3 and R23 calibrations agree well with those of $z\sim0$ galaxies with large EW$_{0}$(H$\beta$).
    \item We report MZRs with the R1000 data at $z=1$--10, confirming the decrease in metallicities from $z\sim0$, $z=1$--2, $z=2$--4, to $z=4$--10 at a given $M_{*}$.
    We do not identify a significant evolution of the MZR slope ($\beta=0.34$--0.37), which suggests that a dominant feedback mechanism does not change drastically in general.
    \item Based on our stack-based calibrations, we select \nempg\ promising EMPG candidates from our \nsrc\ galaxies. Our EMPG candidates have $12+\log(\mathrm{O/H})=\metminempg$--\metmaxempg\ at $z=\zminempg$--\zmaxempg, most of which exhibit lower metallicities than the parent sample and the simulated MZR at $z=4$--10.
    \item Compared to literature EMPG candidates above the MZR, our EMPG candidates generally have lower sSFRs while their SFR$_{20}$ values are mostly higher than SFR$_{100}$.
    These results are consistent with a stochastic SFH with cycles of gas consumption/ejection and metal-poor inflow, as predicted by zoom-in simulations.
    \item We find that two of our EMPG candidates have red optical colours ($\mathrm{F277W}-\mathrm{F444W}=0.88$--1.16 mag) together with relatively blue UV colours ($\mathrm{F150W}-\mathrm{F200W}=0.01$--0.18 mag), which are included in the updated census of LRDs by \citep{Rinaldi2026}. These two LRDs exhibit significant H$\alpha$ broadening and high Ly$\alpha$/H$\beta$ ratios close to the Case B value. These are therefore Extremely Metal Poor LRDs, similar to a few others that have been discovered recently, which are likely tracing the earliest phases of black hole formation.
\end{itemize}

\section*{Acknowledgments}

We thank Tadayuki Kodama, Takahiro Morishita, Yoshihisa Asada, Yuma Sugahara, Thomas Stanton, Karina Caputi, Nicholas Choustikov, Callum Witten, Ali Ahmad Khostovan, Leonardo Clarke, Mingyu Li, and Fengwu Sun for useful discussions.
YI is supported by JSPS KAKENHI Grant No. 24KJ0202.
RM, DP, FDE, JS, XJ, GCJ, IJ, and RGP acknowledge support by the Science and Technology Facilities Council (STFC), by the ERC through Advanced Grant 695671 ``QUENCH'', and by the UKRI Frontier Research grant RISEandFALL.
RM also acknowledges funding from a research professorship from the Royal Society.
QD acknowledges support from a PhD studentship awarded by Trinity College, University of Cambridge.
WM thanks the Science and Technology Facilities Council
(STFC) Center for Doctoral Training (CDT) in Data Intensive Science at the University of Cambridge (STFC grant number 2742968) for a PhD studentship. WM and ST acknowledge support by the Royal Society Research Grant G125142.
DP and IJ acknowledge support by the Huo Family Foundation through a P.C. Ho PhD Studentship.
JAAT acknowledges support from the Simons Foundation and JWST program 3215. Support for program 3215 was provided by NASA through a grant from the Space Telescope Science Institute, which is operated by the Association of Universities for Research in Astronomy, Inc., under NASA contract NAS 5-03127.
TJL gratefully acknowledges support from the Swiss National Science Foundation through a SNSF Mobility Fellowship and from the NASA/JWST Program OASIS (PID \#5997).
The work presented here is supported by the Carlsberg Foundation, grant CF25-0662 (DL).
ST acknowledges support by the Royal Society Research Grant G125142.
H\"U acknowledges support by the Max Planck Society through the Lise Meitner Excellence Program. H\"U acknowledges funding by the European Union (ERC APEX, 101164796). Views and opinions expressed are however those of the authors only and do not necessarily reflect those of the European Union or the European Research Council Executive Agency. Neither the European Union nor the granting authority can be held responsible for them.
WMB gratefully acknowledges support from DARK via the DARK fellowship. This work was supported by a research grant (VIL54489) from VILLUM FONDEN.
AJB acknowledges funding from the "FirstGalaxies" Advanced Grant from the European Research Council (ERC) under the European Union’s Horizon 2020 research and innovation programme (Grant agreement No. 789056).
SC acknowledges support by European Union’s HE ERC Starting Grant No. 101040227 - WINGS.
ECL acknowledges support of an STFC Webb Fellowship (ST/W001438/1).
SG is supported by a Woolf Fisher Scholarship from the Woolf Fisher Trust of New Zealand and Cambridge Commonwealth and European Trust.
MK acknowledges support from a PhD studentship awarded by the University of Cambridge Harding Distinguished Postgraduate Scholars Programme, UK Science and Technology Facilities Council (STFC) Center for Doctoral Training (CDT) in Data Intensive Science, and Girton College Cambridge.
BER acknowledges support from the NIRCam Science Team contract to the University of Arizona, NAS5-02105, and JWST Program 3215.
The research of CCW is supported by NOIRLab, which is managed by the Association of Universities for Research in Astronomy (AURA) under a cooperative agreement with the National Science Foundation.
This work is based on observations made with the NASA/ESA/CSA James Webb Space Telescope. The data were obtained from the Mikulski Archive for Space Telescopes at the Space Telescope Science Institute, which is operated by the Association of Universities for Research in Astronomy, Inc., under NASA contract NAS 5-03127 for \textit{JWST}. These observations are associated with programmes \#1180, 1181, 1210, 1286, 1287, 3215, 4540, 1283, 1895, 1963, 2079, 2514, 3577, 3990, 5398, 5997, 6434, and 6541.
The authors acknowledge use of the lux supercomputer at UC Santa Cruz, funded by NSF MRI grant AST 1828315.

\section*{Data Availability}

The bulk of the \textit{JWST}/NIRSpec data and all the NIRCam data used in this paper are released by the JADES NIRSpec DR4 \citep{Curtis-Lake2025_DR4,Scholtz2025_DR4} and the JADES NIRCam DR5 \citep{Robertson2026_DR5,Johnson2026}, which are available on the JADES MAST website (\url{https://archive.stsci.edu/hlsp/jades}; MAST DOI: \href{https://dx.doi.org/10.17909/8tdj-8n28}{10.17909/8tdj-8n28}) and the JADES Public Online Database (\url{https://jades.herts.ac.uk/search/}).
The rest of datasets will also be public in the MAST archive.
Our analysed data will be made available upon reasonable request.



\bibliographystyle{mnras}
\bibliography{example} 

\begin{thebibliography}{}
\makeatletter
\relax
\def\mn@urlcharsother{\let\do\@makeother \do\$\do\&\do\#\do\^\do\_\do\%\do\~}
\def\mn@doi{\begingroup\mn@urlcharsother \@ifnextchar [ {\mn@doi@} {\mn@doi@[]}}
\def\mn@doi@[#1]#2{\def\@tempa{#1}\ifx\@tempa\@empty \href {http://dx.doi.org/#2} {doi:#2}\else \href {http://dx.doi.org/#2} {#1}\fi \endgroup}
\def\mn@eprint#1#2{\mn@eprint@#1:#2::\@nil}
\def\mn@eprint@arXiv#1{\href {http://arxiv.org/abs/#1} {{\tt arXiv:#1}}}
\def\mn@eprint@dblp#1{\href {http://dblp.uni-trier.de/rec/bibtex/#1.xml} {dblp:#1}}
\def\mn@eprint@#1:#2:#3:#4\@nil{\def\@tempa {#1}\def\@tempb {#2}\def\@tempc {#3}\ifx \@tempc \@empty \let \@tempc \@tempb \let \@tempb \@tempa \fi \ifx \@tempb \@empty \def\@tempb {arXiv}\fi \@ifundefined {mn@eprint@\@tempb}{\@tempb:\@tempc}{\expandafter \expandafter \csname mn@eprint@\@tempb\endcsname \expandafter{\@tempc}}}

\bibitem[\protect\citeauthoryear{{Alves de Oliveira} et~al.,}{{Alves de Oliveira} et~al.}{2018}]{Oliveira2018}
{Alves de Oliveira} C.,  et~al., 2018, in Observatory Operations: Strategies, Processes, and Systems VII. p. 107040Q (\mn@eprint {arXiv} {1805.06922}), \mn@doi{10.1117/12.2313839}

\bibitem[\protect\citeauthoryear{{Ando}, {Harikane}, {Katz}, {Inayoshi}  \& {Tanaka}}{{Ando} et~al.}{2026}]{Ando2026}
{Ando} M.,  {Harikane} Y.,  {Katz} H.,  {Inayoshi} K.,   {Tanaka} T.~S.,  2026, arXiv e-prints, \href {https://ui.adsabs.harvard.edu/abs/2026arXiv260603522A} {p. arXiv:2606.03522}

\bibitem[\protect\citeauthoryear{{Andrews} \& {Martini}}{{Andrews} \& {Martini}}{2013}]{Andrews2013}
{Andrews} B.~H.,  {Martini} P.,  2013, \mn@doi [\apj] {10.1088/0004-637X/765/2/140}, \href {https://ui.adsabs.harvard.edu/abs/2013ApJ...765..140A} {765, 140}

\bibitem[\protect\citeauthoryear{{Asada} et~al.,}{{Asada} et~al.}{2026}]{Asada2026}
{Asada} Y.,  et~al., 2026, \mn@doi [arXiv e-prints] {10.48550/arXiv.2601.20045}, \href {https://ui.adsabs.harvard.edu/abs/2026arXiv260120045A} {p. arXiv:2601.20045}

\bibitem[\protect\citeauthoryear{{Asplund}, {Amarsi}  \& {Grevesse}}{{Asplund} et~al.}{2021}]{Asplund2021}
{Asplund} M.,  {Amarsi} A.~M.,   {Grevesse} N.,  2021, \mn@doi [\aap] {10.1051/0004-6361/202140445}, \href {https://ui.adsabs.harvard.edu/abs/2021A&A...653A.141A} {653, A141}

\bibitem[\protect\citeauthoryear{{Bagley} et~al.,}{{Bagley} et~al.}{2024}]{Bagley2024}
{Bagley} M.~B.,  et~al., 2024, \mn@doi [\apjl] {10.3847/2041-8213/ad2f31}, \href {https://ui.adsabs.harvard.edu/abs/2024ApJ...965L...6B} {965, L6}

\bibitem[\protect\citeauthoryear{{Baker} \& {Maiolino}}{{Baker} \& {Maiolino}}{2023}]{BakerMaiolino2023}
{Baker} W.~M.,  {Maiolino} R.,  2023, \mn@doi [\mnras] {10.1093/mnras/stad802}, \href {https://ui.adsabs.harvard.edu/abs/2023MNRAS.521.4173B} {521, 4173}

\bibitem[\protect\citeauthoryear{{Baker} et~al.,}{{Baker} et~al.}{2023}]{Baker2023}
{Baker} W.~M.,  et~al., 2023, \mn@doi [\mnras] {10.1093/mnras/stac3594}, \href {https://ui.adsabs.harvard.edu/abs/2023MNRAS.519.1149B} {519, 1149}

\bibitem[\protect\citeauthoryear{Baldwin, Phillips  \& Terlevich}{Baldwin et~al.}{1981}]{Baldwin1981}
Baldwin A.,  Phillips M.~M.,   Terlevich R.,  1981, \mn@doi [PASP] {10.1086/130930}, 93, 817

\bibitem[\protect\citeauthoryear{Berg, Erb, Henry, Skillman  \& McQuinn}{Berg et~al.}{2019}]{Berg2019}
Berg D.~A.,  Erb D.~K.,  Henry R. B.~C.,  Skillman E.~D.,   McQuinn K. B.~W.,  2019, \mn@doi [ApJ] {10.3847/1538-4357/ab020a}, 874, 93

\bibitem[\protect\citeauthoryear{{Berg}, {Chisholm}, {Erb}, {Skillman}, {Pogge}  \& {Olivier}}{{Berg} et~al.}{2021}]{Berg2021}
{Berg} D.~A.,  {Chisholm} J.,  {Erb} D.~K.,  {Skillman} E.~D.,  {Pogge} R.~W.,   {Olivier} G.~M.,  2021, \mn@doi [\apj] {10.3847/1538-4357/ac141b}, \href {https://ui.adsabs.harvard.edu/abs/2021ApJ...922..170B} {922, 170}

\bibitem[\protect\citeauthoryear{{Bezanson} et~al.,}{{Bezanson} et~al.}{2024}]{Bezanson2024}
{Bezanson} R.,  et~al., 2024, \mn@doi [\apj] {10.3847/1538-4357/ad66cf}, \href {https://ui.adsabs.harvard.edu/abs/2024ApJ...974...92B} {974, 92}

\bibitem[\protect\citeauthoryear{{Bian}, {Kewley}  \& {Dopita}}{{Bian} et~al.}{2018}]{Bian2018}
{Bian} F.,  {Kewley} L.~J.,   {Dopita} M.~A.,  2018, \mn@doi [\apj] {10.3847/1538-4357/aabd74}, \href {https://ui.adsabs.harvard.edu/abs/2018ApJ...859..175B} {859, 175}

\bibitem[\protect\citeauthoryear{{Brown}, {Martini}  \& {Andrews}}{{Brown} et~al.}{2016}]{Brown2016}
{Brown} J.~S.,  {Martini} P.,   {Andrews} B.~H.,  2016, \mn@doi [\mnras] {10.1093/mnras/stw392}, \href {https://ui.adsabs.harvard.edu/abs/2016MNRAS.458.1529B} {458, 1529}

\bibitem[\protect\citeauthoryear{{Bunker} et~al.,}{{Bunker} et~al.}{2024}]{Bunker2024}
{Bunker} A.~J.,  et~al., 2024, \mn@doi [\aap] {10.1051/0004-6361/202347094}, \href {https://ui.adsabs.harvard.edu/abs/2024A&A...690A.288B} {690, A288}

\bibitem[\protect\citeauthoryear{{Burnham} \& {Anderson}}{{Burnham} \& {Anderson}}{2004}]{Burnham2004}
{Burnham} K.~P.,  {Anderson} D.~R.,  2004, \mn@doi [Sociological Methods \& Research] {10.1177/0049124104268644}, 33, 261

\bibitem[\protect\citeauthoryear{{Cai} \& {Yu}}{{Cai} \& {Yu}}{2026}]{Cai2026}
{Cai} S.,  {Yu} Z.,  2026, \mn@doi [arXiv e-prints] {10.48550/arXiv.2601.17498}, \href {https://ui.adsabs.harvard.edu/abs/2026arXiv260117498C} {p. arXiv:2601.17498}

\bibitem[\protect\citeauthoryear{{Cai} et~al.,}{{Cai} et~al.}{2025}]{Cai2025}
{Cai} S.,  et~al., 2025, \mn@doi [arXiv e-prints] {10.48550/arXiv.2507.17820}, \href {https://ui.adsabs.harvard.edu/abs/2025arXiv250717820C} {p. arXiv:2507.17820}

\bibitem[\protect\citeauthoryear{Calzetti, Armus, Bohlin, Kinney, Koornneef  \& Storchi‐Bergmann}{Calzetti et~al.}{2000}]{Calzetti2000}
Calzetti D.,  Armus L.,  Bohlin R.~C.,  Kinney A.~L.,  Koornneef J.,   Storchi‐Bergmann T.,  2000, \mn@doi [ApJ] {10.1086/308692}, 533, 682

\bibitem[\protect\citeauthoryear{{Cameron}, {Katz}, {Rey}  \& {Saxena}}{{Cameron} et~al.}{2023}]{Cameron2023}
{Cameron} A.~J.,  {Katz} H.,  {Rey} M.~P.,   {Saxena} A.,  2023, \mn@doi [\mnras] {10.1093/mnras/stad1579}, \href {https://ui.adsabs.harvard.edu/abs/2023MNRAS.523.3516C} {523, 3516}

\bibitem[\protect\citeauthoryear{{Cameron} et~al.,}{{Cameron} et~al.}{2026}]{Cameron2026}
{Cameron} A.~J.,  et~al., 2026, \mn@doi [arXiv e-prints] {10.48550/arXiv.2601.15964}, \href {https://ui.adsabs.harvard.edu/abs/2026arXiv260115964C} {p. arXiv:2601.15964}

\bibitem[\protect\citeauthoryear{{Cappellari}}{{Cappellari}}{2023}]{Cappellari2023}
{Cappellari} M.,  2023, \mn@doi [\mnras] {10.1093/mnras/stad2597}, \href {https://ui.adsabs.harvard.edu/abs/2023MNRAS.526.3273C} {526, 3273}

\bibitem[\protect\citeauthoryear{{Caputi}, {Cooper}, {Rinaldi}, {Navarro-Carrera}  \& {Iani}}{{Caputi} et~al.}{2026}]{Caputi2026}
{Caputi} K.~I.,  {Cooper} R.~A.,  {Rinaldi} P.,  {Navarro-Carrera} R.,   {Iani} E.,  2026, \mn@doi [arXiv e-prints] {10.48550/arXiv.2601.11466}, \href {https://ui.adsabs.harvard.edu/abs/2026arXiv260111466C} {p. arXiv:2601.11466}

\bibitem[\protect\citeauthoryear{{Cardelli}, {Clayton}  \& {Mathis}}{{Cardelli} et~al.}{1989}]{Cardelli1989}
{Cardelli} J.~A.,  {Clayton} G.~C.,   {Mathis} J.~S.,  1989, \mn@doi [\apj] {10.1086/167900}, \href {https://ui.adsabs.harvard.edu/abs/1989ApJ...345..245C} {345, 245}

\bibitem[\protect\citeauthoryear{{Carnall}}{{Carnall}}{2017}]{Carnall2017}
{Carnall} A.~C.,  2017, \mn@doi [arXiv e-prints] {10.48550/arXiv.1705.05165}, \href {https://ui.adsabs.harvard.edu/abs/2017arXiv170505165C} {p. arXiv:1705.05165}

\bibitem[\protect\citeauthoryear{{Carreira} et~al.,}{{Carreira} et~al.}{2026}]{Carreira2026}
{Carreira} C.,  et~al., 2026, \mn@doi [arXiv e-prints] {10.48550/arXiv.2601.15957}, \href {https://ui.adsabs.harvard.edu/abs/2026arXiv260115957C} {p. arXiv:2601.15957}

\bibitem[\protect\citeauthoryear{{Cataldi} et~al.,}{{Cataldi} et~al.}{2025a}]{Cataldi2025}
{Cataldi} E.,  et~al., 2025a, \mn@doi [arXiv e-prints] {10.48550/arXiv.2504.03839}, \href {https://ui.adsabs.harvard.edu/abs/2025arXiv250403839C} {p. arXiv:2504.03839}

\bibitem[\protect\citeauthoryear{{Cataldi} et~al.,}{{Cataldi} et~al.}{2025b}]{Cataldi2025b}
{Cataldi} E.,  et~al., 2025b, \mn@doi [arXiv e-prints] {10.48550/arXiv.2512.07955}, \href {https://ui.adsabs.harvard.edu/abs/2025arXiv251207955C} {p. arXiv:2512.07955}

\bibitem[\protect\citeauthoryear{{Chabrier}}{{Chabrier}}{2003}]{chabrier2003imf}
{Chabrier} G.,  2003, \mn@doi [\pasp] {10.1086/376392}, \href {https://ui.adsabs.harvard.edu/abs/2003PASP..115..763C} {115, 763}

\bibitem[\protect\citeauthoryear{{Chakraborty} et~al.,}{{Chakraborty} et~al.}{2025}]{Chakraborty2025}
{Chakraborty} P.,  et~al., 2025, \mn@doi [\apj] {10.3847/1538-4357/adc7b5}, \href {https://ui.adsabs.harvard.edu/abs/2025ApJ...985...24C} {985, 24}

\bibitem[\protect\citeauthoryear{{Charlot} \& {Fall}}{{Charlot} \& {Fall}}{2000}]{charlot2000}
{Charlot} S.,  {Fall} S.~M.,  2000, \mn@doi [\apj] {10.1086/309250}, \href {https://ui.adsabs.harvard.edu/abs/2000ApJ...539..718C} {539, 718}

\bibitem[\protect\citeauthoryear{{Chemerynska} et~al.,}{{Chemerynska} et~al.}{2024}]{Chemerynska2024}
{Chemerynska} I.,  et~al., 2024, \mn@doi [\apjl] {10.3847/2041-8213/ad8dc9}, \href {https://ui.adsabs.harvard.edu/abs/2024ApJ...976L..15C} {976, L15}

\bibitem[\protect\citeauthoryear{{Choi}, {Dotter}, {Conroy}, {Cantiello}, {Paxton}  \& {Johnson}}{{Choi} et~al.}{2016}]{choi2016}
{Choi} J.,  {Dotter} A.,  {Conroy} C.,  {Cantiello} M.,  {Paxton} B.,   {Johnson} B.~D.,  2016, \mn@doi [\apj] {10.3847/0004-637X/823/2/102}, \href {https://ui.adsabs.harvard.edu/abs/2016ApJ...823..102C} {823, 102}

\bibitem[\protect\citeauthoryear{{Chon}, {Hosokawa}, {Omukai}  \& {Schneider}}{{Chon} et~al.}{2024}]{Chon2024}
{Chon} S.,  {Hosokawa} T.,  {Omukai} K.,   {Schneider} R.,  2024, \mn@doi [\mnras] {10.1093/mnras/stae1027}, \href {https://ui.adsabs.harvard.edu/abs/2024MNRAS.530.2453C} {530, 2453}

\bibitem[\protect\citeauthoryear{{Choustikov} et~al.,}{{Choustikov} et~al.}{2026}]{Choustikov2026}
{Choustikov} N.,  et~al., 2026, \mn@doi [The Open Journal of Astrophysics] {10.33232/001c.158199}, \href {https://ui.adsabs.harvard.edu/abs/2026OJAp....958199C} {9, 58199}

\bibitem[\protect\citeauthoryear{{Clarke}, {Shapley}, {Sanders}, {Topping}, {Brammer}, {Bento}, {Reddy}  \& {Kehoe}}{{Clarke} et~al.}{2024}]{Clarke2024}
{Clarke} L.,  {Shapley} A.~E.,  {Sanders} R.~L.,  {Topping} M.~W.,  {Brammer} G.~B.,  {Bento} T.,  {Reddy} N.~A.,   {Kehoe} E.,  2024, \mn@doi [\apj] {10.3847/1538-4357/ad8ba4}, \href {https://ui.adsabs.harvard.edu/abs/2024ApJ...977..133C} {977, 133}

\bibitem[\protect\citeauthoryear{{Clarke}, {Shapley}, {Lam}, {Topping}, {Brammer}, {Sanders}, {Reddy}  \& {Karthikeyan}}{{Clarke} et~al.}{2025}]{Clarke2025}
{Clarke} L.,  {Shapley} A.~E.,  {Lam} N.,  {Topping} M.~W.,  {Brammer} G.~B.,  {Sanders} R.~L.,  {Reddy} N.~A.,   {Karthikeyan} S.,  2025, \mn@doi [arXiv e-prints] {10.48550/arXiv.2510.06681}, \href {https://ui.adsabs.harvard.edu/abs/2025arXiv251006681C} {p. arXiv:2510.06681}

\bibitem[\protect\citeauthoryear{{Coe} et~al.,}{{Coe} et~al.}{2012}]{Coe2012}
{Coe} D.,  et~al., 2012, \mn@doi [\apj] {10.1088/0004-637X/757/1/22}, \href {https://ui.adsabs.harvard.edu/abs/2012ApJ...757...22C} {757, 22}

\bibitem[\protect\citeauthoryear{{Cullen} et~al.,}{{Cullen} et~al.}{2025}]{Cullen2025}
{Cullen} F.,  et~al., 2025, \mn@doi [\mnras] {10.1093/mnras/staf838}, \href {https://ui.adsabs.harvard.edu/abs/2025MNRAS.540.2176C} {540, 2176}

\bibitem[\protect\citeauthoryear{{Curti}, {Cresci}, {Mannucci}, {Marconi}, {Maiolino}  \& {Esposito}}{{Curti} et~al.}{2017}]{Curti2017}
{Curti} M.,  {Cresci} G.,  {Mannucci} F.,  {Marconi} A.,  {Maiolino} R.,   {Esposito} S.,  2017, \mn@doi [\mnras] {10.1093/mnras/stw2766}, \href {https://ui.adsabs.harvard.edu/abs/2017MNRAS.465.1384C} {465, 1384}

\bibitem[\protect\citeauthoryear{{Curti}, {Mannucci}, {Cresci}  \& {Maiolino}}{{Curti} et~al.}{2020}]{Curti2020}
{Curti} M.,  {Mannucci} F.,  {Cresci} G.,   {Maiolino} R.,  2020, \mn@doi [\mnras] {10.1093/mnras/stz2910}, \href {https://ui.adsabs.harvard.edu/abs/2020MNRAS.491..944C} {491, 944}

\bibitem[\protect\citeauthoryear{{Curti} et~al.,}{{Curti} et~al.}{2024}]{Curti2024}
{Curti} M.,  et~al., 2024, \mn@doi [\aap] {10.1051/0004-6361/202346698}, \href {https://ui.adsabs.harvard.edu/abs/2024A&A...684A..75C} {684, A75}

\bibitem[\protect\citeauthoryear{{Curtis-Lake} et~al.,}{{Curtis-Lake} et~al.}{2023}]{Curtis-Lake2023}
{Curtis-Lake} E.,  et~al., 2023, \mn@doi [Nature Astronomy] {10.1038/s41550-023-01918-w}, \href {https://ui.adsabs.harvard.edu/abs/2023NatAs...7..622C} {7, 622}

\bibitem[\protect\citeauthoryear{{Curtis-Lake} et~al.,}{{Curtis-Lake} et~al.}{2025}]{Curtis-Lake2025_DR4}
{Curtis-Lake} E.,  et~al., 2025, \mn@doi [arXiv e-prints] {10.48550/arXiv.2510.01033}, \href {https://ui.adsabs.harvard.edu/abs/2025arXiv251001033C} {p. arXiv:2510.01033}

\bibitem[\protect\citeauthoryear{{D'Eugenio} et~al.,}{{D'Eugenio} et~al.}{2025a}]{DEugenio2025_DH}
{D'Eugenio} F.,  et~al., 2025a, \mn@doi [arXiv e-prints] {10.48550/arXiv.2510.11626}, \href {https://ui.adsabs.harvard.edu/abs/2025arXiv251011626D} {p. arXiv:2510.11626}

\bibitem[\protect\citeauthoryear{{D'Eugenio} et~al.,}{{D'Eugenio} et~al.}{2025b}]{DEugenio2025}
{D'Eugenio} F.,  et~al., 2025b, \mn@doi [\apjs] {10.3847/1538-4365/ada148}, \href {https://ui.adsabs.harvard.edu/abs/2025ApJS..277....4D} {277, 4}

\bibitem[\protect\citeauthoryear{{Dalcanton}}{{Dalcanton}}{2007}]{Dalcanton2007}
{Dalcanton} J.~J.,  2007, \mn@doi [\apj] {10.1086/508913}, \href {https://ui.adsabs.harvard.edu/abs/2007ApJ...658..941D} {658, 941}

\bibitem[\protect\citeauthoryear{{Dav{\'e}}, {Finlator}  \& {Oppenheimer}}{{Dav{\'e}} et~al.}{2012}]{Dave2012}
{Dav{\'e}} R.,  {Finlator} K.,   {Oppenheimer} B.~D.,  2012, \mn@doi [\mnras] {10.1111/j.1365-2966.2011.20148.x}, \href {https://ui.adsabs.harvard.edu/abs/2012MNRAS.421...98D} {421, 98}

\bibitem[\protect\citeauthoryear{{Dorner} et~al.,}{{Dorner} et~al.}{2016}]{Dorner2016}
{Dorner} B.,  et~al., 2016, \mn@doi [\aap] {10.1051/0004-6361/201628263}, \href {https://ui.adsabs.harvard.edu/abs/2016A&A...592A.113D} {592, A113}

\bibitem[\protect\citeauthoryear{{Dors}, {Maiolino}, {Cardaci}, {H{\"a}gele}, {Krabbe}, {P{\'e}rez-Montero}  \& {Armah}}{{Dors} et~al.}{2020}]{Dors2020}
{Dors} O.~L.,  {Maiolino} R.,  {Cardaci} M.~V.,  {H{\"a}gele} G.~F.,  {Krabbe} A.~C.,  {P{\'e}rez-Montero} E.,   {Armah} M.,  2020, \mn@doi [\mnras] {10.1093/mnras/staa1781}, \href {https://ui.adsabs.harvard.edu/abs/2020MNRAS.496.3209D} {496, 3209}

\bibitem[\protect\citeauthoryear{{Dotter}}{{Dotter}}{2016}]{dotter2016}
{Dotter} A.,  2016, \mn@doi [\apjs] {10.3847/0067-0049/222/1/8}, \href {https://ui.adsabs.harvard.edu/abs/2016ApJS..222....8D} {222, 8}

\bibitem[\protect\citeauthoryear{{Duan} et~al.,}{{Duan} et~al.}{2026}]{Duan2026}
{Duan} Q.,  et~al., 2026, \mn@doi [arXiv e-prints] {10.48550/arXiv.2605.21599}, \href {https://ui.adsabs.harvard.edu/abs/2026arXiv260521599D} {p. arXiv:2605.21599}

\bibitem[\protect\citeauthoryear{{Eisenstein} et~al.,}{{Eisenstein} et~al.}{2023}]{Eisenstein2023b}
{Eisenstein} D.~J.,  et~al., 2023, \mn@doi [arXiv e-prints] {10.48550/arXiv.2310.12340}, \href {https://ui.adsabs.harvard.edu/abs/2023arXiv231012340E} {p. arXiv:2310.12340}

\bibitem[\protect\citeauthoryear{{Eldridge}, {Stanway}, {Xiao}, {McClelland}, {Taylor}, {Ng}, {Greis}  \& {Bray}}{{Eldridge} et~al.}{2017}]{Eldridge2017}
{Eldridge} J.~J.,  {Stanway} E.~R.,  {Xiao} L.,  {McClelland} L.~A.~S.,  {Taylor} G.,  {Ng} M.,  {Greis} S.~M.~L.,   {Bray} J.~C.,  2017, \mn@doi [\pasa] {10.1017/pasa.2017.51}, \href {https://ui.adsabs.harvard.edu/abs/2017PASA...34...58E} {34, e058}

\bibitem[\protect\citeauthoryear{{Ellison}, {Patton}, {Simard}  \& {McConnachie}}{{Ellison} et~al.}{2008}]{Ellison2008}
{Ellison} S.~L.,  {Patton} D.~R.,  {Simard} L.,   {McConnachie} A.~W.,  2008, \mn@doi [\apjl] {10.1086/527296}, \href {https://ui.adsabs.harvard.edu/abs/2008ApJ...672L.107E} {672, L107}

\bibitem[\protect\citeauthoryear{{Falc{\'o}n-Barroso}, {S{\'a}nchez-Bl{\'a}zquez}, {Vazdekis}, {Ricciardelli}, {Cardiel}, {Cenarro}, {Gorgas}  \& {Peletier}}{{Falc{\'o}n-Barroso} et~al.}{2011}]{miles_2011}
{Falc{\'o}n-Barroso} J.,  {S{\'a}nchez-Bl{\'a}zquez} P.,  {Vazdekis} A.,  {Ricciardelli} E.,  {Cardiel} N.,  {Cenarro} A.~J.,  {Gorgas} J.,   {Peletier} R.~F.,  2011, \mn@doi [\aap] {10.1051/0004-6361/201116842}, \href {https://ui.adsabs.harvard.edu/abs/2011A&A...532A..95F} {532, A95}

\bibitem[\protect\citeauthoryear{{Ferland} et~al.,}{{Ferland} et~al.}{2013}]{Ferland2013}
{Ferland} G.~J.,  et~al., 2013, \rmxaa, \href {https://ui.adsabs.harvard.edu/abs/2013RMxAA..49..137F} {49, 137}

\bibitem[\protect\citeauthoryear{{Ferruit} et~al.,}{{Ferruit} et~al.}{2022}]{Ferruit2022}
{Ferruit} P.,  et~al., 2022, \mn@doi [\aap] {10.1051/0004-6361/202142673}, \href {https://ui.adsabs.harvard.edu/abs/2022A&A...661A..81F} {661, A81}

\bibitem[\protect\citeauthoryear{{Flury} \& {Moran}}{{Flury} \& {Moran}}{2020}]{Flury2020}
{Flury} S.~R.,  {Moran} E.~C.,  2020, \mn@doi [\mnras] {10.1093/mnras/staa1563}, \href {https://ui.adsabs.harvard.edu/abs/2020MNRAS.496.2191F} {496, 2191}

\bibitem[\protect\citeauthoryear{{Foreman-Mackey}, {Hogg}, {Lang}  \& {Goodman}}{{Foreman-Mackey} et~al.}{2013}]{ForemanMackey2013}
{Foreman-Mackey} D.,  {Hogg} D.~W.,  {Lang} D.,   {Goodman} J.,  2013, \mn@doi [\pasp] {10.1086/670067}, \href {https://ui.adsabs.harvard.edu/abs/2013PASP..125..306F} {125, 306}

\bibitem[\protect\citeauthoryear{{Froese Fischer} \& {Tachiev}}{{Froese Fischer} \& {Tachiev}}{2004}]{FroeseFischer2004}
{Froese Fischer} C.,  {Tachiev} G.,  2004, \mn@doi [Atomic Data and Nuclear Data Tables] {10.1016/j.adt.2004.02.001}, \href {https://ui.adsabs.harvard.edu/abs/2004ADNDT..87....1F} {87, 1}

\bibitem[\protect\citeauthoryear{{Fujimoto} et~al.,}{{Fujimoto} et~al.}{2025a}]{Fujimoto2025b}
{Fujimoto} S.,  et~al., 2025a, \mn@doi [arXiv e-prints] {10.48550/arXiv.2512.11790}, \href {https://ui.adsabs.harvard.edu/abs/2025arXiv251211790F} {p. arXiv:2512.11790}

\bibitem[\protect\citeauthoryear{{Fujimoto} et~al.,}{{Fujimoto} et~al.}{2025b}]{Fujimoto2025}
{Fujimoto} S.,  et~al., 2025b, \mn@doi [\apj] {10.3847/1538-4357/ade9a1}, \href {https://ui.adsabs.harvard.edu/abs/2025ApJ...989...46F} {989, 46}

\bibitem[\protect\citeauthoryear{{Gallazzi}, {Charlot}, {Brinchmann}, {White}  \& {Tremonti}}{{Gallazzi} et~al.}{2005}]{gallazzi2005}
{Gallazzi} A.,  {Charlot} S.,  {Brinchmann} J.,  {White} S. D.~M.,   {Tremonti} C.~A.,  2005, \mn@doi [\mnras] {10.1111/j.1365-2966.2005.09321.x}, \href {https://ui.adsabs.harvard.edu/abs/2005MNRAS.362...41G} {362, 41}

\bibitem[\protect\citeauthoryear{{Gardner} et~al.,}{{Gardner} et~al.}{2023}]{jwst_new}
{Gardner} J.~P.,  et~al., 2023, \mn@doi [\pasp] {10.1088/1538-3873/acd1b5}, \href {https://ui.adsabs.harvard.edu/abs/2023PASP..135f8001G} {135, 068001}

\bibitem[\protect\citeauthoryear{{Geris} et~al.,}{{Geris} et~al.}{2026}]{Geris2026}
{Geris} S.,  et~al., 2026, \mn@doi [\mnras] {10.1093/mnras/staf1979}, \href {https://ui.adsabs.harvard.edu/abs/2026MNRAS.545f1979G} {545, staf1979}

\bibitem[\protect\citeauthoryear{{Gim{\'e}nez-Alc{\'a}zar}, {Amor{\'\i}n}  \& {Vilchez}}{{Gim{\'e}nez-Alc{\'a}zar} et~al.}{2026}]{Gimenez-Alcazar2026}
{Gim{\'e}nez-Alc{\'a}zar} A.,  {Amor{\'\i}n} R.,   {Vilchez} J.~M.,  2026, \mn@doi [arXiv e-prints] {10.48550/arXiv.2605.05327}, \href {https://ui.adsabs.harvard.edu/abs/2026arXiv260505327G} {p. arXiv:2605.05327}

\bibitem[\protect\citeauthoryear{{Guo} et~al.,}{{Guo} et~al.}{2016}]{Guo2016}
{Guo} Y.,  et~al., 2016, \mn@doi [\apj] {10.3847/0004-637X/822/2/103}, \href {https://ui.adsabs.harvard.edu/abs/2016ApJ...822..103G} {822, 103}

\bibitem[\protect\citeauthoryear{{Hao}, {Kennicutt}, {Johnson}, {Calzetti}, {Dale}  \& {Moustakas}}{{Hao} et~al.}{2011}]{Hao2011}
{Hao} C.-N.,  {Kennicutt} R.~C.,  {Johnson} B.~D.,  {Calzetti} D.,  {Dale} D.~A.,   {Moustakas} J.,  2011, \mn@doi [\apj] {10.1088/0004-637X/741/2/124}, \href {https://ui.adsabs.harvard.edu/abs/2011ApJ...741..124H} {741, 124}

\bibitem[\protect\citeauthoryear{{Harikane} et~al.,}{{Harikane} et~al.}{2025}]{Harikane2025b}
{Harikane} Y.,  et~al., 2025, \mn@doi [arXiv e-prints] {10.48550/arXiv.2505.09186}, \href {https://ui.adsabs.harvard.edu/abs/2025arXiv250509186H} {p. arXiv:2505.09186}

\bibitem[\protect\citeauthoryear{{Heintz} et~al.,}{{Heintz} et~al.}{2023}]{Heintz2023}
{Heintz} K.~E.,  et~al., 2023, \mn@doi [Nature Astronomy] {10.1038/s41550-023-02078-7}, \href {https://ui.adsabs.harvard.edu/abs/2023NatAs...7.1517H} {7, 1517}

\bibitem[\protect\citeauthoryear{{Heintz} et~al.,}{{Heintz} et~al.}{2025}]{Heintz2025}
{Heintz} K.~E.,  et~al., 2025, \mn@doi [\aap] {10.1051/0004-6361/202450243}, \href {https://ui.adsabs.harvard.edu/abs/2025A&A...693A..60H} {693, A60}

\bibitem[\protect\citeauthoryear{{Hirano}, {Hosokawa}, {Yoshida}, {Omukai}  \& {Yorke}}{{Hirano} et~al.}{2015}]{Hirano2015}
{Hirano} S.,  {Hosokawa} T.,  {Yoshida} N.,  {Omukai} K.,   {Yorke} H.~W.,  2015, \mn@doi [\mnras] {10.1093/mnras/stv044}, \href {https://ui.adsabs.harvard.edu/abs/2015MNRAS.448..568H} {448, 568}

\bibitem[\protect\citeauthoryear{{Hirschmann} et~al.,}{{Hirschmann} et~al.}{2023}]{Hirschmann2023}
{Hirschmann} M.,  et~al., 2023, \mn@doi [\mnras] {10.1093/mnras/stad2955}, \href {https://ui.adsabs.harvard.edu/abs/2023MNRAS.526.3610H} {526, 3610}

\bibitem[\protect\citeauthoryear{{Hsiao} et~al.,}{{Hsiao} et~al.}{2025}]{Hsiao2025}
{Hsiao} T. Y.-Y.,  et~al., 2025, \mn@doi [arXiv e-prints] {10.48550/arXiv.2505.03873}, \href {https://ui.adsabs.harvard.edu/abs/2025arXiv250503873H} {p. arXiv:2505.03873}

\bibitem[\protect\citeauthoryear{{Hsiao} et~al.,}{{Hsiao} et~al.}{2026}]{Hsiao2026}
{Hsiao} T. Y.-Y.,  et~al., 2026, arXiv e-prints, \href {https://ui.adsabs.harvard.edu/abs/2026arXiv260506770H} {p. arXiv:2605.06770}

\bibitem[\protect\citeauthoryear{{Illingworth} et~al.,}{{Illingworth} et~al.}{2016}]{Illingworth2016}
{Illingworth} G.,  et~al., 2016, \mn@doi [arXiv e-prints] {10.48550/arXiv.1606.00841}, \href {https://ui.adsabs.harvard.edu/abs/2016arXiv160600841I} {p. arXiv:1606.00841}

\bibitem[\protect\citeauthoryear{{Isobe} et~al.,}{{Isobe} et~al.}{2022}]{Isobe2022}
{Isobe} Y.,  et~al., 2022, \mn@doi [\apj] {10.3847/1538-4357/ac3509}, \href {https://ui.adsabs.harvard.edu/abs/2022ApJ...925..111I} {925, 111}

\bibitem[\protect\citeauthoryear{{Isobe}, {Ouchi}, {Nakajima}, {Harikane}, {Ono}, {Xu}, {Zhang}  \& {Umeda}}{{Isobe} et~al.}{2023}]{Isobe2023b}
{Isobe} Y.,  {Ouchi} M.,  {Nakajima} K.,  {Harikane} Y.,  {Ono} Y.,  {Xu} Y.,  {Zhang} Y.,   {Umeda} H.,  2023, \mn@doi [\apj] {10.3847/1538-4357/acf376}, \href {https://ui.adsabs.harvard.edu/abs/2023ApJ...956..139I} {956, 139}

\bibitem[\protect\citeauthoryear{{Isobe} et~al.,}{{Isobe} et~al.}{2025}]{Isobe2025}
{Isobe} Y.,  et~al., 2025, \mn@doi [\mnras] {10.1093/mnrasl/slaf056}, \href {https://ui.adsabs.harvard.edu/abs/2025MNRAS.541L..71I} {541, L71}

\bibitem[\protect\citeauthoryear{{Isobe} et~al.,}{{Isobe} et~al.}{2026}]{Isobe2026}
{Isobe} Y.,  et~al., 2026, \mn@doi [\mnras] {10.1093/mnras/stag123}, \href {https://ui.adsabs.harvard.edu/abs/2026MNRAS.547ag123I} {547, stag123}

\bibitem[\protect\citeauthoryear{{Ivey} et~al.,}{{Ivey} et~al.}{2026a}]{Ivey2026}
{Ivey} L.~R.,  et~al., 2026a, \mn@doi [arXiv e-prints] {10.48550/arXiv.2604.09177}, \href {https://ui.adsabs.harvard.edu/abs/2026arXiv260409177I} {p. arXiv:2604.09177}

\bibitem[\protect\citeauthoryear{{Ivey} et~al.,}{{Ivey} et~al.}{2026b}]{Ivey2026a}
{Ivey} L.~R.,  et~al., 2026b, \mn@doi [\mnras] {10.1093/mnras/stag094}, \href {https://ui.adsabs.harvard.edu/abs/2026MNRAS.546ag094I} {546, stag094}

\bibitem[\protect\citeauthoryear{Izotov, Stasi{\'{n}}ska, Meynet, Guseva  \& Thuan}{Izotov et~al.}{2006}]{Izotov2006}
Izotov Y.~I.,  Stasi{\'{n}}ska G.,  Meynet G.,  Guseva N.~G.,   Thuan T.~X.,  2006, \mn@doi [A{\&}A] {10.1051/0004-6361:20053763}, 448, 955

\bibitem[\protect\citeauthoryear{{Izotov}, {Thuan}  \& {Guseva}}{{Izotov} et~al.}{2012}]{Izotov2012}
{Izotov} Y.~I.,  {Thuan} T.~X.,   {Guseva} N.~G.,  2012, \mn@doi [\aap] {10.1051/0004-6361/201219733}, \href {https://ui.adsabs.harvard.edu/abs/2012A&A...546A.122I} {546, A122}

\bibitem[\protect\citeauthoryear{{Izotov}, {Thuan}, {Guseva}  \& {Liss}}{{Izotov} et~al.}{2018}]{Izotov2018}
{Izotov} Y.~I.,  {Thuan} T.~X.,  {Guseva} N.~G.,   {Liss} S.~E.,  2018, \mn@doi [\mnras] {10.1093/mnras/stx2478}, \href {https://ui.adsabs.harvard.edu/abs/2018MNRAS.473.1956I} {473, 1956}

\bibitem[\protect\citeauthoryear{{Jakobsen} et~al.,}{{Jakobsen} et~al.}{2022}]{Jakobsen2022}
{Jakobsen} P.,  et~al., 2022, \mn@doi [\aap] {10.1051/0004-6361/202142663}, \href {https://ui.adsabs.harvard.edu/abs/2022A&A...661A..80J} {661, A80}

\bibitem[\protect\citeauthoryear{{Jeon}, {Jeong}, {Zhang}  \& {Bromm}}{{Jeon} et~al.}{2026}]{Jeon2026}
{Jeon} J.,  {Jeong} T.~B.,  {Zhang} S.,   {Bromm} V.,  2026, \mn@doi [arXiv e-prints] {10.48550/arXiv.2604.19075}, \href {https://ui.adsabs.harvard.edu/abs/2026arXiv260419075J} {p. arXiv:2604.19075}

\bibitem[\protect\citeauthoryear{{Ji}, {Belokurov}, {Maiolino}, {Monty}, {Isobe}, {Kravtsov}, {McClymont}  \& {{\"U}bler}}{{Ji} et~al.}{2025}]{Ji2025b}
{Ji} X.,  {Belokurov} V.,  {Maiolino} R.,  {Monty} S.,  {Isobe} Y.,  {Kravtsov} A.,  {McClymont} W.,   {{\"U}bler} H.,  2025, \mn@doi [arXiv e-prints] {10.48550/arXiv.2505.12505}, \href {https://ui.adsabs.harvard.edu/abs/2025arXiv250512505J} {p. arXiv:2505.12505}

\bibitem[\protect\citeauthoryear{{Ji} et~al.,}{{Ji} et~al.}{2026}]{Ji2026}
{Ji} X.,  et~al., 2026, \mn@doi [arXiv e-prints] {10.48550/arXiv.2604.03370}, \href {https://ui.adsabs.harvard.edu/abs/2026arXiv260403370J} {p. arXiv:2604.03370}

\bibitem[\protect\citeauthoryear{{Johnson}, {Leja}, {Conroy}  \& {Speagle}}{{Johnson} et~al.}{2019}]{Johnson2019}
{Johnson} B.~D.,  {Leja} J.~L.,  {Conroy} C.,   {Speagle} J.~S.,  2019, {Prospector: Stellar population inference from spectra and SEDs}, Astrophysics Source Code Library, record ascl:1905.025

\bibitem[\protect\citeauthoryear{{Johnson}, {Leja}, {Conroy}  \& {Speagle}}{{Johnson} et~al.}{2021}]{Johnson2021}
{Johnson} B.~D.,  {Leja} J.,  {Conroy} C.,   {Speagle} J.~S.,  2021, \mn@doi [\apjs] {10.3847/1538-4365/abef67}, \href {https://ui.adsabs.harvard.edu/abs/2021ApJS..254...22J} {254, 22}

\bibitem[\protect\citeauthoryear{{Johnson} et~al.,}{{Johnson} et~al.}{2026}]{Johnson2026}
{Johnson} B.~D.,  et~al., 2026, \mn@doi [arXiv e-prints] {10.48550/arXiv.2601.15954}, \href {https://ui.adsabs.harvard.edu/abs/2026arXiv260115954J} {p. arXiv:2601.15954}

\bibitem[\protect\citeauthoryear{{Jones} et~al.,}{{Jones} et~al.}{2024}]{Jones2024}
{Jones} G.~C.,  et~al., 2024, \mn@doi [\aap] {10.1051/0004-6361/202347099}, \href {https://ui.adsabs.harvard.edu/abs/2024A&A...683A.238J} {683, A238}

\bibitem[\protect\citeauthoryear{{Jones} et~al.,}{{Jones} et~al.}{2025a}]{Jones2025}
{Jones} G.~C.,  et~al., 2025a, \mn@doi [arXiv e-prints] {10.48550/arXiv.2509.20455}, \href {https://ui.adsabs.harvard.edu/abs/2025arXiv250920455J} {p. arXiv:2509.20455}

\bibitem[\protect\citeauthoryear{{Jones} et~al.,}{{Jones} et~al.}{2025b}]{Jones2025a}
{Jones} G.~C.,  et~al., 2025b, \mn@doi [\mnras] {10.1093/mnras/stae2670}, \href {https://ui.adsabs.harvard.edu/abs/2025MNRAS.536.2355J} {536, 2355}

\bibitem[\protect\citeauthoryear{{Juod{\v{z}}balis} et~al.,}{{Juod{\v{z}}balis} et~al.}{2025}]{Juodzbalis2025}
{Juod{\v{z}}balis} I.,  et~al., 2025, \mn@doi [arXiv e-prints] {10.48550/arXiv.2504.03551}, \href {https://ui.adsabs.harvard.edu/abs/2025arXiv250403551J} {p. arXiv:2504.03551}

\bibitem[\protect\citeauthoryear{Juod{\v z}balis et~al.,}{Juod{\v z}balis et~al.}{2026}]{Juodzbalis2026}
Juod{\v z}balis I.,  et~al., 2026, \mn@doi [Nature] {10.1038/s41586-026-10579-4}, 653, 1017

\bibitem[\protect\citeauthoryear{{Kannan} et~al.,}{{Kannan} et~al.}{2025}]{Kannan2025}
{Kannan} R.,  et~al., 2025, \mn@doi [arXiv e-prints] {10.48550/arXiv.2502.20437}, \href {https://ui.adsabs.harvard.edu/abs/2025arXiv250220437K} {p. arXiv:2502.20437}

\bibitem[\protect\citeauthoryear{{Kashino} \& {Inoue}}{{Kashino} \& {Inoue}}{2019}]{Kashino2019}
{Kashino} D.,  {Inoue} A.~K.,  2019, \mn@doi [\mnras] {10.1093/mnras/stz881}, \href {https://ui.adsabs.harvard.edu/abs/2019MNRAS.486.1053K} {486, 1053}

\bibitem[\protect\citeauthoryear{{Kewley} \& {Dopita}}{{Kewley} \& {Dopita}}{2002}]{Kewley2002}
{Kewley} L.~J.,  {Dopita} M.~A.,  2002, \mn@doi [\apjs] {10.1086/341326}, \href {https://ui.adsabs.harvard.edu/abs/2002ApJS..142...35K} {142, 35}

\bibitem[\protect\citeauthoryear{{Kewley}, {Nicholls}  \& {Sutherland}}{{Kewley} et~al.}{2019}]{Kewley2019_review}
{Kewley} L.~J.,  {Nicholls} D.~C.,   {Sutherland} R.~S.,  2019, \mn@doi [\araa] {10.1146/annurev-astro-081817-051832}, \href {https://ui.adsabs.harvard.edu/abs/2019ARA&A..57..511K} {57, 511}

\bibitem[\protect\citeauthoryear{{Kisielius}, {Storey}, {Ferland}  \& {Keenan}}{{Kisielius} et~al.}{2009}]{Kisielius2009}
{Kisielius} R.,  {Storey} P.~J.,  {Ferland} G.~J.,   {Keenan} F.~P.,  2009, \mn@doi [\mnras] {10.1111/j.1365-2966.2009.14989.x}, \href {https://ui.adsabs.harvard.edu/abs/2009MNRAS.397..903K} {397, 903}

\bibitem[\protect\citeauthoryear{{Kobayashi} \& {Taylor}}{{Kobayashi} \& {Taylor}}{2023}]{Kobayashi2023}
{Kobayashi} C.,  {Taylor} P.,  2023, Chemo-Dynamical Evolution of Galaxies.
Springer Nature, Netherlands, \mn@doi{10.1007/978-981-15-8818-1\_106-1}

\bibitem[\protect\citeauthoryear{{Kojima} et~al.,}{{Kojima} et~al.}{2020}]{Kojima2020}
{Kojima} T.,  et~al., 2020, \mn@doi [\apj] {10.3847/1538-4357/aba047}, \href {https://ui.adsabs.harvard.edu/abs/2020ApJ...898..142K} {898, 142}

\bibitem[\protect\citeauthoryear{{Koller} et~al.,}{{Koller} et~al.}{2026}]{Koller2026}
{Koller} M.,  et~al., 2026, \mn@doi [arXiv e-prints] {10.48550/arXiv.2604.07076}, \href {https://ui.adsabs.harvard.edu/abs/2026arXiv260407076K} {p. arXiv:2604.07076}

\bibitem[\protect\citeauthoryear{{Korhonen Cuestas}, {Strom}, {Miller}, {Steidel}, {Trainor}, {Rudie}  \& {Nu{\~n}ez}}{{Korhonen Cuestas} et~al.}{2025}]{KorhonenCuestas2025}
{Korhonen Cuestas} N.~A.,  {Strom} A.~L.,  {Miller} T.~B.,  {Steidel} C.~C.,  {Trainor} R.~F.,  {Rudie} G.~C.,   {Nu{\~n}ez} E.~H.,  2025, \mn@doi [\apj] {10.3847/1538-4357/adc5f7}, \href {https://ui.adsabs.harvard.edu/abs/2025ApJ...984..188K} {984, 188}

\bibitem[\protect\citeauthoryear{{Kriek} et~al.,}{{Kriek} et~al.}{2006}]{Kriek2006}
{Kriek} M.,  et~al., 2006, \mn@doi [\apj] {10.1086/504103}, \href {https://ui.adsabs.harvard.edu/abs/2006ApJ...645...44K} {645, 44}

\bibitem[\protect\citeauthoryear{{Kunth} \& {{\"O}stlin}}{{Kunth} \& {{\"O}stlin}}{2000}]{Kunth2000}
{Kunth} D.,  {{\"O}stlin} G.,  2000, \mn@doi [\aapr] {10.1007/s001590000005}, \href {https://ui.adsabs.harvard.edu/abs/2000A&ARv..10....1K} {10, 1}

\bibitem[\protect\citeauthoryear{{Lam}, {Clarke}, {Shapley}, {Sanders}, {Topping}, {Brammer}, {Reddy}  \& {Karthikeyan}}{{Lam} et~al.}{2026}]{Lam2026}
{Lam} N.,  {Clarke} L.,  {Shapley} A.~E.,  {Sanders} R.~L.,  {Topping} M.~W.,  {Brammer} G.~B.,  {Reddy} N.~A.,   {Karthikeyan} S.,  2026, \mn@doi [arXiv e-prints] {10.48550/arXiv.2605.30513}, \href {https://ui.adsabs.harvard.edu/abs/2026arXiv260530513L} {p. arXiv:2605.30513}

\bibitem[\protect\citeauthoryear{{Langan}, {Ceverino}  \& {Finlator}}{{Langan} et~al.}{2020}]{Langan2020}
{Langan} I.,  {Ceverino} D.,   {Finlator} K.,  2020, \mn@doi [\mnras] {10.1093/mnras/staa880}, \href {https://ui.adsabs.harvard.edu/abs/2020MNRAS.494.1988L} {494, 1988}

\bibitem[\protect\citeauthoryear{{Langeroodi} \& {Hjorth}}{{Langeroodi} \& {Hjorth}}{2023}]{Langeroodi2023}
{Langeroodi} D.,  {Hjorth} J.,  2023, \mn@doi [arXiv e-prints] {10.48550/arXiv.2307.06336}, \href {https://ui.adsabs.harvard.edu/abs/2023arXiv230706336L} {p. arXiv:2307.06336}

\bibitem[\protect\citeauthoryear{{Langeroodi} \& {Hjorth}}{{Langeroodi} \& {Hjorth}}{2026}]{Langeroodi2026}
{Langeroodi} D.,  {Hjorth} J.,  2026, \mn@doi [\apjl] {10.3847/2041-8213/ae346f}, \href {https://ui.adsabs.harvard.edu/abs/2026ApJ...997L..30L} {997, L30}

\bibitem[\protect\citeauthoryear{{Langeroodi} et~al.,}{{Langeroodi} et~al.}{2023}]{Langeroodi2023b}
{Langeroodi} D.,  et~al., 2023, \mn@doi [\apj] {10.3847/1538-4357/acdbc1}, \href {https://ui.adsabs.harvard.edu/abs/2023ApJ...957...39L} {957, 39}

\bibitem[\protect\citeauthoryear{{Laseter} et~al.,}{{Laseter} et~al.}{2024}]{Laseter2024}
{Laseter} I.~H.,  et~al., 2024, \mn@doi [\aap] {10.1051/0004-6361/202347133}, \href {https://ui.adsabs.harvard.edu/abs/2024A&A...681A..70L} {681, A70}

\bibitem[\protect\citeauthoryear{{Laseter}, {Maseda}, {Bunker}, {Cameron}, {Curti}  \& {Simmonds}}{{Laseter} et~al.}{2025}]{Laseter2025}
{Laseter} I.~H.,  {Maseda} M.~V.,  {Bunker} A.~J.,  {Cameron} A.~J.,  {Curti} M.,   {Simmonds} C.,  2025, \mn@doi [arXiv e-prints] {10.48550/arXiv.2510.15024}, \href {https://ui.adsabs.harvard.edu/abs/2025arXiv251015024L} {p. arXiv:2510.15024}

\bibitem[\protect\citeauthoryear{{Leja}, {Carnall}, {Johnson}, {Conroy}  \& {Speagle}}{{Leja} et~al.}{2019}]{Leja2019}
{Leja} J.,  {Carnall} A.~C.,  {Johnson} B.~D.,  {Conroy} C.,   {Speagle} J.~S.,  2019, \mn@doi [\apj] {10.3847/1538-4357/ab133c}, \href {https://ui.adsabs.harvard.edu/abs/2019ApJ...876....3L} {876, 3}

\bibitem[\protect\citeauthoryear{{Lennon} \& {Burke}}{{Lennon} \& {Burke}}{1994}]{Lennon1994}
{Lennon} D.~J.,  {Burke} V.~M.,  1994, \aaps, \href {https://ui.adsabs.harvard.edu/abs/1994A&AS..103..273L} {103, 273}

\bibitem[\protect\citeauthoryear{{Lequeux}, {Peimbert}, {Rayo}, {Serrano}  \& {Torres-Peimbert}}{{Lequeux} et~al.}{1979}]{Lequeux1979}
{Lequeux} J.,  {Peimbert} M.,  {Rayo} J.~F.,  {Serrano} A.,   {Torres-Peimbert} S.,  1979, \aap, \href {https://ui.adsabs.harvard.edu/abs/1979A&A....80..155L} {80, 155}

\bibitem[\protect\citeauthoryear{{Lewis} et~al.,}{{Lewis} et~al.}{2025}]{Lewis2025}
{Lewis} Z.,  et~al., 2025, \mn@doi [arXiv e-prints] {10.48550/arXiv.2512.03134}, \href {https://ui.adsabs.harvard.edu/abs/2025arXiv251203134L} {p. arXiv:2512.03134}

\bibitem[\protect\citeauthoryear{{Lilly}, {Carollo}, {Pipino}, {Renzini}  \& {Peng}}{{Lilly} et~al.}{2013}]{Lilly2013}
{Lilly} S.~J.,  {Carollo} C.~M.,  {Pipino} A.,  {Renzini} A.,   {Peng} Y.,  2013, \mn@doi [\apj] {10.1088/0004-637X/772/2/119}, \href {https://ui.adsabs.harvard.edu/abs/2013ApJ...772..119L} {772, 119}

\bibitem[\protect\citeauthoryear{{Lin} et~al.,}{{Lin} et~al.}{2026}]{Lin2026}
{Lin} X.,  et~al., 2026, \mn@doi [\apj] {10.3847/1538-4357/ae2bdf}, \href {https://ui.adsabs.harvard.edu/abs/2026ApJ...997..364L} {997, 364}

\bibitem[\protect\citeauthoryear{{Liu} et~al.,}{{Liu} et~al.}{2025}]{Liu2025}
{Liu} B.,  et~al., 2025, \mn@doi [arXiv e-prints] {10.48550/arXiv.2506.06139}, \href {https://ui.adsabs.harvard.edu/abs/2025arXiv250606139L} {p. arXiv:2506.06139}

\bibitem[\protect\citeauthoryear{{Luridiana}, {Morisset}  \& {Shaw}}{{Luridiana} et~al.}{2015}]{Luridiana2015}
{Luridiana} V.,  {Morisset} C.,   {Shaw} R.~A.,  2015, \mn@doi [\aap] {10.1051/0004-6361/201323152}, \href {https://ui.adsabs.harvard.edu/abs/2015A&A...573A..42L} {573, A42}

\bibitem[\protect\citeauthoryear{{Ma}, {Hopkins}, {Faucher-Gigu{\`e}re}, {Zolman}, {Muratov}, {Kere{\v{s}}}  \& {Quataert}}{{Ma} et~al.}{2016}]{Ma2016}
{Ma} X.,  {Hopkins} P.~F.,  {Faucher-Gigu{\`e}re} C.-A.,  {Zolman} N.,  {Muratov} A.~L.,  {Kere{\v{s}}} D.,   {Quataert} E.,  2016, \mn@doi [\mnras] {10.1093/mnras/stv2659}, \href {https://ui.adsabs.harvard.edu/abs/2016MNRAS.456.2140M} {456, 2140}

\bibitem[\protect\citeauthoryear{{Madau} \& {Dickinson}}{{Madau} \& {Dickinson}}{2014}]{Madau2014}
{Madau} P.,  {Dickinson} M.,  2014, \mn@doi [\araa] {10.1146/annurev-astro-081811-125615}, \href {https://ui.adsabs.harvard.edu/abs/2014ARA&A..52..415M} {52, 415}

\bibitem[\protect\citeauthoryear{{Maheson} et~al.,}{{Maheson} et~al.}{2025}]{Maheson2025}
{Maheson} G.,  et~al., 2025, \mn@doi [arXiv e-prints] {10.48550/arXiv.2504.15346}, \href {https://ui.adsabs.harvard.edu/abs/2025arXiv250415346M} {p. arXiv:2504.15346}

\bibitem[\protect\citeauthoryear{{Maiolino} \& {Mannucci}}{{Maiolino} \& {Mannucci}}{2019}]{Maiolino2019}
{Maiolino} R.,  {Mannucci} F.,  2019, \mn@doi [\aapr] {10.1007/s00159-018-0112-2}, \href {https://ui.adsabs.harvard.edu/abs/2019A&ARv..27....3M} {27, 3}

\bibitem[\protect\citeauthoryear{{Maiolino} et~al.,}{{Maiolino} et~al.}{2008}]{Maiolino2008}
{Maiolino} R.,  et~al., 2008, \mn@doi [\aap] {10.1051/0004-6361:200809678}, \href {https://ui.adsabs.harvard.edu/abs/2008A&A...488..463M} {488, 463}

\bibitem[\protect\citeauthoryear{{Maiolino} et~al.,}{{Maiolino} et~al.}{2025}]{Maiolino2025}
{Maiolino} R.,  et~al., 2025, \mn@doi [arXiv e-prints] {10.48550/arXiv.2505.22567}, \href {https://ui.adsabs.harvard.edu/abs/2025arXiv250522567M} {p. arXiv:2505.22567}

\bibitem[\protect\citeauthoryear{{Maiolino} et~al.,}{{Maiolino} et~al.}{2026}]{Maiolino2026}
{Maiolino} R.,  et~al., 2026, \mn@doi [arXiv e-prints] {10.48550/arXiv.2603.20362}, \href {https://ui.adsabs.harvard.edu/abs/2026arXiv260320362M} {p. arXiv:2603.20362}

\bibitem[\protect\citeauthoryear{{Malmquist}}{{Malmquist}}{1922}]{Malmquist1922}
{Malmquist} K.~G.,  1922, Meddelanden fran Lunds Astronomiska Observatorium Serie I, \href {https://ui.adsabs.harvard.edu/abs/1922MeLuF.100....1M} {100, 1}

\bibitem[\protect\citeauthoryear{{Mannucci}, {Cresci}, {Maiolino}, {Marconi}  \& {Gnerucci}}{{Mannucci} et~al.}{2010}]{Mannucci2010}
{Mannucci} F.,  {Cresci} G.,  {Maiolino} R.,  {Marconi} A.,   {Gnerucci} A.,  2010, \mn@doi [\mnras] {10.1111/j.1365-2966.2010.17291.x}, \href {https://ui.adsabs.harvard.edu/abs/2010MNRAS.408.2115M} {408, 2115}

\bibitem[\protect\citeauthoryear{{Marconi} et~al.,}{{Marconi} et~al.}{2024}]{Marconi2024}
{Marconi} A.,  et~al., 2024, \mn@doi [\aap] {10.1051/0004-6361/202449240}, \href {https://ui.adsabs.harvard.edu/abs/2024A&A...689A..78M} {689, A78}

\bibitem[\protect\citeauthoryear{{Marino} et~al.,}{{Marino} et~al.}{2013}]{Marino2013}
{Marino} R.~A.,  et~al., 2013, \mn@doi [\aap] {10.1051/0004-6361/201321956}, \href {https://ui.adsabs.harvard.edu/abs/2013A&A...559A.114M} {559, A114}

\bibitem[\protect\citeauthoryear{{Markov} et~al.,}{{Markov} et~al.}{2025a}]{Markov2025b}
{Markov} V.,  et~al., 2025a, \mn@doi [arXiv e-prints] {10.48550/arXiv.2504.12378}, \href {https://ui.adsabs.harvard.edu/abs/2025arXiv250412378M} {p. arXiv:2504.12378}

\bibitem[\protect\citeauthoryear{{Markov}, {Gallerani}, {Ferrara}, {Pallottini}, {Parlanti}, {Mascia}, {Sommovigo}  \& {Kohandel}}{{Markov} et~al.}{2025b}]{Markov2025}
{Markov} V.,  {Gallerani} S.,  {Ferrara} A.,  {Pallottini} A.,  {Parlanti} E.,  {Mascia} F.~D.,  {Sommovigo} L.,   {Kohandel} M.,  2025b, \mn@doi [Nature Astronomy] {10.1038/s41550-024-02426-1}, \href {https://ui.adsabs.harvard.edu/abs/2025NatAs...9..458M} {9, 458}

\bibitem[\protect\citeauthoryear{{Martinez} et~al.,}{{Martinez} et~al.}{2025}]{Martinez2025}
{Martinez} Z.,  et~al., 2025, \mn@doi [\apj] {10.3847/1538-4357/ae17c6}, \href {https://ui.adsabs.harvard.edu/abs/2025ApJ...995..204M} {995, 204}

\bibitem[\protect\citeauthoryear{{Mazzolari} et~al.,}{{Mazzolari} et~al.}{2024}]{Mazzolari2024}
{Mazzolari} G.,  et~al., 2024, \mn@doi [\aap] {10.1051/0004-6361/202450407}, \href {https://ui.adsabs.harvard.edu/abs/2024A&A...691A.345M} {691, A345}

\bibitem[\protect\citeauthoryear{{McClymont} et~al.,}{{McClymont} et~al.}{2025a}]{McClymont2025a}
{McClymont} W.,  et~al., 2025a, \mn@doi [arXiv e-prints] {10.48550/arXiv.2503.00106}, \href {https://ui.adsabs.harvard.edu/abs/2025arXiv250300106M} {p. arXiv:2503.00106}

\bibitem[\protect\citeauthoryear{{McClymont} et~al.,}{{McClymont} et~al.}{2025b}]{McClymont2025c}
{McClymont} W.,  et~al., 2025b, \mn@doi [arXiv e-prints] {10.48550/arXiv.2507.08787}, \href {https://ui.adsabs.harvard.edu/abs/2025arXiv250708787M} {p. arXiv:2507.08787}

\bibitem[\protect\citeauthoryear{{McClymont} et~al.,}{{McClymont} et~al.}{2025c}]{McClymont2025}
{McClymont} W.,  et~al., 2025c, \mn@doi [\mnras] {10.1093/mnras/staf745}, \href {https://ui.adsabs.harvard.edu/abs/2025MNRAS.540..190M} {540, 190}

\bibitem[\protect\citeauthoryear{{McGaugh}}{{McGaugh}}{1991}]{McGaugh1991}
{McGaugh} S.~S.,  1991, \mn@doi [\apj] {10.1086/170569}, \href {https://ui.adsabs.harvard.edu/abs/1991ApJ...380..140M} {380, 140}

\bibitem[\protect\citeauthoryear{{Moreschini} et~al.,}{{Moreschini} et~al.}{2026}]{Moreschini2026}
{Moreschini} B.,  et~al., 2026, \mn@doi [arXiv e-prints] {10.48550/arXiv.2601.08939}, \href {https://ui.adsabs.harvard.edu/abs/2026arXiv260108939M} {p. arXiv:2601.08939}

\bibitem[\protect\citeauthoryear{{Morishita} et~al.,}{{Morishita} et~al.}{2024}]{Morishita2024}
{Morishita} T.,  et~al., 2024, \mn@doi [\apj] {10.3847/1538-4357/ad5290}, \href {https://ui.adsabs.harvard.edu/abs/2024ApJ...971...43M} {971, 43}

\bibitem[\protect\citeauthoryear{{Morishita}, {Liu}, {Stiavelli}, {Treu}, {Bergamini}  \& {Zhang}}{{Morishita} et~al.}{2025a}]{Morishita2025}
{Morishita} T.,  {Liu} Z.,  {Stiavelli} M.,  {Treu} T.,  {Bergamini} P.,   {Zhang} Y.,  2025a, \mn@doi [arXiv e-prints] {10.48550/arXiv.2507.10521}, \href {https://ui.adsabs.harvard.edu/abs/2025arXiv250710521M} {p. arXiv:2507.10521}

\bibitem[\protect\citeauthoryear{{Morishita} et~al.,}{{Morishita} et~al.}{2025b}]{Morishita2025_beacon}
{Morishita} T.,  et~al., 2025b, \mn@doi [\apj] {10.3847/1538-4357/adbbdc}, \href {https://ui.adsabs.harvard.edu/abs/2025ApJ...983..152M} {983, 152}

\bibitem[\protect\citeauthoryear{{Mowla} et~al.,}{{Mowla} et~al.}{2024}]{Mowla2024}
{Mowla} L.,  et~al., 2024, \mn@doi [\nat] {10.1038/s41586-024-08293-0}, \href {https://ui.adsabs.harvard.edu/abs/2024Natur.636..332M} {636, 332}

\bibitem[\protect\citeauthoryear{{Nagao}, {Maiolino}  \& {Marconi}}{{Nagao} et~al.}{2006}]{Nagao2006}
{Nagao} T.,  {Maiolino} R.,   {Marconi} A.,  2006, \mn@doi [\aap] {10.1051/0004-6361:20065216}, \href {https://ui.adsabs.harvard.edu/abs/2006A&A...459...85N} {459, 85}

\bibitem[\protect\citeauthoryear{{Nakajima} \& {Maiolino}}{{Nakajima} \& {Maiolino}}{2022}]{NakajimaMaiolino2022}
{Nakajima} K.,  {Maiolino} R.,  2022, \mn@doi [\mnras] {10.1093/mnras/stac1242}, \href {https://ui.adsabs.harvard.edu/abs/2022MNRAS.513.5134N} {513, 5134}

\bibitem[\protect\citeauthoryear{{Nakajima} \& {Ouchi}}{{Nakajima} \& {Ouchi}}{2014}]{Nakajima2014}
{Nakajima} K.,  {Ouchi} M.,  2014, \mn@doi [\mnras] {10.1093/mnras/stu902}, \href {https://ui.adsabs.harvard.edu/abs/2014MNRAS.442..900N} {442, 900}

\bibitem[\protect\citeauthoryear{{Nakajima} et~al.,}{{Nakajima} et~al.}{2022}]{Nakajima2022}
{Nakajima} K.,  et~al., 2022, \mn@doi [\apjs] {10.3847/1538-4365/ac7710}, \href {https://ui.adsabs.harvard.edu/abs/2022ApJS..262....3N} {262, 3}

\bibitem[\protect\citeauthoryear{{Nakajima}, {Ouchi}, {Isobe}, {Harikane}, {Zhang}, {Ono}, {Umeda}  \& {Oguri}}{{Nakajima} et~al.}{2023}]{Nakajima2023}
{Nakajima} K.,  {Ouchi} M.,  {Isobe} Y.,  {Harikane} Y.,  {Zhang} Y.,  {Ono} Y.,  {Umeda} H.,   {Oguri} M.,  2023, \mn@doi [\apjs] {10.3847/1538-4365/acd556}, \href {https://ui.adsabs.harvard.edu/abs/2023ApJS..269...33N} {269, 33}

\bibitem[\protect\citeauthoryear{{Nakajima} et~al.,}{{Nakajima} et~al.}{2025}]{Nakajima2025}
{Nakajima} K.,  et~al., 2025, \mn@doi [arXiv e-prints] {10.48550/arXiv.2506.11846}, \href {https://ui.adsabs.harvard.edu/abs/2025arXiv250611846N} {p. arXiv:2506.11846}

\bibitem[\protect\citeauthoryear{{Nishigaki} et~al.,}{{Nishigaki} et~al.}{2025}]{Nishigaki2025b}
{Nishigaki} M.,  et~al., 2025, arXiv e-prints, \href {https://ui.adsabs.harvard.edu/abs/2025arXiv251212983N} {p. arXiv:2512.12983}

\bibitem[\protect\citeauthoryear{{Oesch} et~al.,}{{Oesch} et~al.}{2023}]{Oesch2023}
{Oesch} P.~A.,  et~al., 2023, \mn@doi [\mnras] {10.1093/mnras/stad2411}, \href {https://ui.adsabs.harvard.edu/abs/2023MNRAS.525.2864O} {525, 2864}

\bibitem[\protect\citeauthoryear{{Osterbrock}}{{Osterbrock}}{1989}]{Osterbrock1989}
{Osterbrock} D.~E.,  1989, {Astrophysics of gaseous nebulae and active galactic nuclei}

\bibitem[\protect\citeauthoryear{{{\"O}stlin} et~al.,}{{{\"O}stlin} et~al.}{2025}]{Ostlin2025}
{{\"O}stlin} G.,  et~al., 2025, \mn@doi [\aap] {10.1051/0004-6361/202451723}, \href {https://ui.adsabs.harvard.edu/abs/2025A&A...696A..57O} {696, A57}

\bibitem[\protect\citeauthoryear{{Pagel}, {Edmunds}, {Blackwell}, {Chun}  \& {Smith}}{{Pagel} et~al.}{1979}]{Pagel1979}
{Pagel} B.~E.~J.,  {Edmunds} M.~G.,  {Blackwell} D.~E.,  {Chun} M.~S.,   {Smith} G.,  1979, \mn@doi [\mnras] {10.1093/mnras/189.1.95}, \href {https://ui.adsabs.harvard.edu/abs/1979MNRAS.189...95P} {189, 95}

\bibitem[\protect\citeauthoryear{{Pallottini}, {Ferrara}, {Gallerani}, {Sommovigo}, {Carniani}, {Vallini}, {Kohandel}  \& {Venturi}}{{Pallottini} et~al.}{2025}]{Pallottini2025}
{Pallottini} A.,  {Ferrara} A.,  {Gallerani} S.,  {Sommovigo} L.,  {Carniani} S.,  {Vallini} L.,  {Kohandel} M.,   {Venturi} G.,  2025, \mn@doi [\aap] {10.1051/0004-6361/202451742}, \href {https://ui.adsabs.harvard.edu/abs/2025A&A...699A...6P} {699, A6}

\bibitem[\protect\citeauthoryear{{Pascalau} et~al.,}{{Pascalau} et~al.}{2026}]{Pascalau2026}
{Pascalau} R.~G.,  et~al., 2026, \mn@doi [arXiv e-prints] {10.48550/arXiv.2603.00232}, \href {https://ui.adsabs.harvard.edu/abs/2026arXiv260300232P} {p. arXiv:2603.00232}

\bibitem[\protect\citeauthoryear{{Peimbert}}{{Peimbert}}{1967}]{Peimbert1967}
{Peimbert} M.,  1967, \mn@doi [\apj] {10.1086/149385}, \href {https://ui.adsabs.harvard.edu/abs/1967ApJ...150..825P} {150, 825}

\bibitem[\protect\citeauthoryear{{Pettini} \& {Pagel}}{{Pettini} \& {Pagel}}{2004}]{Pettini2004}
{Pettini} M.,  {Pagel} B. E.~J.,  2004, \mn@doi [\mnras] {10.1111/j.1365-2966.2004.07591.x}, \href {https://ui.adsabs.harvard.edu/abs/2004MNRAS.348L..59P} {348, L59}

\bibitem[\protect\citeauthoryear{{Planck Collaboration} et~al.,}{{Planck Collaboration} et~al.}{2020}]{Planck2020}
{Planck Collaboration} et~al., 2020, \mn@doi [\aap] {10.1051/0004-6361/201833910}, \href {https://ui.adsabs.harvard.edu/abs/2020A&A...641A...6P} {641, A6}

\bibitem[\protect\citeauthoryear{{Pollock} et~al.,}{{Pollock} et~al.}{2026}]{Pollock2026}
{Pollock} C.~L.,  et~al., 2026, \mn@doi [\aap] {10.1051/0004-6361/202556032}, \href {https://ui.adsabs.harvard.edu/abs/2026A&A...708A.203P} {708, A203}

\bibitem[\protect\citeauthoryear{{Price} et~al.,}{{Price} et~al.}{2025}]{Price2025}
{Price} S.~H.,  et~al., 2025, \mn@doi [\apj] {10.3847/1538-4357/adaec1}, \href {https://ui.adsabs.harvard.edu/abs/2025ApJ...982...51P} {982, 51}

\bibitem[\protect\citeauthoryear{{Pusk{\'a}s} et~al.,}{{Pusk{\'a}s} et~al.}{2025a}]{Puskas2025b}
{Pusk{\'a}s} D.,  et~al., 2025a, \mn@doi [arXiv e-prints] {10.48550/arXiv.2510.14743}, \href {https://ui.adsabs.harvard.edu/abs/2025arXiv251014743P} {p. arXiv:2510.14743}

\bibitem[\protect\citeauthoryear{{Pusk{\'a}s} et~al.,}{{Pusk{\'a}s} et~al.}{2025b}]{Puskas2025a}
{Pusk{\'a}s} D.,  et~al., 2025b, \mn@doi [\mnras] {10.1093/mnras/staf813}, \href {https://ui.adsabs.harvard.edu/abs/2025MNRAS.540.2146P} {540, 2146}

\bibitem[\protect\citeauthoryear{{Reddy} et~al.,}{{Reddy} et~al.}{2018}]{Reddy2018}
{Reddy} N.~A.,  et~al., 2018, \mn@doi [\apj] {10.3847/1538-4357/aaed1e}, \href {https://ui.adsabs.harvard.edu/abs/2018ApJ...869...92R} {869, 92}

\bibitem[\protect\citeauthoryear{{Reddy} et~al.,}{{Reddy} et~al.}{2020}]{Reddy2020}
{Reddy} N.~A.,  et~al., 2020, \mn@doi [\apj] {10.3847/1538-4357/abb674}, \href {https://ui.adsabs.harvard.edu/abs/2020ApJ...902..123R} {902, 123}

\bibitem[\protect\citeauthoryear{{Revalski} et~al.,}{{Revalski} et~al.}{2024}]{Revalski2024}
{Revalski} M.,  et~al., 2024, \mn@doi [\apj] {10.3847/1538-4357/ad382c}, \href {https://ui.adsabs.harvard.edu/abs/2024ApJ...966..228R} {966, 228}

\bibitem[\protect\citeauthoryear{{Rinaldi} et~al.,}{{Rinaldi} et~al.}{2025}]{Rinaldi2025}
{Rinaldi} P.,  et~al., 2025, \mn@doi [\apj] {10.3847/1538-4357/ae089c}, \href {https://ui.adsabs.harvard.edu/abs/2025ApJ...994...86R} {994, 86}

\bibitem[\protect\citeauthoryear{{Rinaldi} et~al.,}{{Rinaldi} et~al.}{2026}]{Rinaldi2026}
{Rinaldi} P.,  et~al., 2026, \mn@doi [arXiv e-prints] {10.48550/arXiv.2604.07138}, \href {https://ui.adsabs.harvard.edu/abs/2026arXiv260407138R} {p. arXiv:2604.07138}

\bibitem[\protect\citeauthoryear{{Robertson} et~al.,}{{Robertson} et~al.}{2026}]{Robertson2026_DR5}
{Robertson} B.~E.,  et~al., 2026, \mn@doi [arXiv e-prints] {10.48550/arXiv.2601.15956}, \href {https://ui.adsabs.harvard.edu/abs/2026arXiv260115956R} {p. arXiv:2601.15956}

\bibitem[\protect\citeauthoryear{{Rosales-Ortega} et~al.,}{{Rosales-Ortega} et~al.}{2026}]{Rosales-Ortega2026}
{Rosales-Ortega} F.~F.,  et~al., 2026, arXiv e-prints, \href {https://ui.adsabs.harvard.edu/abs/2026arXiv260416273R} {p. arXiv:2604.16273}

\bibitem[\protect\citeauthoryear{{Rusta} et~al.,}{{Rusta} et~al.}{2026}]{Rusta2026}
{Rusta} E.,  et~al., 2026, \mn@doi [arXiv e-prints] {10.48550/arXiv.2603.20363}, \href {https://ui.adsabs.harvard.edu/abs/2026arXiv260320363R} {p. arXiv:2603.20363}

\bibitem[\protect\citeauthoryear{{Rynkun}, {Gaigalas}  \& {J{\"o}nsson}}{{Rynkun} et~al.}{2019}]{Rynkun2019}
{Rynkun} P.,  {Gaigalas} G.,   {J{\"o}nsson} P.,  2019, \mn@doi [\aap] {10.1051/0004-6361/201834931}, \href {https://ui.adsabs.harvard.edu/abs/2019A&A...623A.155R} {623, A155}

\bibitem[\protect\citeauthoryear{{Salpeter}}{{Salpeter}}{1955}]{Salpeter1955}
{Salpeter} E.~E.,  1955, \mn@doi [\apj] {10.1086/145971}, \href {https://ui.adsabs.harvard.edu/abs/1955ApJ...121..161S} {121, 161}

\bibitem[\protect\citeauthoryear{{S{\'a}nchez Almeida}, {P{\'e}rez-Montero}, {Morales-Luis}, {Mu{\~n}oz-Tu{\~n}{\'o}n}, {Garc{\'\i}a-Benito}, {Nuza}  \& {Kitaura}}{{S{\'a}nchez Almeida} et~al.}{2016}]{Sanchez2016}
{S{\'a}nchez Almeida} J.,  {P{\'e}rez-Montero} E.,  {Morales-Luis} A.~B.,  {Mu{\~n}oz-Tu{\~n}{\'o}n} C.,  {Garc{\'\i}a-Benito} R.,  {Nuza} S.~E.,   {Kitaura} F.~S.,  2016, \mn@doi [\apj] {10.3847/0004-637X/819/2/110}, \href {https://ui.adsabs.harvard.edu/abs/2016ApJ...819..110S} {819, 110}

\bibitem[\protect\citeauthoryear{{S{\'a}nchez-Bl{\'a}zquez} et~al.,}{{S{\'a}nchez-Bl{\'a}zquez} et~al.}{2006}]{sanchez2006}
{S{\'a}nchez-Bl{\'a}zquez} P.,  et~al., 2006, \mn@doi [\mnras] {10.1111/j.1365-2966.2006.10699.x}, \href {https://ui.adsabs.harvard.edu/abs/2006MNRAS.371..703S} {371, 703}

\bibitem[\protect\citeauthoryear{{Sanders} et~al.,}{{Sanders} et~al.}{2016}]{Sanders2016}
{Sanders} R.~L.,  et~al., 2016, \mn@doi [\apj] {10.3847/0004-637X/816/1/23}, \href {https://ui.adsabs.harvard.edu/abs/2016ApJ...816...23S} {816, 23}

\bibitem[\protect\citeauthoryear{{Sanders} et~al.,}{{Sanders} et~al.}{2020}]{Sanders2020}
{Sanders} R.~L.,  et~al., 2020, \mn@doi [\mnras] {10.1093/mnras/stz3032}, \href {https://ui.adsabs.harvard.edu/abs/2020MNRAS.491.1427S} {491, 1427}

\bibitem[\protect\citeauthoryear{{Sanders} et~al.,}{{Sanders} et~al.}{2021}]{Sanders2021}
{Sanders} R.~L.,  et~al., 2021, \mn@doi [\apj] {10.3847/1538-4357/abf4c1}, \href {https://ui.adsabs.harvard.edu/abs/2021ApJ...914...19S} {914, 19}

\bibitem[\protect\citeauthoryear{{Sanders}, {Shapley}, {Topping}, {Reddy}  \& {Brammer}}{{Sanders} et~al.}{2024}]{Sanders2024}
{Sanders} R.~L.,  {Shapley} A.~E.,  {Topping} M.~W.,  {Reddy} N.~A.,   {Brammer} G.~B.,  2024, \mn@doi [\apj] {10.3847/1538-4357/ad15fc}, \href {https://ui.adsabs.harvard.edu/abs/2024ApJ...962...24S} {962, 24}

\bibitem[\protect\citeauthoryear{{Sanders} et~al.,}{{Sanders} et~al.}{2025a}]{Sanders2025}
{Sanders} R.~L.,  et~al., 2025a, \mn@doi [arXiv e-prints] {10.48550/arXiv.2508.10099}, \href {https://ui.adsabs.harvard.edu/abs/2025arXiv250810099S} {p. arXiv:2508.10099}

\bibitem[\protect\citeauthoryear{{Sanders} et~al.,}{{Sanders} et~al.}{2025b}]{Sanders2025b}
{Sanders} R.~L.,  et~al., 2025b, \mn@doi [\apj] {10.3847/1538-4357/adf066}, \href {https://ui.adsabs.harvard.edu/abs/2025ApJ...989..209S} {989, 209}

\bibitem[\protect\citeauthoryear{{Sarkar} et~al.,}{{Sarkar} et~al.}{2025}]{Sarkar2025}
{Sarkar} A.,  et~al., 2025, \mn@doi [\apj] {10.3847/1538-4357/ad8f32}, \href {https://ui.adsabs.harvard.edu/abs/2025ApJ...978..136S} {978, 136}

\bibitem[\protect\citeauthoryear{{Scarlata}, {Hayes}, {Panagia}, {Mehta}, {Haardt}  \& {Bagley}}{{Scarlata} et~al.}{2024}]{Scarlata2024}
{Scarlata} C.,  {Hayes} M.,  {Panagia} N.,  {Mehta} V.,  {Haardt} F.,   {Bagley} M.,  2024, \mn@doi [arXiv e-prints] {10.48550/arXiv.2404.09015}, \href {https://ui.adsabs.harvard.edu/abs/2024arXiv240409015S} {p. arXiv:2404.09015}

\bibitem[\protect\citeauthoryear{{Scholte} et~al.,}{{Scholte} et~al.}{2025}]{Scholte2025}
{Scholte} D.,  et~al., 2025, \mn@doi [\mnras] {10.1093/mnras/staf834}, \href {https://ui.adsabs.harvard.edu/abs/2025MNRAS.540.1800S} {540, 1800}

\bibitem[\protect\citeauthoryear{{Scholte} et~al.,}{{Scholte} et~al.}{2026}]{Scholte2026}
{Scholte} D.,  et~al., 2026, \mn@doi [arXiv e-prints] {10.48550/arXiv.2601.02463}, \href {https://ui.adsabs.harvard.edu/abs/2026arXiv260102463S} {p. arXiv:2601.02463}

\bibitem[\protect\citeauthoryear{{Scholtz} et~al.,}{{Scholtz} et~al.}{2025a}]{Scholtz2025_DR4}
{Scholtz} J.,  et~al., 2025a, \mn@doi [arXiv e-prints] {10.48550/arXiv.2510.01034}, \href {https://ui.adsabs.harvard.edu/abs/2025arXiv251001034S} {p. arXiv:2510.01034}

\bibitem[\protect\citeauthoryear{{Scholtz} et~al.,}{{Scholtz} et~al.}{2025b}]{Scholtz2025}
{Scholtz} J.,  et~al., 2025b, \mn@doi [\aap] {10.1051/0004-6361/202348804}, \href {https://ui.adsabs.harvard.edu/abs/2025A&A...697A.175S} {697, A175}

\bibitem[\protect\citeauthoryear{{Shapley}, {Sanders}, {Reddy}, {Topping}  \& {Brammer}}{{Shapley} et~al.}{2023}]{Shapley2023}
{Shapley} A.~E.,  {Sanders} R.~L.,  {Reddy} N.~A.,  {Topping} M.~W.,   {Brammer} G.~B.,  2023, \mn@doi [\apj] {10.3847/1538-4357/acea5a}, \href {https://ui.adsabs.harvard.edu/abs/2023ApJ...954..157S} {954, 157}

\bibitem[\protect\citeauthoryear{{Shirazi} \& {Brinchmann}}{{Shirazi} \& {Brinchmann}}{2012}]{Shirazi2012}
{Shirazi} M.,  {Brinchmann} J.,  2012, \mn@doi [\mnras] {10.1111/j.1365-2966.2012.20439.x}, \href {https://ui.adsabs.harvard.edu/abs/2012MNRAS.421.1043S} {421, 1043}

\bibitem[\protect\citeauthoryear{{Shivaei} et~al.,}{{Shivaei} et~al.}{2025}]{Shivaei2025}
{Shivaei} I.,  et~al., 2025, arXiv e-prints, \href {https://ui.adsabs.harvard.edu/abs/2025arXiv250901795S} {p. arXiv:2509.01795}

\bibitem[\protect\citeauthoryear{{Simmonds} et~al.,}{{Simmonds} et~al.}{2024}]{Simmonds2024b}
{Simmonds} C.,  et~al., 2024, \mn@doi [\mnras] {10.1093/mnras/stae2537}, \href {https://ui.adsabs.harvard.edu/abs/2024MNRAS.535.2998S} {535, 2998}

\bibitem[\protect\citeauthoryear{{Simmonds} et~al.,}{{Simmonds} et~al.}{2025}]{Simmonds2025}
{Simmonds} C.,  et~al., 2025, arXiv e-prints, \href {https://ui.adsabs.harvard.edu/abs/2025arXiv250804410S} {p. arXiv:2508.04410}

\bibitem[\protect\citeauthoryear{{Stacy}, {Bromm}  \& {Lee}}{{Stacy} et~al.}{2016}]{Stacy2016}
{Stacy} A.,  {Bromm} V.,   {Lee} A.~T.,  2016, \mn@doi [\mnras] {10.1093/mnras/stw1728}, \href {https://ui.adsabs.harvard.edu/abs/2016MNRAS.462.1307S} {462, 1307}

\bibitem[\protect\citeauthoryear{{Stanton} et~al.,}{{Stanton} et~al.}{2026}]{Stanton2026}
{Stanton} T.~M.,  et~al., 2026, \mn@doi [\mnras] {10.1093/mnras/stag449}, \href {https://ui.adsabs.harvard.edu/abs/2026MNRAS.547ag449S} {547, stag449}

\bibitem[\protect\citeauthoryear{{Stanway} \& {Eldridge}}{{Stanway} \& {Eldridge}}{2018}]{Stanway2018}
{Stanway} E.~R.,  {Eldridge} J.~J.,  2018, \mn@doi [\mnras] {10.1093/mnras/sty1353}, \href {https://ui.adsabs.harvard.edu/abs/2018MNRAS.479...75S} {479, 75}

\bibitem[\protect\citeauthoryear{{Storey} \& {Hummer}}{{Storey} \& {Hummer}}{1995}]{Storey1995}
{Storey} P.~J.,  {Hummer} D.~G.,  1995, \mn@doi [\mnras] {10.1093/mnras/272.1.41}, \href {https://ui.adsabs.harvard.edu/abs/1995MNRAS.272...41S} {272, 41}

\bibitem[\protect\citeauthoryear{{Sun} et~al.,}{{Sun} et~al.}{2025}]{Sun2025b}
{Sun} F.,  et~al., 2025, \mn@doi [arXiv e-prints] {10.48550/arXiv.2503.15587}, \href {https://ui.adsabs.harvard.edu/abs/2025arXiv250315587S} {p. arXiv:2503.15587}

\bibitem[\protect\citeauthoryear{{Tacchella} et~al.,}{{Tacchella} et~al.}{2022}]{Tacchella2022}
{Tacchella} S.,  et~al., 2022, \mn@doi [\apj] {10.3847/1538-4357/ac4cad}, \href {https://ui.adsabs.harvard.edu/abs/2022ApJ...927..170T} {927, 170}

\bibitem[\protect\citeauthoryear{{Tacchella} et~al.,}{{Tacchella} et~al.}{2023}]{Tacchella2023a}
{Tacchella} S.,  et~al., 2023, \mn@doi [\mnras] {10.1093/mnras/stad1408}, \href {https://ui.adsabs.harvard.edu/abs/2023MNRAS.522.6236T} {522, 6236}

\bibitem[\protect\citeauthoryear{{Tang} et~al.,}{{Tang} et~al.}{2026}]{Tang2026}
{Tang} M.,  et~al., 2026, \mn@doi [arXiv e-prints] {10.48550/arXiv.2604.03563}, \href {https://ui.adsabs.harvard.edu/abs/2026arXiv260403563T} {p. arXiv:2604.03563}

\bibitem[\protect\citeauthoryear{{Tayal} \& {Zatsarinny}}{{Tayal} \& {Zatsarinny}}{2010}]{Tayal2010}
{Tayal} S.~S.,  {Zatsarinny} O.,  2010, \mn@doi [\apjs] {10.1088/0067-0049/188/1/32}, \href {https://ui.adsabs.harvard.edu/abs/2010ApJS..188...32T} {188, 32}

\bibitem[\protect\citeauthoryear{{Topping} et~al.,}{{Topping} et~al.}{2024}]{Topping2024a}
{Topping} M.~W.,  et~al., 2024, \mn@doi [\mnras] {10.1093/mnras/stae682}, \href {https://ui.adsabs.harvard.edu/abs/2024MNRAS.529.3301T} {529, 3301}

\bibitem[\protect\citeauthoryear{{Topping} et~al.,}{{Topping} et~al.}{2025}]{Topping2025b}
{Topping} M.~W.,  et~al., 2025, \mn@doi [\mnras] {10.1093/mnras/staf903}, \href {https://ui.adsabs.harvard.edu/abs/2025MNRAS.541.1707T} {541, 1707}

\bibitem[\protect\citeauthoryear{{Torralba} et~al.,}{{Torralba} et~al.}{2026}]{Torralba2026}
{Torralba} A.,  et~al., 2026, \mn@doi [\aap] {10.1051/0004-6361/202555596}, \href {https://ui.adsabs.harvard.edu/abs/2026A&A...705A.147T} {705, A147}

\bibitem[\protect\citeauthoryear{{Torrey} et~al.,}{{Torrey} et~al.}{2019}]{Torrey2019}
{Torrey} P.,  et~al., 2019, \mn@doi [\mnras] {10.1093/mnras/stz243}, \href {https://ui.adsabs.harvard.edu/abs/2019MNRAS.484.5587T} {484, 5587}

\bibitem[\protect\citeauthoryear{{Tremonti} et~al.,}{{Tremonti} et~al.}{2004}]{Tremonti2004}
{Tremonti} C.~A.,  et~al., 2004, \mn@doi [\apj] {10.1086/423264}, \href {https://ui.adsabs.harvard.edu/abs/2004ApJ...613..898T} {613, 898}

\bibitem[\protect\citeauthoryear{{Trussler} et~al.,}{{Trussler} et~al.}{2026}]{Trussler2026}
{Trussler} J. A.~A.,  et~al., 2026, \mn@doi [arXiv e-prints] {10.48550/arXiv.2603.15761}, \href {https://ui.adsabs.harvard.edu/abs/2026arXiv260315761T} {p. arXiv:2603.15761}

\bibitem[\protect\citeauthoryear{{{\"U}bler} et~al.,}{{{\"U}bler} et~al.}{2026}]{Ubler2026}
{{\"U}bler} H.,  et~al., 2026, \mn@doi [arXiv e-prints] {10.48550/arXiv.2603.20360}, \href {https://ui.adsabs.harvard.edu/abs/2026arXiv260320360U} {p. arXiv:2603.20360}

\bibitem[\protect\citeauthoryear{{Ucci} et~al.,}{{Ucci} et~al.}{2023}]{Ucci2023}
{Ucci} G.,  et~al., 2023, \mn@doi [\mnras] {10.1093/mnras/stac2654}, \href {https://ui.adsabs.harvard.edu/abs/2023MNRAS.518.3557U} {518, 3557}

\bibitem[\protect\citeauthoryear{{Vanzella} et~al.,}{{Vanzella} et~al.}{2023}]{Vanzella2023}
{Vanzella} E.,  et~al., 2023, \mn@doi [\aap] {10.1051/0004-6361/202346981}, \href {https://ui.adsabs.harvard.edu/abs/2023A&A...678A.173V} {678, A173}

\bibitem[\protect\citeauthoryear{{Vanzella} et~al.,}{{Vanzella} et~al.}{2024}]{Vanzella2024}
{Vanzella} E.,  et~al., 2024, \mn@doi [\aap] {10.1051/0004-6361/202451696}, \href {https://ui.adsabs.harvard.edu/abs/2024A&A...691A.251V} {691, A251}

\bibitem[\protect\citeauthoryear{{Vanzella} et~al.,}{{Vanzella} et~al.}{2025}]{Vanzella2025}
{Vanzella} E.,  et~al., 2025, \mn@doi [arXiv e-prints] {10.48550/arXiv.2509.07073}, \href {https://ui.adsabs.harvard.edu/abs/2025arXiv250907073V} {p. arXiv:2509.07073}

\bibitem[\protect\citeauthoryear{{Veilleux} \& {Osterbrock}}{{Veilleux} \& {Osterbrock}}{1987}]{VO87}
{Veilleux} S.,  {Osterbrock} D.~E.,  1987, \mn@doi [\apjs] {10.1086/191166}, \href {https://ui.adsabs.harvard.edu/abs/1987ApJS...63..295V} {63, 295}

\bibitem[\protect\citeauthoryear{{Wang} et~al.,}{{Wang} et~al.}{2023}]{wang2023}
{Wang} B.,  et~al., 2023, \mn@doi [\apjl] {10.3847/2041-8213/acba99}, \href {https://ui.adsabs.harvard.edu/abs/2023ApJ...944L..58W} {944, L58}

\bibitem[\protect\citeauthoryear{{Whitaker} et~al.,}{{Whitaker} et~al.}{2019}]{Whitaker2019}
{Whitaker} K.~E.,  et~al., 2019, \mn@doi [\apjs] {10.3847/1538-4365/ab3853}, \href {https://ui.adsabs.harvard.edu/abs/2019ApJS..244...16W} {244, 16}

\bibitem[\protect\citeauthoryear{{Williams} et~al.,}{{Williams} et~al.}{2023}]{Williams2023}
{Williams} C.~C.,  et~al., 2023, \mn@doi [\apjs] {10.3847/1538-4365/acf130}, \href {https://ui.adsabs.harvard.edu/abs/2023ApJS..268...64W} {268, 64}

\bibitem[\protect\citeauthoryear{{Williams} et~al.,}{{Williams} et~al.}{2025}]{Williams2025}
{Williams} C.~C.,  et~al., 2025, \mn@doi [\apj] {10.3847/1538-4357/ad97bc}, \href {https://ui.adsabs.harvard.edu/abs/2025ApJ...979..140W} {979, 140}

\bibitem[\protect\citeauthoryear{{Willott} et~al.,}{{Willott} et~al.}{2025}]{Willott2025}
{Willott} C.~J.,  et~al., 2025, \mn@doi [\apj] {10.3847/1538-4357/addf49}, \href {https://ui.adsabs.harvard.edu/abs/2025ApJ...988...26W} {988, 26}

\bibitem[\protect\citeauthoryear{{Witten} et~al.,}{{Witten} et~al.}{2025}]{Witten2025}
{Witten} C.,  et~al., 2025, \mn@doi [\mnras] {10.1093/mnras/staf001}, \href {https://ui.adsabs.harvard.edu/abs/2025MNRAS.537..112W} {537, 112}

\bibitem[\protect\citeauthoryear{{Yanagisawa} et~al.,}{{Yanagisawa} et~al.}{2024}]{Yanagisawa2024b}
{Yanagisawa} H.,  et~al., 2024, \mn@doi [\apj] {10.3847/1538-4357/ad7097}, \href {https://ui.adsabs.harvard.edu/abs/2024ApJ...974..180Y} {974, 180}

\bibitem[\protect\citeauthoryear{{Zaritsky}, {Kennicutt}  \& {Huchra}}{{Zaritsky} et~al.}{1994}]{Zaritsky1994}
{Zaritsky} D.,  {Kennicutt} Jr. R.~C.,   {Huchra} J.~P.,  1994, \mn@doi [\apj] {10.1086/173544}, \href {https://ui.adsabs.harvard.edu/abs/1994ApJ...420...87Z} {420, 87}

\bibitem[\protect\citeauthoryear{{Zucchi}, {Ji}, {Madau}, {Maiolino}, {Juod{\v{z}}balis}, {D'Eugenio}, {Geris}  \& {Isobe}}{{Zucchi} et~al.}{2026}]{Zucchi2026}
{Zucchi} G.,  {Ji} X.,  {Madau} P.,  {Maiolino} R.,  {Juod{\v{z}}balis} I.,  {D'Eugenio} F.,  {Geris} S.,   {Isobe} Y.,  2026, \mn@doi [\aap] {10.1051/0004-6361/202557687}, \href {https://ui.adsabs.harvard.edu/abs/2026A&A...707A..52Z} {707, A52}

\bibitem[\protect\citeauthoryear{{de Graaff} et~al.,}{{de Graaff} et~al.}{2024}]{deGraaff2024}
{de Graaff} A.,  et~al., 2024, \mn@doi [\aap] {10.1051/0004-6361/202347755}, \href {https://ui.adsabs.harvard.edu/abs/2024A&A...684A..87D} {684, A87}

\makeatother
\end{thebibliography}




\appendix
\section{Properties of the stacked spectra} \label{apsec:stkprop}

Section \ref{subsec:subsamp} constructs the subsamples used for our stacks, whose fundamental properties are listed in Table \ref{tab:fund}.
We present the measurements of the emission line fluxes and the nebular properties of the stacks in Section \ref{subsec:emis}, which are summarised in Table \ref{tab:emis}.
Section \ref{subsec:stkidx} describes strong-line indices of the stacks used for our new metallicity calibrations, which are shown in Table \ref{tab:idxstk}.
Note that we present the coefficients of our calibrations in the main text (Table \ref{tab:calib}).

\begin{table*}
	\centering
	\caption{Fundamental properties of our stacks. The value with the errors represents the median value with the 16th-84th percentile range of the stacked sources. The \metsl\ values for the $Z$-bin stacks (the left three columns) are based on \citet{Cataldi2025}'s calibration (see Section \ref{subsec:subsamp}). We use 1488 unique sources for these stacks.}
	\label{tab:fund}
	\begin{tabular}{lccccccc}
		\hline
        & \multicolumn{3}{c}{\metsl} & \multicolumn{4}{c}{$\log(M_{*}/M_{\odot})$}\\
        Property & $<$7.0 & 7.0--7.3 & 7.3--7.5 & $<$7.5 & 7.5--8.0 & 8.0--8.5 & 8.5--9.0\\
        \hline
        \# of spectra & 31 & 105 & 135 & 133 & 322 & 517 & 482\\
        $z$ & $5.30^{+0.74}_{-2.28}$ & $4.15^{+1.83}_{-1.74}$ & $3.97^{+1.92}_{-1.11}$ & $3.65^{+1.93}_{-1.66}$ & $3.61^{+2.38}_{-1.60}$ & $3.47^{+2.47}_{-1.51}$ & $3.28^{+1.66}_{-1.25}$\\
        $\log(M_{*}/M_{\odot})$ & $7.47^{+0.64}_{-0.33}$ & $7.68^{+0.51}_{-0.40}$ & $7.93^{+0.56}_{-0.44}$ & $7.29^{+0.16}_{-0.24}$ & $7.80^{+0.14}_{-0.19}$ & $8.26^{+0.17}_{-0.16}$ & $8.73^{+0.18}_{-0.15}$\\
        $\log(\mathrm{SFR}_{\mathrm{H\alpha}}/M_{\odot}\,\mathrm{yr^{-1}})$ & $-0.32^{+0.37}_{-0.36}$ & $-0.30^{+0.39}_{-0.33}$ & $-0.10^{+0.36}_{-0.42}$ & $-0.45^{+0.24}_{-0.27}$ & $-0.15^{+0.28}_{-0.40}$ & $-0.05^{+0.46}_{-0.51}$ & $-0.04^{+0.51}_{-0.44}$\\
        $\log(\mathrm{SFR_{20}/SFR_{100}})$ & $0.27^{+0.13}_{-0.30}$ & $0.31^{+0.11}_{-0.28}$ & $0.28^{+0.14}_{-0.44}$ & $0.33^{+0.10}_{-0.10}$ & $0.30^{+0.11}_{-0.18}$ & $0.16^{+0.21}_{-0.31}$ & $-0.07^{+0.29}_{-0.28}$\\
        Valid $\lambda_{\mathrm{rest}}$ range (\AA) & 2917--7500 & 2775--7500 & 2418--7500 & 3155--7500 & 2498--7500 & 2231--7500 & 2154--7500\\
        \hline
	\end{tabular}
\end{table*}

\begin{table*}
	\centering
	\caption{Emission line fluxes and nebular properties of our stacks. The fluxes are not corrected for dust extinction but normalised by H$\beta$. The upper limits are $3\sigma$.}
	\label{tab:emis}
	\begin{tabular}{lccccccc}
		\hline
        & \multicolumn{3}{c}{\metsl} & \multicolumn{4}{c}{$\log(M_{*}/M_{\odot})$}\\
        Property & $<$7.0 & 7.0--7.3 & 7.3--7.5 & $<$7.5 & 7.5--8.0 & 8.0--8.5 & 8.5--9.0\\
        \hline
        [\oii]$\lambda\lambda$3726,3729 & $<$19.9 & $21.5\pm2.9$ & $44.0\pm2.6$ & $33.8\pm3.2$ & $53.2\pm1.7$ & $93.9\pm1.7$ & $175.1\pm2.3$\\
        
        [\neiii]$\lambda$3869 & $25.2\pm4.1$ & $27.4\pm2.1$ & $36.3\pm1.6$ & $34.0\pm2.2$ & $40.5\pm1.2$ & $42.6\pm0.9$ & $45.0\pm1.2$\\
        
        H$\delta$ & $15.9\pm3.4$ & $21.5\pm2.0$ & $21.5\pm1.3$ & $20.2\pm2.0$ & $22.1\pm1.0$ & $23.8\pm0.8$ & $19.8\pm1.0$\\
        
        H$\gamma$ & $32.2\pm3.0$ & $38.5\pm1.9$ & $43.0\pm1.5$ & $43.9\pm1.8$ & $41.5\pm1.0$ & $42.5\pm0.8$ & $41.5\pm1.1$\\
        
        [\oiii]$\lambda$4363 & $9.2\pm2.9$ & $10.6\pm1.5$ & $15.1\pm1.2$ & $13.5\pm1.5$ & $10.7\pm0.7$ & $12.1\pm0.6$ & $8.5\pm0.8$\\
        
        \heii\,$\lambda$4686 & $<$8.2 & $<$3.7 & $<$2.1 & $<$3.6 & $2.1\pm0.6$ & $1.5\pm0.5$ & $<$1.9\\
        
        H$\beta$ & $100.0\pm3.7$ & $100.0\pm2.1$ & $100.0\pm1.4$ & $100.0\pm1.9$ & $100.0\pm1.0$ & $100.0\pm0.9$ & $100.0\pm1.2$\\

        [\oiii]$\lambda$4959 & $90.2\pm3.7$ & $119.0\pm2.3$ & $174.0\pm1.6$ & $140.5\pm1.9$ & $172.5\pm1.2$ & $197.3\pm1.0$ & $193.0\pm1.2$\\
        
        [\oiii]$\lambda$5007 & $232.1\pm4.6$ & $364.3\pm3.4$ & $501.1\pm2.4$ & $406.1\pm2.8$ & $518.5\pm2.1$ & $579.9\pm2.0$ & $561.3\pm2.1$\\
        
        H$\alpha$ & $255.0\pm5.3$ & $259.4\pm2.7$ & $278.0\pm2.2$ & $267.9\pm2.5$ & $288.7\pm1.8$ & $304.9\pm1.8$ & $317.2\pm1.9$\\
        
        [\nii]$\lambda$6583 & $<$8.0 & $<$3.7 & $3.5\pm0.9$ & $<$3.3 & $5.4\pm0.6$ & $7.2\pm0.5$ & $11.4\pm0.5$\\
        
        [\sii]$\lambda$6716 & $<$7.5 & $5.8\pm1.3$ & $6.2\pm0.9$ & $6.6\pm1.0$ & $6.1\pm0.6$ & $8.8\pm0.5$ & $19.6\pm0.7$\\
        
        [\sii]$\lambda$6731 & $<$8.9 & $<$4.0 & $<$2.7 & $3.3\pm1.1$ & $3.6\pm0.6$ & $5.9\pm0.5$ & $12.9\pm0.6$\\
        
        [\ariii]$\lambda$7135 & $<$7.5 & $<$4.1 & $2.9\pm0.8$ & $<$3.1 & $3.6\pm0.5$ & $4.6\pm0.4$ & $6.2\pm0.6$\\
        
        $E(B-V)$ & $0.04^{+0.14}_{-0.04}$ & $0.01^{+0.07}_{-0.01}$ & $0.03^{+0.03}_{-0.03}$ & $0.00^{+0.03}_{-0.00}$ & $0.07^{+0.03}_{-0.04}$ & $0.11^{+0.02}_{-0.02}$ & $0.14^{+0.02}_{-0.02}$\\
        
        $T_{\mathrm{e}}$[\oiii]\,($10^{4}$\,K) & $2.28^{+0.49}_{-0.52}$ & $1.85^{+0.17}_{-0.15}$ & $1.91^{+0.08}_{-0.09}$ & $2.00^{+0.15}_{-0.13}$ & $1.58^{+0.05}_{-0.05}$ & $1.61^{+0.04}_{-0.04}$ & $1.40^{+0.06}_{-0.06}$\\
        
        $\log(n_{\mathrm{e}}[\text{\textsc{S\,ii}}]/\mathrm{cm^{-3}})$ & $\cdots$ & $<$0.60 & $<$0.60 & $0.60^{+1.67}_{-0.00}$ & $0.60^{+1.22}_{-0.00}$ & $0.60^{+1.45}_{-0.00}$ & $0.60^{+0.64}_{-0.00}$\\
        
        \mette & $6.99^{+0.28}_{-0.16}$ & $7.40^{+0.08}_{-0.08}$ & $7.53^{+0.04}_{-0.04}$ & $7.39^{+0.06}_{-0.06}$ & $7.73^{+0.04}_{-0.04}$ & $7.78^{+0.03}_{-0.02}$ & $7.96^{+0.05}_{-0.04}$\\
        \hline
	\end{tabular}
\end{table*}

\begin{table*}
	\centering
	\caption{Strong-line indices of our stacks. The upper and lower limits are $3\sigma$.}
	\label{tab:idxstk}
	\begin{tabular}{lccccccc}
		\hline
        & \multicolumn{3}{c}{\metsl} & \multicolumn{4}{c}{$\log(M_{*}/M_{\odot})$}\\
        Property & $<$7.0 & 7.0--7.3 & 7.3--7.5 & $<$7.5 & 7.5--8.0 & 8.0--8.5 & 8.5--9.0\\
        \hline
        log(R3) & $0.36^{+0.02}_{-0.02}$ & $0.56^{+0.01}_{-0.01}$ & $0.70^{+0.01}_{-0.01}$ & $0.61^{+0.01}_{-0.01}$ & $0.71^{+0.00}_{-0.00}$ & $0.76^{+0.00}_{-0.00}$ & $0.74^{+0.01}_{-0.01}$\\
        log(R2) & $<-0.73$ & $-0.67^{+0.06}_{-0.06}$ & $-0.38^{+0.03}_{-0.03}$ & $-0.47^{+0.04}_{-0.04}$ & $-0.32^{+0.01}_{-0.01}$ & $-0.09^{+0.01}_{-0.01}$ & $0.16^{+0.01}_{-0.01}$\\
        log(R23) & $0.51^{+0.04}_{-0.02}$ & $0.70^{+0.01}_{-0.01}$ & $0.85^{+0.01}_{-0.01}$ & $0.76^{+0.01}_{-0.01}$ & $0.87^{+0.00}_{-0.00}$ & $0.93^{+0.00}_{-0.00}$ & $0.95^{+0.01}_{-0.01}$\\
        log(O32) & $>1.09$ & $1.23^{+0.05}_{-0.06}$ & $1.07^{+0.02}_{-0.03}$ & $1.08^{+0.04}_{-0.04}$ & $1.03^{+0.01}_{-0.01}$ & $0.85^{+0.01}_{-0.01}$ & $0.58^{+0.01}_{-0.01}$\\
        $\hat{\mathrm{R}}_{\mathrm{Laseter}}$ & $<-0.02$ & $0.18^{+0.03}_{-0.03}$ & $0.44^{+0.01}_{-0.01}$ & $0.31^{+0.02}_{-0.02}$ & $0.48^{+0.01}_{-0.01}$ & $0.62^{+0.01}_{-0.01}$ & $0.73^{+0.01}_{-0.01}$\\
        $\hat{\mathrm{R}}_{\mathrm{Chakraborty}}$ & $<0.23$ & $0.43^{+0.01}_{-0.02}$ & $0.62^{+0.01}_{-0.01}$ & $0.51^{+0.01}_{-0.01}$ & $0.64^{+0.01}_{-0.01}$ & $0.73^{+0.00}_{-0.00}$ & $0.76^{+0.01}_{-0.01}$\\
        $\tilde{\mathrm{R}}_{\mathrm{Cataldi}}$ & $<-0.01$ & $0.19^{+0.03}_{-0.03}$ & $0.44^{+0.01}_{-0.01}$ & $0.32^{+0.02}_{-0.02}$ & $0.48^{+0.01}_{-0.01}$ & $0.62^{+0.01}_{-0.01}$ & $0.73^{+0.01}_{-0.01}$\\
        log(Ne3) & $-0.10^{+0.07}_{-0.09}$ & $-0.15^{+0.04}_{-0.04}$ & $-0.07^{+0.02}_{-0.02}$ & $-0.11^{+0.03}_{-0.03}$ & $0.00^{+0.02}_{-0.02}$ & $0.02^{+0.01}_{-0.01}$ & $0.06^{+0.02}_{-0.02}$\\
        log(RO2Ne3) & $0.20^{+0.28}_{-0.14}$ & $0.35^{+0.05}_{-0.06}$ & $0.56^{+0.03}_{-0.03}$ & $0.52^{+0.05}_{-0.05}$ & $0.59^{+0.02}_{-0.02}$ & $0.69^{+0.01}_{-0.02}$ & $0.95^{+0.02}_{-0.02}$\\
        $\widehat{\mathrm{RNe}}$ & $<-0.20$ & $-0.25^{+0.04}_{-0.05}$ & $-0.07^{+0.02}_{-0.03}$ & $-0.15^{+0.03}_{-0.04}$ & $0.03^{+0.02}_{-0.02}$ & $0.14^{+0.01}_{-0.01}$ & $0.29^{+0.02}_{-0.02}$\\
        log(Ne3O2) & $>0.15$ & $0.11^{+0.06}_{-0.07}$ & $-0.05^{+0.03}_{-0.03}$ & $0.01^{+0.05}_{-0.05}$ & $-0.05^{+0.02}_{-0.02}$ & $-0.23^{+0.01}_{-0.01}$ & $-0.45^{+0.01}_{-0.01}$\\
        log(Ar3) & $<-1.54$ & $<-1.81$ & $-1.98^{+0.10}_{-0.13}$ & $<-1.93$ & $-1.91^{+0.06}_{-0.07}$ & $-1.83^{+0.04}_{-0.04}$ & $-1.72^{+0.04}_{-0.04}$\\
        log(N2) & $<-1.50$ & $<-1.85$ & $-1.90^{+0.10}_{-0.13}$ & $<-1.91$ & $-1.73^{+0.05}_{-0.05}$ & $-1.62^{+0.03}_{-0.03}$ & $-1.45^{+0.02}_{-0.02}$\\
        log(O3N2) & $>1.87$ & $>2.41$ & $2.60^{+0.10}_{-0.13}$ & $>2.52$ & $2.44^{+0.05}_{-0.05}$ & $2.38^{+0.03}_{-0.03}$ & $2.19^{+0.02}_{-0.02}$\\
        log(S2) & $<-1.20$ & $-1.65^{+0.28}_{-0.11}$ & $-1.66^{+0.20}_{-0.07}$ & $-1.43^{+0.06}_{-0.07}$ & $-1.48^{+0.04}_{-0.04}$ & $-1.34^{+0.02}_{-0.02}$ & $-1.01^{+0.01}_{-0.01}$\\
        log(O3S2) & $>1.56$ & $2.21^{+0.09}_{-1.08}$ & $2.36^{+0.06}_{-0.38}$ & $2.04^{+0.06}_{-0.07}$ & $2.19^{+0.04}_{-0.04}$ & $2.09^{+0.02}_{-0.02}$ & $1.75^{+0.01}_{-0.01}$\\
        \hline
	\end{tabular}
\end{table*}

\section{Systematics of metallicity calibrations onto MZR} \label{apsec:difslope}
\begin{figure}
	\centering
    \includegraphics[width=\columnwidth]{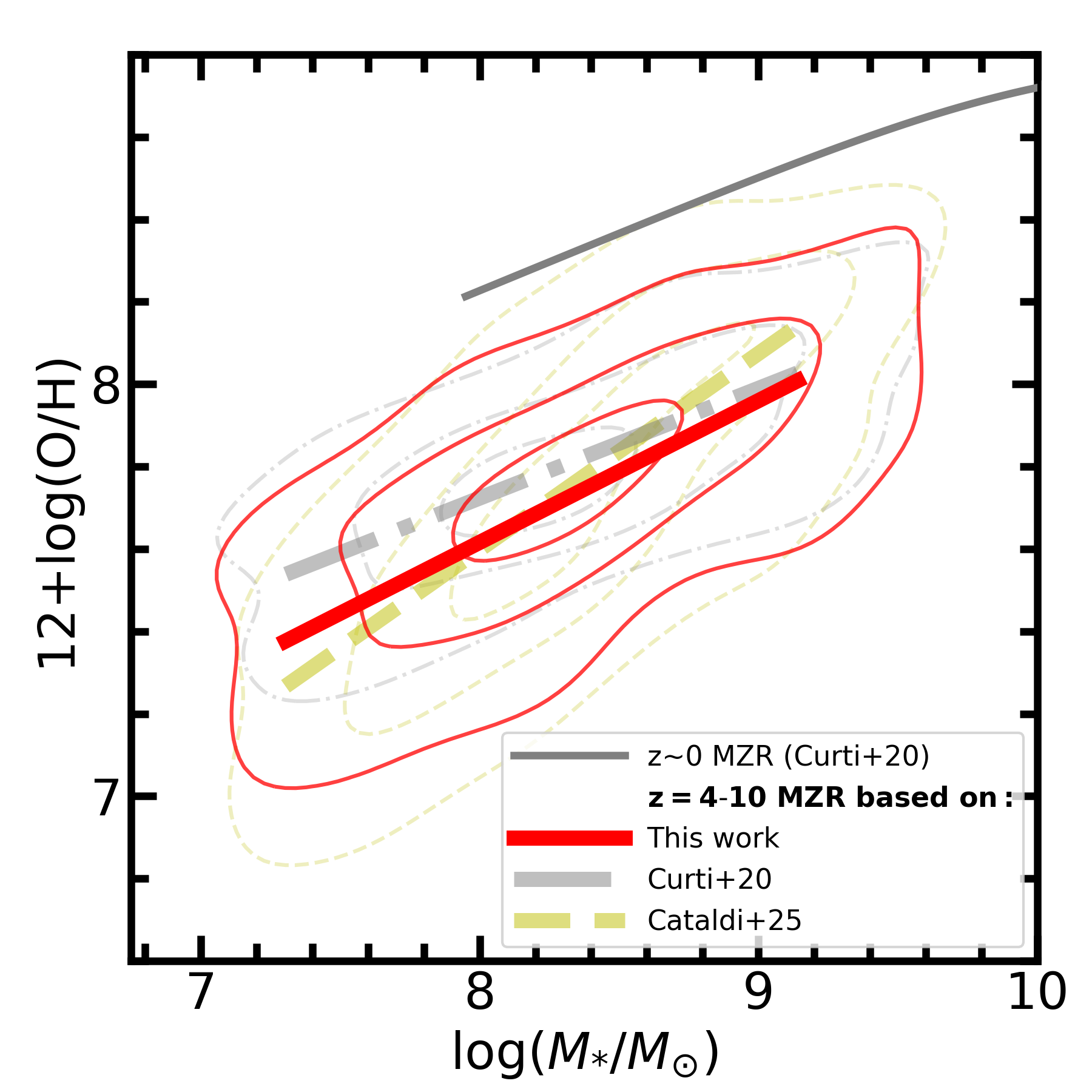}
    \caption{MZRs at $z=4$--10 using our stack-based calibrations (red solid line), those based on high-$z$ individual sources \citep[yellow dashed line;][]{Cataldi2025}, and those based on $z\sim0$ galaxies \citep[grey dashdot line;][]{Curti2020}, together with the $z\sim0$ MZR \citep{Curti2020}.
    }
    \label{fig:mzrcomp}
\end{figure}

Figure \ref{fig:mzrcomp} compares $z=4$--10 MZR slopes based on our calibrations and the calibrations based on individual auroral-line emitters at $z\sim0$ \citep{Curti2020} and high-$z$ \citep{Cataldi2025}.
The slope based on our calibrations ($\beta=0.34^{+0.14}_{-0.18}$) is slightly steeper than that based on \citet{Curti2020}'s calibrations (0.27) within 1$\sigma$ but flatter than that based on \citet{Cataldi2025}'s calibrations (0.48).
We confirm that \citet{Sanders2025}'s calibrations provide a similarly steep MZR slope to \citet{Cataldi2025}'s calibration.
This comparison highlights the impact of choosing different metallicity calibrations on the measurement of the MZR slope (see also Appendix \ref{apsec:mzrref}).

\section{Properties of EMPG candidates} \label{apsec:empgprop}
\begin{table*}
	\centering
	\caption{Properties of our \nempg\ EMPG candidates. The \metsl\ values are based on our calibration. Flag `M' (`P'): Selected from the R1000 (R100) data. Flag `MP': Selected from both R1000 and R100 data, whose reported spectroscopic measurements are based on the R1000 data. $^{\dagger}$: EMP LRDs (Section \ref{subsec:lrd}).}
	\label{tab:empg}
	\begin{tabular}{lccccccc}
		\hline
        ID & $z$ & \metsl & $E(B-V)$ & $\log(M_{*}/M_{\odot})$ & $\log({\mathrm{SFR_{H\alpha}}}/M_{\odot}\,\mathrm{yr^{-1}})$ & $\log(\mathrm{SFR_{20}/SFR_{100}})$ & Flag\\
        \hline
goods-s-ultradeep\_127079 & 5.262 & $6.72^{+0.13}_{-0.13}$ & $0.00\pm0.29$ & $7.06^{+0.38}_{-0.26}$ & $-0.82\pm0.09$ & $0.06^{+0.36}_{-0.65}$ & P\\
goods-s-mediumjwst\_10011541 & 7.433 & $6.80^{+0.08}_{-0.08}$ & $\cdots$ & $7.85^{+0.27}_{-0.14}$ & $0.11\pm0.06$ & $0.24^{+0.31}_{-0.37}$ & M\\
goods-s-mediumjwst\_56812 & 5.760 & $6.95^{+0.11}_{-0.11}$ & $0.00\pm0.06$ & $7.09^{+0.21}_{-0.18}$ & $-0.19\pm0.08$ & $0.34^{+0.25}_{-0.27}$ & MP\\
goods-s-deepjwst\_30010919 & 7.610 & $6.95^{+0.12}_{-0.11}$ & $\cdots$ & $7.24^{+0.39}_{-0.22}$ & $-0.12\pm0.09$ & $0.28^{+0.32}_{-0.51}$ & P\\
goods-s-mediumhst\_13561 & 4.138 & $6.99^{+0.14}_{-0.14}$ & $0.00\pm0.62$ & $8.10^{+0.28}_{-0.36}$ & $0.11\pm0.11$ & $0.08^{+0.40}_{-0.41}$ & P\\
goods-s-mediumjwst\_32880 & 5.609 & $6.99^{+0.10}_{-0.09}$ & $0.00\pm0.35$ & $7.61^{+0.13}_{-0.14}$ & $0.04\pm0.07$ & $0.41^{+0.18}_{-0.15}$ & MP\\
goods-n-mediumjwst\_5088 & 6.992 & $7.01^{+0.07}_{-0.06}$ & $0.00\pm0.61$ & $7.75^{+0.21}_{-0.17}$ & $0.46\pm0.04$ & $0.37^{+0.26}_{-0.26}$ & P\\
goods-s-mediumjwst\_20057765 & 8.913 & $7.03^{+0.12}_{-0.11}$ & $\cdots$ & $7.73^{+0.27}_{-0.15}$ & $0.45\pm0.09$ & $0.37^{+0.23}_{-0.35}$ & P\\
goods-n-mediumjwst\_10001897 & 5.298 & $7.04^{+0.08}_{-0.08}$ & $0.00\pm0.30$ & $8.42^{+0.20}_{-0.33}$ & $0.05\pm0.06$ & $-0.43^{+0.49}_{-0.71}$ & MP\\
goods-s-deephst\_10001892 & 4.772 & $7.05^{+0.22}_{-0.18}$ & $0.00\pm0.26$ & $6.72^{+0.37}_{-0.25}$ & $-1.01\pm0.14$ & $0.14^{+0.66}_{-1.07}$ & P\\
goods-s-mediumjwst\_59998 & 5.888 & $7.05^{+0.12}_{-0.11}$ & $0.00\pm0.70$ & $7.52^{+0.11}_{-0.09}$ & $0.07\pm0.08$ & $0.42^{+0.13}_{-0.15}$ & P\\
goods-s-mediumjwst\_193756 & 5.905 & $7.05^{+0.21}_{-0.17}$ & $0.69\pm0.00$ & $7.69^{+0.29}_{-0.18}$ & $0.56\pm0.13$ & $0.30^{+0.35}_{-0.42}$ & MP\\
goods-s-ultradeep\_20198852 & 8.267 & $7.07^{+0.10}_{-0.09}$ & $0.00\pm0.28$ & $7.50^{+0.19}_{-0.15}$ & $-0.33\pm0.07$ & $0.31^{+0.26}_{-0.29}$ & MP\\
goods-s-deepjwst\_72574 & 4.891 & $7.07^{+0.11}_{-0.10}$ & $0.04\pm0.00$ & $7.19^{+0.19}_{-0.18}$ & $-0.46\pm0.08$ & $0.21^{+0.23}_{-0.25}$ & P\\
goods-s-mediumjwst\_20083087 & 9.060 & $7.08^{+0.18}_{-0.15}$ & $\cdots$ & $8.50^{+0.24}_{-0.38}$ & $0.47\pm0.12$ & $-0.22^{+0.64}_{-0.76}$ & P\\
goods-s-mediumjwst\_43481 & 4.423 & $7.08^{+0.09}_{-0.09}$ & $0.00\pm0.69$ & $7.39^{+0.33}_{-0.27}$ & $-0.09\pm0.06$ & $0.28^{+0.26}_{-0.37}$ & P\\
goods-s-mediumhst\_13639 & 5.789 & $7.08^{+0.18}_{-0.14}$ & $0.00\pm0.23$ & $7.95^{+0.25}_{-0.21}$ & $0.07\pm0.11$ & $0.21^{+0.27}_{-0.39}$ & P\\
goods-s-mediumjwst\_186158 & 4.368 & $7.09^{+0.19}_{-0.15}$ & $0.00\pm0.16$ & $7.23^{+0.26}_{-0.14}$ & $-0.36\pm0.12$ & $0.35^{+0.18}_{-0.22}$ & P\\
darkhorse\_174741 & 3.199 & $7.10^{+0.17}_{-0.15}$ & $0.00\pm0.28$ & $7.79^{+0.16}_{-0.19}$ & $-0.32\pm0.11$ & $0.31^{+0.16}_{-0.22}$ & M\\
goods-s-mediumjwst\_30149608 & 7.035 & $7.10^{+0.17}_{-0.14}$ & $0.00\pm0.20$ & $7.28^{+0.19}_{-0.13}$ & $-0.03\pm0.11$ & $0.39^{+0.22}_{-0.27}$ & M\\
goods-s-ultradeep\_100227 & 4.431 & $7.10^{+0.11}_{-0.09}$ & $0.08\pm0.00$ & $7.37^{+0.25}_{-0.14}$ & $-0.60\pm0.07$ & $0.29^{+0.15}_{-0.27}$ & P\\
goods-s-ultradeep\_113056 & 5.984 & $7.11^{+0.06}_{-0.05}$ & $0.23\pm0.04$ & $7.55^{+0.21}_{-0.18}$ & $0.06\pm0.02$ & $0.40^{+0.34}_{-0.36}$ & MP\\
darkhorse\_168163 & 3.332 & $7.12^{+0.07}_{-0.06}$ & $0.00\pm0.11$ & $7.26^{+0.32}_{-0.19}$ & $-0.35\pm0.04$ & $0.26^{+0.22}_{-0.26}$ & M\\
goods-s-mediumjwst\_30142058 & 7.270 & $7.13^{+0.16}_{-0.13}$ & $\cdots$ & $7.84^{+0.34}_{-0.22}$ & $0.31\pm0.10$ & $0.46^{+0.51}_{-0.52}$ & P\\
goods-s-mediumjwst\_20192042 & 8.836 & $7.15^{+0.15}_{-0.13}$ & $\cdots$ & $8.22^{+0.16}_{-0.20}$ & $0.54\pm0.10$ & $0.19^{+0.26}_{-0.25}$ & P\\
oasis\_212846 & 5.049 & $7.15^{+0.19}_{-0.14}$ & $0.16\pm0.00$ & $7.07^{+0.35}_{-0.24}$ & $-0.51\pm0.03$ & $0.23^{+0.30}_{-0.42}$ & P\\
goods-s-mediumjwst\_10011141 & 1.244 & $7.15^{+0.08}_{-0.07}$ & $0.09\pm0.00$ & $6.87^{+0.39}_{-0.36}$ & $-0.84\pm0.04$ & $0.32^{+0.27}_{-0.33}$ & M\\
darkhorse\_345635 & 5.470 & $7.16^{+0.16}_{-0.13}$ & $0.10\pm0.00$ & $7.47^{+0.34}_{-0.29}$ & $-0.29\pm0.09$ & $0.23^{+0.43}_{-0.46}$ & M\\
goods-s-mediumjwst\_267649 & 6.575 & $7.16^{+0.17}_{-0.13}$ & $0.00\pm0.39$ & $8.11^{+0.29}_{-0.18}$ & $0.06\pm0.10$ & $0.20^{+0.33}_{-0.48}$ & P\\
goods-s-mediumjwst\_135134 & 7.432 & $7.17^{+0.19}_{-0.14}$ & $\cdots$ & $8.16^{+0.36}_{-0.34}$ & $0.04\pm0.10$ & $0.37^{+0.69}_{-0.66}$ & MP\\
goods-s-ultradeep\_30080593 & 5.960 & $7.17^{+0.09}_{-0.07}$ & $0.00\pm0.15$ & $7.26^{+0.20}_{-0.16}$ & $-0.29\pm0.05$ & $0.33^{+0.18}_{-0.24}$ & M\\
goods-s-ultradeep\_201906 & 5.517 & $7.19^{+0.08}_{-0.07}$ & $0.00\pm0.02$ & $7.20^{+0.20}_{-0.15}$ & $-0.37\pm0.04$ & $0.44^{+0.29}_{-0.30}$ & P\\
goods-n-mediumhst\_45979 & 2.583 & $7.19^{+0.13}_{-0.10}$ & $0.04\pm0.04$ & $7.29^{+0.29}_{-0.22}$ & $-0.23\pm0.07$ & $0.33^{+0.22}_{-0.29}$ & M\\
goods-s-mediumjwst\_51871 & 5.779 & $7.19^{+0.14}_{-0.11}$ & $0.00\pm0.24$ & $7.53^{+0.23}_{-0.21}$ & $-0.36\pm0.08$ & $0.16^{+0.30}_{-0.39}$ & M\\
goods-s-mediumjwst\_90155 & 5.577 & $7.19^{+0.10}_{-0.09}$ & $0.00\pm0.28$ & $7.27^{+0.07}_{-0.05}$ & $0.10\pm0.06$ & $0.50^{+0.07}_{-0.08}$ & MP\\
goods-s-mediumjwst\_182875 & 3.425 & $7.19^{+0.15}_{-0.12}$ & $0.00\pm0.03$ & $7.48^{+0.11}_{-0.12}$ & $-0.30\pm0.08$ & $0.38^{+0.13}_{-0.11}$ & M\\
goods-s-deepjwst\_20050575 & 8.414 & $7.19^{+0.08}_{-0.07}$ & $\cdots$ & $7.42^{+0.15}_{-0.11}$ & $-0.05\pm0.04$ & $0.41^{+0.17}_{-0.20}$ & P\\
goods-s-deepjwst\_72975 & 5.587 & $7.19^{+0.21}_{-0.17}$ & $0.13\pm0.00$ & $6.66^{+0.17}_{-0.15}$ & $-0.67\pm0.12$ & $0.39^{+0.20}_{-0.23}$ & P\\
goods-s-ultradeep\_199524 & 5.942 & $7.19^{+0.15}_{-0.12}$ & $0.00\pm0.03$ & $7.29^{+0.38}_{-0.23}$ & $-0.76\pm0.09$ & $0.15^{+0.34}_{-0.52}$ & P\\
darkhorse\_377126 & 6.055 & $7.20^{+0.21}_{-0.17}$ & $0.00\pm0.24$ & $7.05^{+0.16}_{-0.13}$ & $-0.30\pm0.10$ & $0.38^{+0.18}_{-0.21}$ & M\\
goods-s-mediumjwst\_73690$^{\dagger}$ & 5.497 & $7.22^{+0.10}_{-0.08}$ & $0.50\pm0.00$ & $9.08^{+0.14}_{-0.15}$ & $0.87\pm0.05$ & $0.51^{+0.27}_{-0.26}$ & P\\
goods-s-mediumjwst\_90864 & 7.146 & $7.23^{+0.10}_{-0.09}$ & $0.00\pm0.32$ & $7.73^{+0.18}_{-0.16}$ & $0.16\pm0.05$ & $0.43^{+0.25}_{-0.27}$ & P\\
goods-n-mediumjwst\_59788 & 5.381 & $7.23^{+0.21}_{-0.16}$ & $0.15\pm0.00$ & $7.83^{+0.19}_{-0.17}$ & $0.00\pm0.11$ & $0.31^{+0.22}_{-0.29}$ & MP\\
darkhorse\_29724 & 4.053 & $7.24^{+0.13}_{-0.11}$ & $0.08\pm0.00$ & $7.14^{+0.20}_{-0.12}$ & $-0.43\pm0.07$ & $0.39^{+0.12}_{-0.20}$ & M\\
goods-s-ultradeep\_109389 & 5.786 & $7.26^{+0.08}_{-0.06}$ & $0.00\pm0.08$ & $7.58^{+0.26}_{-0.22}$ & $-0.34\pm0.03$ & $0.22^{+0.25}_{-0.34}$ & MP\\
goods-s-deephst\_10056849 & 5.814 & $7.27^{+0.12}_{-0.11}$ & $0.00\pm0.53$ & $7.58^{+0.12}_{-0.10}$ & $-0.00\pm0.06$ & $0.42^{+0.12}_{-0.13}$ & M\\
goods-n-mediumjwst\_39353$^{\dagger}$ & 4.846 & $7.28^{+0.10}_{-0.09}$ & $0.55\pm0.00$ & $9.53^{+0.09}_{-0.09}$ & $0.97\pm0.04$ & $0.48^{+0.20}_{-0.20}$ & P\\
goods-s-mediumjwst\_171875 & 5.738 & $7.28^{+0.11}_{-0.09}$ & $0.46\pm0.00$ & $8.06^{+0.11}_{-0.11}$ & $0.82\pm0.05$ & $0.49^{+0.18}_{-0.17}$ & MP\\
goods-s-deephst\_10005217 & 4.887 & $7.28^{+0.04}_{-0.04}$ & $0.00\pm0.05$ & $8.40^{+0.28}_{-0.24}$ & $0.21\pm0.01$ & $0.18^{+0.41}_{-0.41}$ & P\\
goods-s-deephst\_10005447 & 6.621 & $7.30^{+0.18}_{-0.13}$ & $0.00\pm0.04$ & $7.18^{+0.15}_{-0.11}$ & $-0.36\pm0.08$ & $0.38^{+0.21}_{-0.21}$ & P\\
        \hline
	\end{tabular}
\end{table*}

\begin{table*}
	\centering
	\caption{Properties of the EMPG candidates from the literature (Section \ref{subsec:litempg}). $^{\mathrm{a}}$: LRD. $^{\mathrm{b}}$: Metallicity based on our Ne3 calibration. $^{\mathrm{c}}$: $T_{\mathrm{e}}$-based metallicity. The other metallicities are based on our R3 calibration. $^{\mathrm{d}}$: Dynamical mass as an upper limit on $M_{*}$. $^{\mathrm{e}}$: Derived from the reported Balmer line fluxes. References: 1): \citet{Cai2025}, 2): \citet{Ivey2026}, 3): \citet{Cai2026}, 4): \citet{Vanzella2025}, 5): \citet{Morishita2025}, 6): \citet{Caputi2026}, 7): \citet{Vanzella2024}, 8): \citet{Nakajima2025}, 9): \citet{Vanzella2023}, 10): \citet{Maiolino2025}; \citet{Juodzbalis2026}, 11): \citet{Willott2025}, 12): \citet{Ubler2026}; \citet{Maiolino2026}; \citet{Rusta2026}, 13): \citet{Hsiao2025}, 14): \citet{Chemerynska2024}; \citet{Price2025}, 15): \citet{Fujimoto2025}; \citet{Fujimoto2025b}, 16): \citet{Asada2026}, 17): \citet{Koller2026}, 18): \citet{Cullen2025}, 19): \citet{Mowla2024}}
	\label{tab:empglit}
	\begin{tabular}{lcccccc}
		\hline
        Name & $z$ & log(R3) & \met & $\log(M_{*}/M_{\odot})$ & $\log({\mathrm{SFR_{H\alpha}}}/M_{\odot}\,\mathrm{yr^{-1}})$ & Ref.\\
        \hline
MPG-CR3 & 3.19 & $<0.12$ & $<6.82$ & $5.79$ & $-0.04^{+0.05}_{-0.05}$$^{\mathrm{e}}$ & 1)\\
The Cliff$^{\mathrm{a}}$ & 3.55 & $0.20^{+0.07}_{-0.08}$ & $6.89^{+0.07}_{-0.08}$ & $<8.41$$^{\mathrm{d}}$ & $\cdots$ & 2)\\
CAPERS-39810 & 3.65 & $0.19^{+0.12}_{-0.18}$ & $6.89^{+0.13}_{-0.18}$ & $8.02^{+0.22}_{-0.34}$ & $0.22^{+0.01}_{-0.01}$$^{\mathrm{e}}$ & 3)\\
LAP2 & 4.19 & $<0.11$ & $<6.80$ & $<4.30$ & $-2.49^{+0.07}_{-0.09}$$^{\mathrm{e}}$ & 4)\\
AMORE6-A+B & 5.73 & $<-0.47$ & $<6.21$ & $5.60^{+0.20}_{-0.10}$ & $-0.63^{+0.07}_{-0.08}$ & 5)\\
Pseudo-LRD-NOM & 5.96 & $<-0.42$ & $<6.26$ & $8.68^{+0.72}_{-0.72}$ & $>0.25$ & 6)\\
T2c & 6.15 & $0.35^{+0.08}_{-0.10}$ & $7.05^{+0.09}_{-0.11}$ & $<4.30$ & $-1.52^{+0.06}_{-0.07}$$^{\mathrm{e}}$ & 7)\\
LAP1-B & 6.62 & $-0.16^{+0.15}_{-0.23}$ & $6.52^{+0.15}_{-0.23}$ & $<3.43$ & $-2.23^{+0.05}_{-0.06}$$^{\mathrm{e}}$ & 8)\\
LAP1 & 6.64 & $-0.26^{+0.10}_{-0.13}$ & $6.43^{+0.10}_{-0.13}$ & $<4.00$ & $-1.79^{+0.03}_{-0.04}$$^{\mathrm{e}}$ & 9)\\
QSO1(central)$^{\mathrm{a}}$ & 7.04 & $-0.26^{+0.11}_{-0.15}$ & $6.43^{+0.11}_{-0.15}$ & $<7.30$$^{\mathrm{d}}$ & $\cdots$ & 10)\\
QSO1(200pc)$^{\mathrm{a}}$ & 7.04 & $<-0.39$ & $<6.30$ & $<7.30$$^{\mathrm{d}}$ & $\cdots$ & 10)\\
CANUCS-A370-z8-LAE & 8.20 & $0.12^{+0.05}_{-0.06}$ & $6.81^{+0.05}_{-0.06}$ & $7.67^{+0.21}_{-0.27}$ & $-0.04^{+0.05}_{-0.05}$ & 11)\\
Hebe & 10.60 & $\cdots$ & $<7.0$$^{\mathrm{b}}$ & 4.30--5.78 & $0.31^{+0.07}_{-0.08}$$^{\mathrm{e}}$ & 12)\\
SAPPHIRES-1449 & 5.92 & $0.29^{+0.10}_{-0.12}$ & $6.98^{+0.10}_{-0.13}$ & $7.44^{+0.19}_{-0.18}$ & $0.78^{+0.09}_{-0.11}$$^{\mathrm{e}}$ & 13)\\
SAPPHIRES-7695 & 5.80 & $-0.08^{+0.12}_{-0.16}$ & $6.61^{+0.12}_{-0.16}$ & $7.20^{+0.11}_{-0.15}$ & $0.64^{+0.09}_{-0.11}$$^{\mathrm{e}}$ & 13)\\
SAPPHIRES-9770 & 6.29 & $0.31^{+0.12}_{-0.16}$ & $7.01^{+0.13}_{-0.17}$ & $7.14^{+0.32}_{-0.37}$ & $0.67^{+0.11}_{-0.15}$$^{\mathrm{e}}$ & 13)\\
SAPPHIRES-20130 & 6.79 & $0.32^{+0.08}_{-0.09}$ & $7.02^{+0.08}_{-0.09}$ & $7.76^{+0.19}_{-0.18}$ & $0.92^{+0.07}_{-0.08}$$^{\mathrm{e}}$ & 13)\\
SAPPHIRES-21782 & 5.81 & $0.17^{+0.10}_{-0.13}$ & $6.86^{+0.10}_{-0.13}$ & $6.81^{+0.23}_{-0.22}$ & $0.32^{+0.06}_{-nan}$$^{\mathrm{e}}$ & 13)\\
SAPPHIRES-24804 & 6.67 & $0.11^{+0.11}_{-0.16}$ & $6.81^{+0.12}_{-0.16}$ & $7.49^{+0.20}_{-0.27}$ & $0.64^{+0.10}_{-0.12}$$^{\mathrm{e}}$ & 13)\\
SAPPHIRES-37141 & 5.76 & $0.28^{+0.10}_{-0.13}$ & $6.98^{+0.11}_{-0.14}$ & $7.64^{+0.12}_{-0.14}$ & $0.95^{+0.09}_{-0.11}$$^{\mathrm{e}}$ & 13)\\
UNCOVER-18924 & 7.70 & $0.57^{+0.05}_{-0.06}$ & $7.32^{+0.08}_{-0.08}$ & $5.88^{+0.13}_{-0.08}$ & $-0.57^{+0.05}_{-0.05}$$^{\mathrm{e}}$ & 14)\\
UNCOVER-16155 & 6.87 & $0.55^{+0.03}_{-0.04}$ & $7.29^{+0.05}_{-0.05}$ & $6.61^{+0.07}_{-0.06}$ & $-0.26^{+0.02}_{-0.02}$$^{\mathrm{e}}$ & 14)\\
UNCOVER-23920 & 6.00 & $0.42^{+0.05}_{-0.05}$ & $7.13^{+0.06}_{-0.06}$ & $6.30^{+0.03}_{-0.03}$ & $-0.05^{+0.02}_{-0.03}$$^{\mathrm{e}}$ & 14)\\
UNCOVER-12899 & 6.88 & $>0.36$ & $>7.06$ & $6.54^{+0.14}_{-0.19}$ & $-0.83^{+0.07}_{-0.08}$$^{\mathrm{e}}$ & 14)\\
UNCOVER-8613 & 6.38 & $0.46^{+0.10}_{-0.13}$ & $7.17^{+0.13}_{-0.14}$ & $7.12^{+0.07}_{-0.08}$ & $-0.34^{+0.05}_{-0.05}$$^{\mathrm{e}}$ & 14)\\
UNCOVER-23619 & 6.72 & $0.53^{+0.09}_{-0.11}$ & $7.27^{+0.14}_{-0.14}$ & $6.57^{+0.10}_{-0.06}$ & $-0.05^{+0.06}_{-0.07}$$^{\mathrm{e}}$ & 14)\\
UNCOVER-38335 & 6.23 & $>0.58$ & $>7.34$ & $6.83^{+0.25}_{-0.20}$ & $0.08^{+0.05}_{-0.06}$$^{\mathrm{e}}$ & 14)\\
UNCOVER-27335 & 6.75 & $0.48^{+0.13}_{-0.19}$ & $7.21^{+0.19}_{-0.21}$ & $6.73^{+0.15}_{-0.08}$ & $0.00^{+0.08}_{-0.10}$$^{\mathrm{e}}$ & 14)\\
GLIMPSE-16043 & 6.20 & $0.25^{+0.04}_{-0.04}$ & $6.94^{+0.04}_{-0.05}$ & $5.00$ & $-0.46^{+0.02}_{-0.02}$$^{\mathrm{e}}$ & 15)\\
GLIMPSE-7685 & 6.42 & $0.54^{+0.10}_{-0.13}$ & $7.27^{+0.17}_{-0.17}$ & $6.79^{+0.07}_{-0.07}$ & $0.09^{+0.07}_{-0.07}$ & 16)\\
GLIMPSE-8139 & 6.22 & $>0.83$ & $>7.86$ & $6.80^{+0.08}_{-0.09}$ & $0.07^{+0.04}_{-0.04}$ & 16)\\
GLIMPSE-9081 & 6.22 & $0.77^{+0.08}_{-0.10}$ & $7.86^{+0.00}_{-0.34}$ & $7.00^{+0.04}_{-0.04}$ & $0.18^{+0.04}_{-0.04}$ & 16)\\
GLIMPSE-11789 & 6.52 & $0.42^{+0.08}_{-0.09}$ & $7.13^{+0.10}_{-0.10}$ & $6.84^{+0.06}_{-0.07}$ & $0.36^{+0.08}_{-0.08}$ & 16)\\
GLIMPSE-13186 & 6.16 & $0.19^{+0.09}_{-0.11}$ & $6.88^{+0.09}_{-0.11}$ & $6.40^{+0.18}_{-0.16}$ & $0.13^{+0.05}_{-0.05}$ & 16)\\
GLIMPSE-17070 & 6.03 & $0.36^{+0.08}_{-0.11}$ & $7.06^{+0.10}_{-0.11}$ & $6.34^{+0.06}_{-0.06}$ & $-0.28^{+0.05}_{-0.05}$ & 16)\\
GLIMPSE-19034 & 6.20 & $0.23^{+0.05}_{-0.06}$ & $6.92^{+0.05}_{-0.06}$ & $6.57^{+0.07}_{-0.06}$ & $-0.52^{+0.05}_{-0.05}$ & 16)\\
GLIMPSE-35552 & 6.50 & $0.54^{+0.09}_{-0.11}$ & $7.29^{+0.15}_{-0.14}$ & $6.95^{+0.06}_{-0.05}$ & $-0.38^{+0.09}_{-0.09}$ & 16)\\
GLIMPSE-35957 & 6.11 & $0.67^{+0.06}_{-0.06}$ & $7.51^{+0.17}_{-0.13}$ & $6.51^{+0.06}_{-0.06}$ & $-0.63^{+0.07}_{-0.07}$ & 16)\\
GLIMPSE-37914 & 6.11 & $0.46^{+0.07}_{-0.08}$ & $7.18^{+0.09}_{-0.09}$ & $6.47^{+0.07}_{-0.07}$ & $-0.58^{+0.08}_{-0.08}$ & 16)\\
GLIMPSE-45084 & 6.49 & $0.37^{+0.08}_{-0.09}$ & $7.07^{+0.09}_{-0.10}$ & $6.82^{+0.06}_{-0.06}$ & $-0.05^{+0.05}_{-0.05}$ & 16)\\
GLIMPSE-45706 & 6.42 & $0.25^{+0.07}_{-0.09}$ & $6.94^{+0.07}_{-0.09}$ & $6.16^{+0.08}_{-0.04}$ & $0.48^{+0.05}_{-0.05}$ & 16)\\
GLIMPSE-46431 & 6.52 & $0.35^{+0.13}_{-0.19}$ & $7.05^{+0.15}_{-0.19}$ & $6.28^{+0.20}_{-0.16}$ & $0.42^{+0.07}_{-0.07}$ & 16)\\
GLIMPSE-47757 & 6.00 & $0.41^{+0.03}_{-0.03}$ & $7.11^{+0.03}_{-0.03}$ & $5.62^{+0.06}_{-0.06}$ & $-0.60^{+0.05}_{-0.05}$ & 16)\\
GLIMPSE-54637 & 6.07 & $>0.35$ & $>7.04$ & $6.47^{+0.25}_{-0.17}$ & $-1.62^{+0.24}_{-0.24}$ & 16)\\
GLIMPSE-55241 & 5.50 & $0.70^{+0.01}_{-0.01}$ & $7.58^{+0.02}_{-0.02}$ & $6.76^{+0.09}_{-0.08}$ & $0.07^{+0.08}_{-0.08}$ & 16)\\
SMACS0723\_4590-S1 & 8.45 & $0.41^{+0.05}_{-0.06}$ & $7.11^{+0.07}_{-0.07}$ & $6.77^{+0.05}_{-0.04}$ & $-0.49^{+0.05}_{-0.05}$ & 17)\\
SMACS0723\_4590-S2 & 8.45 & $0.40^{+0.08}_{-0.10}$ & $7.11^{+0.10}_{-0.11}$ & $7.29^{+0.04}_{-0.04}$ & $-0.63^{+0.08}_{-0.08}$ & 17)\\
RX2129\_11027-S1 & 9.51 & $0.50^{+0.07}_{-0.09}$ & $7.23^{+0.11}_{-0.11}$ & $6.83^{+0.19}_{-0.10}$ & $-0.82^{+0.07}_{-0.07}$ & 17)\\
EXCELS-63107 & 8.27 & $0.56^{+0.04}_{-0.04}$ & $6.89^{+0.26}_{-0.21}$$^{\mathrm{c}}$ & $8.57^{+0.32}_{-1.03}$ & $1.11^{+0.03}_{-0.03}$ & 18)\\
Firefly Sparkle & 8.30 & $0.71^{+0.03}_{-0.03}$ & $6.99^{+0.15}_{-0.33}$$^{\mathrm{c}}$ & $7.80^{+0.60}_{-0.30}$ & $-0.16^{+0.03}_{-0.03}$$^{\mathrm{e}}$ & 19)\\
        \hline
	\end{tabular}
\end{table*}

\begin{figure}
	\centering
    \includegraphics[width=\columnwidth]{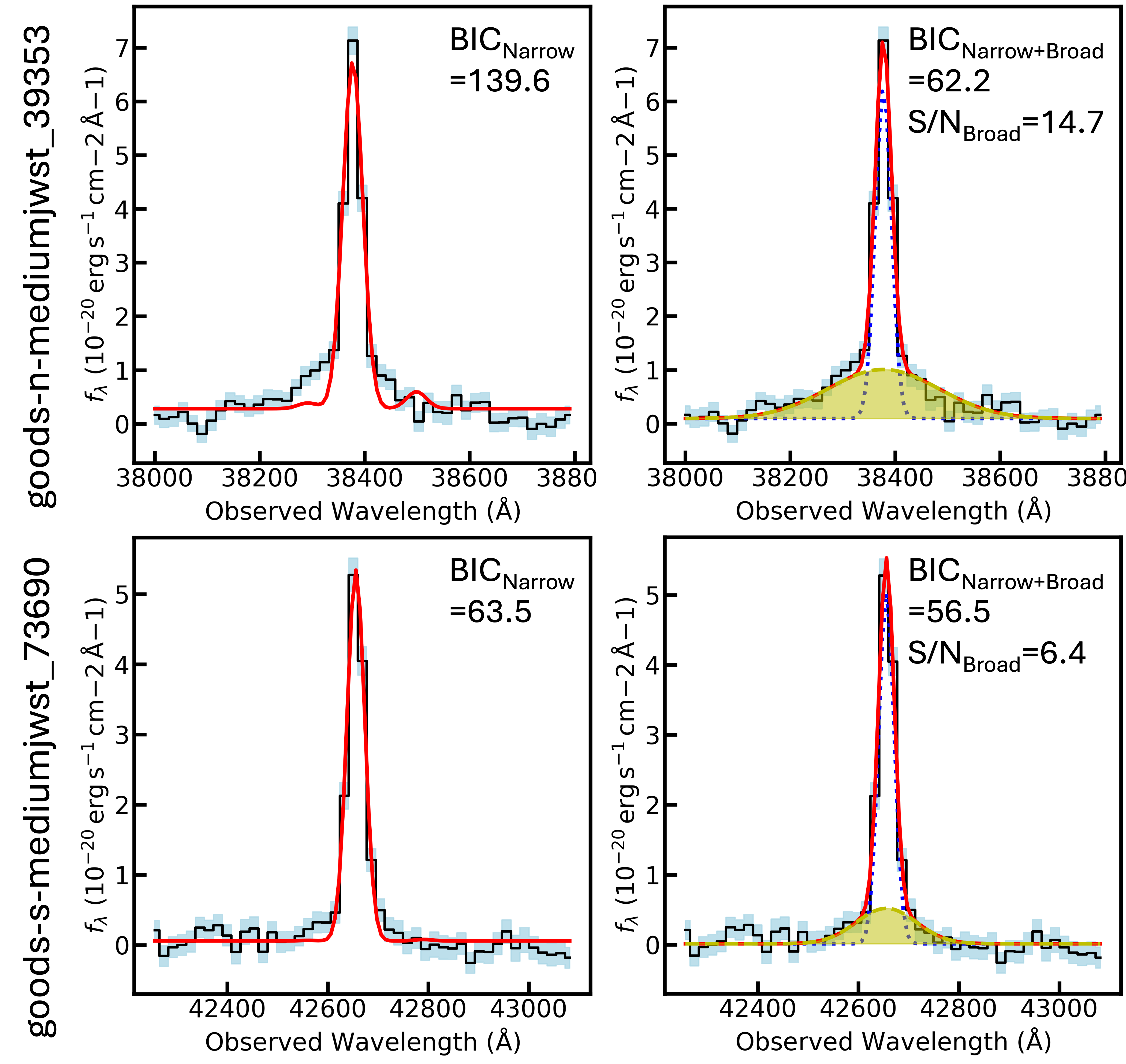}
    \caption{Best-fit results (red curve) of the observed spectra around H$\alpha$ and [\nii]$\lambda\lambda$6548,6583 (black histogram with the lightblue shade) for the two EMP LRDs (Section \ref{subsec:lrd}). (Left) Only narrow components. (Right) A broad component added to H$\alpha$ as shown by the yellow dashed curves with the yellow shades. The blue dotted curves are the narrow components of H$\alpha$. We confirm $\Delta\mathrm{BIC}\equiv\mathrm{BIC_{Narrow}}-\mathrm{BIC_{Narrow+Broad}}=6.9$--77 and $S/N_{\mathrm{Broad}}=6.4$--15, which support the presence of the broad-line components.
    }
    \label{fig:lrd_br}
\end{figure}

In Section \ref{subsec:empgslct}, we select \nempg\ EMPG candidates from our parent sample consisting of the JADES DR4 R100 and R1000 data (Section \ref{subsec:dr4}), the DH R1000 data (Section \ref{subsec:dh}), and the OASIS R100 data (Section \ref{subsec:oasis}).
Table \ref{tab:empg} summarises properties of our EMPG candidates.
Additionally, Section \ref{subsec:litempg} presents our compilation of the EMPG candidates from the literature, whose properties are shown in Table \ref{tab:empglit}.
We emphasise that all the metallicity values listed in these tables are based on our new metallicity calibrations, except for EXCELS-63107 and Firefly Sparkle with auroral line detections.
Note that the photometry of goods-s-ultradeep\_127079 is likely contaminated by a nearby object at lower redshift.
However, this contamination does not affect the spectral analysis using NIRSpec.

Figure \ref{fig:lrd_br} illustrates the fitting results of the two EMP LRDs around H$\alpha$ and [\nii]$\lambda\lambda$6548,6583 (Section \ref{subsec:lrd}).
We confirm $\Delta\mathrm{BIC}\equiv\mathrm{BIC_{Narrow}}-\mathrm{BIC_{Narrow+Broad}}=6.9$--77 and $S/N_{\mathrm{Broad}}=6.4$--15, which satisfy the criteria used by \citet{Juodzbalis2025} to statistically support the presence of H$\alpha$ broadening.

\section{Comparison with MZRs from the literature} \label{apsec:mzrref}

\begin{figure*}
    \centering
    \begin{minipage}{\columnwidth}
      \includegraphics[width=\columnwidth]{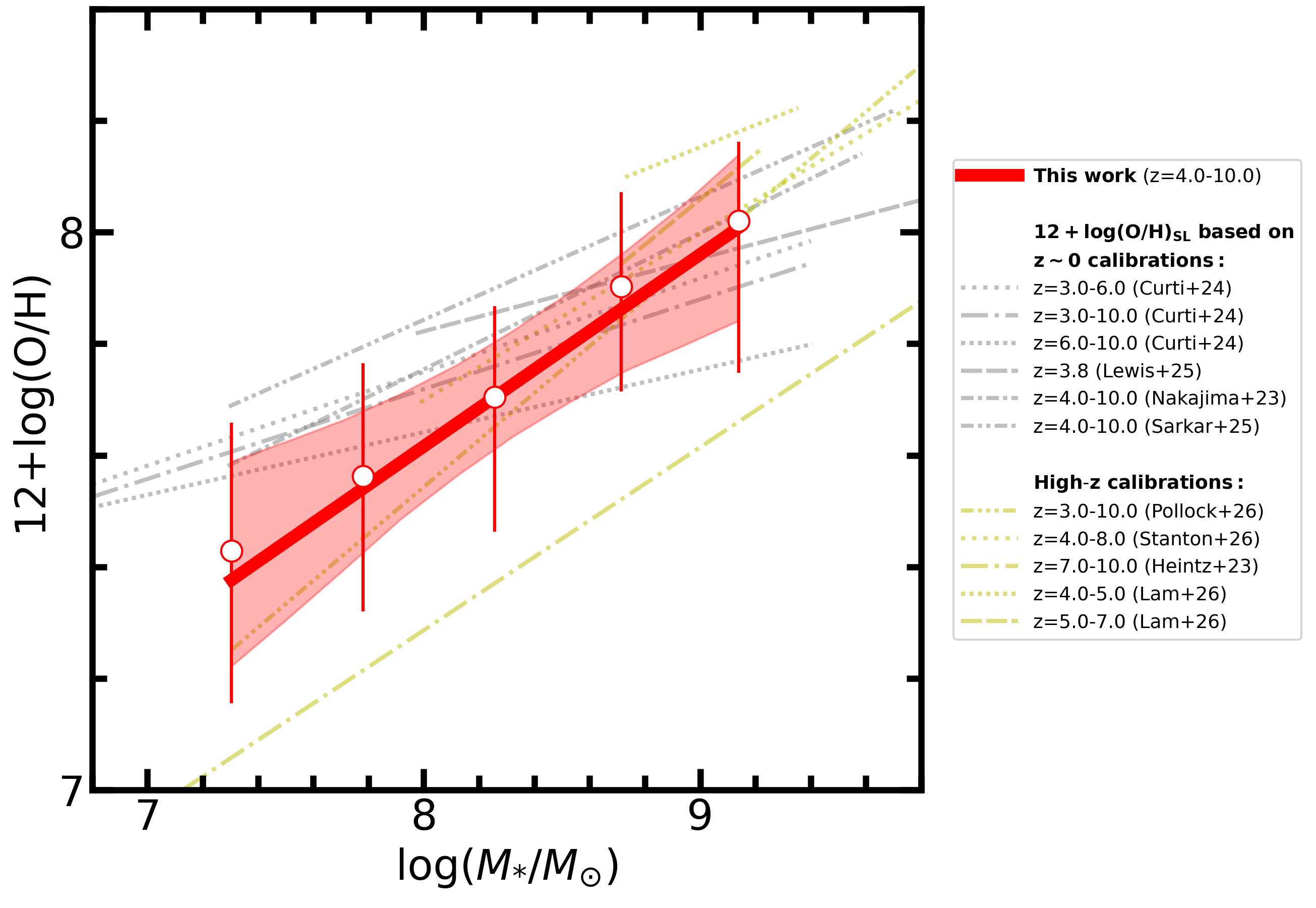}
    \end{minipage}
    \begin{minipage}{\columnwidth}
      \includegraphics[width=\columnwidth]{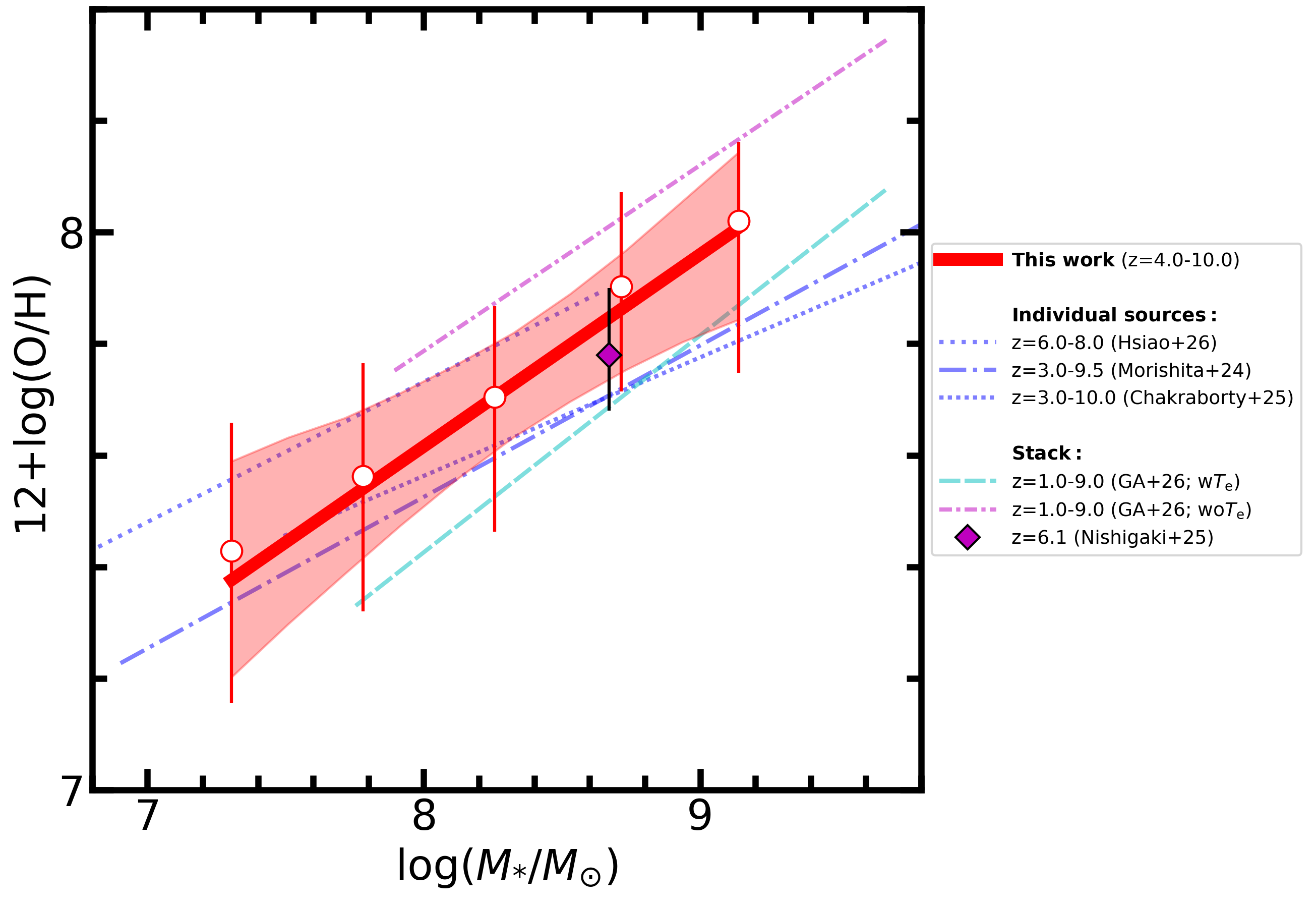}
    \end{minipage}
    \caption{MZR of our JADES+DH sources at $z=$4--10 (red solid line) compared with the literature. (Left) The MZRs whose strong-line metallicities \metsl\ are based on the $z\sim0$ calibrations \citep[grey lines;][]{Curti2024,Lewis2025,Nakajima2023,Sarkar2025} and the high-$z$ calibrations \citep[yellow lines;][see Section \ref{apsec:mzrref} for more details]{Pollock2026,Stanton2026,Heintz2023,Lam2026}. The MZRs based on the $z\sim0$ calibrations are generally shallower than those based on the high-$z$ calibrations. (Right) The $T_{\mathrm{e}}$-based MZRs of individual auroral-line emitters \citep[blue lines;][]{Hsiao2026,Morishita2024,Chakraborty2025} and stacks \citep{Gimenez-Alcazar2026,Nishigaki2025b}. The MZRs of the individual auroral-line emitters tend to show lower metallicities than our MZR.}
    \label{fig:mzr_obs}
\end{figure*}

\begin{figure}
	\centering
    \includegraphics[width=\columnwidth]{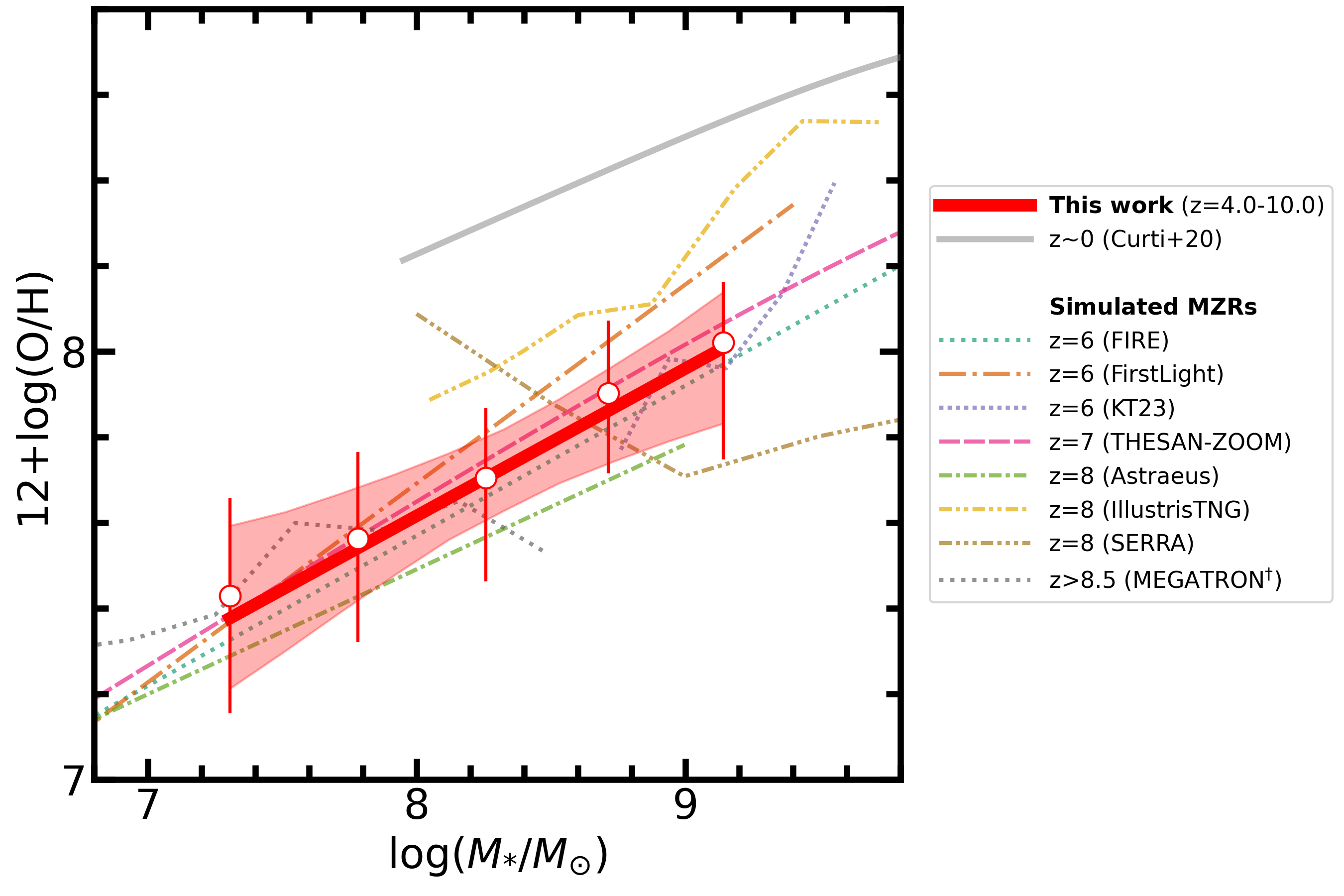}
    \caption{Same as Figure \ref{fig:mzr_obs} but with the simulated MZRs from the literature: FIRE \citep{Ma2016}, FirstLight \citep{Langan2020}, \citeauthor{Kobayashi2023} (\citeyear{Kobayashi2023}; KT23), \textsc{thesan-zoom} \citep{McClymont2025c}, Asteraeus \citep{Ucci2023}, IllustrisTNG \citep{Torrey2019}, SERRA \citep{Pallottini2025}, and MEGATRON \citep{Choustikov2026}. $^{\dagger}$: Bursty Star Formation (SF) model.
    }
    \label{fig:mzr_sim}
\end{figure}

Here we discuss the MZRs from the literature.
We first note that a strict apples-to-apples comparison across the literature is difficult, given the heterogeneity in the datasets, methodologies to estimate metallicity and stellar-mass, and parameterisation schemes adopted by different studies.
To mitigate the impact of these differences, we limit the references on high-$z$ MZRs to those based on NIRSpec data only, and compare them separately for each method of metallicity measurements.

First, we compile the literature whose \metsl\ are mostly derived in a self-consistent manner.
We refer to the MZRs based on the $z\sim0$ calibrations: \citet{Curti2024} based on \citet{Curti2020} and \citet{Laseter2024}' calibrations, \citet{Lewis2025} and \citet{Nakajima2023} based on \citet{Nakajima2023}'s calibrations, and \citet{Sarkar2025} based on \citet{Curti2020}'s calibrations.
We also cite the MZRs mainly based on the high-$z$ calibrations or \citet{Scholte2025}, whose $\mathrm{\hat{R}_{Laseter}}$ calibration quite agrees with the high-$z$ calibrations (see Figure \ref{fig:calib}): \citet{Pollock2026} based on \citet{Sanders2024} and \citet{Laseter2024}'s calibrations \citep[see also][]{Heintz2025}, \citet{Stanton2026} based on \citet{Scholte2025} and \citet{Sanders2024} calibrations, \citet{Heintz2023} based on \citet{Sanders2024}'s calibrations, and \citet{Lam2026} based on \citet{Sanders2025}'s calibrations.

The left panel of Figure \ref{fig:mzr_obs} compares the MZRs based on the $z\sim0$ calibrations and the high-$z$ calibrations.
We find that the MZRs based on the $z\sim0$ calibrations generally exhibit shallower slopes than those based on the high-$z$ calibrations.
This is consistent with the fact that the $z\sim0$ calibrations of \citet{Curti2020} provide a shallower MZR slope than the high-$z$ calibrations of \citet{Cataldi2025} for the same dataset as reported in Appendix \ref{apsec:difslope}.
Among these MZRs, \citet{Pollock2026}'s $z=3$--10 MZR based on \citet{Sanders2024} and \citet{Laseter2024}'s calibrations agrees best with our MZR.

The right panel of Figure \ref{fig:mzr_obs} shows the literature MZRs whose metallicities are based on the direct-$T_{\mathrm{e}}$ method.
\citet{Gimenez-Alcazar2026} report that the stacks of auroral-line emitters (cyan dashed line) exhibit systematically lower metallicities than those without auroral-line detections (magenta dashdot line), suggesting a bias of auroral-line emitters to lower metallicities at a given $M_{*}$.
Indeed, the MZRs of \citet{Morishita2024} and \citet{Chakraborty2025} based on individual auroral-line emitters are generally below our MZR based on our strong-line calibrations.
A similar trend has been observed at $z\sim0$ \citep{Curti2020}, where it is interpreted as a selection effect arising from the preferential detection of [\oiii]$\lambda$4363 in lower-metallicity galaxies at a given $M_{*}$.
However, the MZR reported by \citet{Hsiao2026} based on new observations of the GLIMPSE-D programme towards the lensing field shows marginally higher metallicities than the other $T_{\mathrm{e}}$-based MZRs.
This may indicate that, at the low-$M_{*}$ frontier where the available samples are still small, the derived MZR is somewhat sensitive to sample selection.
Further deep observations will be essential to establish the behaviour of the MZR in this regime more robustly.
It is also worth noting that our MZR is comparable to the stack measurement of \citet{Nishigaki2025b} at $z=6.1$.

In Figure \ref{fig:mzr_sim}, we plot the predicted high-$z$ MZRs of cosmological simulations compiled by \citet{Curti2024}: FIRE \citep{Ma2016}, FirstLight \citep{Langan2020}, the chemo-dynamical simulations of \citet{Kobayashi2023}, Asteraeus \citep{Ucci2023}, IllustrisTNG \citep{Torrey2019}, and SERRA \citep{Pallottini2025}.
Note that the FIRE MZR are rescaled by \citet{Curti2024} to match the $z\sim0$ MZR.
We add recent radiation-hydrodynamic simulations of \textsc{thesan-zoom} \citep{McClymont2025c} and the Bursty Star Formation model of MEGATRON \citep{Choustikov2026}.

Our MZR agrees with most of these simulated MZRs.
In particular, the MZR slopes of \textsc{thesan-zoom} and FIRE are quite comparable to ours.
Our MZR also matches the Bursty Star Formation model of MEGATRON, which predicts systematically lower metallicities than the Efficient Star Formation model.
This model favours stochastic SFHs at the low-$M_{*}$ end of the observations, in line with the large metallicity scatter observed similarly at the low-$M_{*}$ end of the MZR (Section \ref{subsec:empgmzr}).


\bsp	
\label{lastpage}
\end{document}